\documentclass[11pt]{llncs}

\usepackage{llncsdoc}
\usepackage{amsfonts}

\newdimen\proofrulebreadth \proofrulebreadth=.05em
\newdimen\proofdotseparation \proofdotseparation=1.25ex
\newdimen\proofrulebaseline \proofrulebaseline=2ex
\newcount\proofdotnumber \proofdotnumber=3
\let\then\relax
\def\hfi{\hskip0pt plus.0001fil}
\mathchardef\squigto="3A3B
\newif\ifinsideprooftree\insideprooftreefalse
\newif\ifonleftofproofrule\onleftofproofrulefalse
\newif\ifproofdots\proofdotsfalse
\newif\ifdoubleproof\doubleprooffalse
\let\wereinproofbit\relax
\newdimen\shortenproofleft
\newdimen\shortenproofright
\newdimen\proofbelowshift
\newbox\proofabove
\newbox\proofbelow
\newbox\proofrulename
\def\shiftproofbelow{\let\next\relax\afterassignment\setshiftproofbelow\dimen0
}

\def\shiftproofbelowneg{\def\next{\multiply\dimen0 by-1 }%
\afterassignment\setshiftproofbelow\dimen0 }
\def\setshiftproofbelow{\next\proofbelowshift=\dimen0 }
\def\setproofrulebreadth{\proofrulebreadth}

\def\prooftree{
\ifnum  \lastpenalty=1
\then   \unpenalty
\else   \onleftofproofrulefalse
\fi
\ifonleftofproofrule
\else   \ifinsideprooftree
        \then   \hskip.5em plus1fil
        \fi
\fi
\bgroup
\setbox\proofbelow=\hbox{}\setbox\proofrulename=\hbox{}%
\let\justifies\proofover\let\leadsto\proofoverdots\let\Justifies\proofoverdbl
\let\using\proofusing\let\[\prooftree
\ifinsideprooftree\let\]\endprooftree\fi
\proofdotsfalse\doubleprooffalse
\let\thickness\setproofrulebreadth
\let\shiftright\shiftproofbelow \let\shift\shiftproofbelow
\let\shiftleft\shiftproofbelowneg
\let\ifwasinsideprooftree\ifinsideprooftree
\insideprooftreetrue
\setbox\proofabove=\hbox\bgroup$\displaystyle 
\let\wereinproofbit\prooftree
\shortenproofleft=0pt \shortenproofright=0pt \proofbelowshift=0pt
\onleftofproofruletrue\penalty1
}

\def\eproofbit{
\ifx    \wereinproofbit\prooftree
\then   \ifcase \lastpenalty
        \then   \shortenproofright=0pt  
        \or     \unpenalty\hfil         
        \or     \unpenalty\unskip       
        \else   \shortenproofright=0pt  
        \fi
\fi
\global\dimen0=\shortenproofleft
\global\dimen1=\shortenproofright
\global\dimen2=\proofrulebreadth
\global\dimen3=\proofbelowshift
\global\dimen4=\proofdotseparation
$\egroup  
\shortenproofleft=\dimen0
\shortenproofright=\dimen1
\proofrulebreadth=\dimen2
\proofbelowshift=\dimen3
\proofdotseparation=\dimen4
}

\def\proofover{
\eproofbit 
\setbox\proofbelow=\hbox\bgroup 
\let\wereinproofbit\proofover
$\displaystyle
}%
\def\proofoverdbl{
\eproofbit 
\doubleprooftrue
\setbox\proofbelow=\hbox\bgroup 
\let\wereinproofbit\proofoverdbl
$\displaystyle
}%
\def\proofoverdots{
\eproofbit 
\proofdotstrue
\setbox\proofbelow=\hbox\bgroup 
\let\wereinproofbit\proofoverdots
$\displaystyle
}%
\def\proofusing{
\eproofbit 
\setbox\proofrulename=\hbox\bgroup 
\let\wereinproofbit\proofusing
\kern0.3em$
}

\def\endprooftree{
\eproofbit 
  \dimen5 =0pt
\dimen0=\wd\proofabove \advance\dimen0-\shortenproofleft
\advance\dimen0-\shortenproofright
\dimen1=.5\dimen0 \advance\dimen1-.5\wd\proofbelow
\dimen4=\dimen1
\advance\dimen1\proofbelowshift \advance\dimen4-\proofbelowshift
\ifdim  \dimen1<0pt
\then   \advance\shortenproofleft\dimen1
        \advance\dimen0-\dimen1
        \dimen1=0pt
        \ifdim  \shortenproofleft<0pt
        \then   \setbox\proofabove=\hbox{%
                        \kern-\shortenproofleft\unhbox\proofabove}%
                \shortenproofleft=0pt
        \fi
\fi
\ifdim  \dimen4<0pt
\then   \advance\shortenproofright\dimen4
        \advance\dimen0-\dimen4
        \dimen4=0pt
\fi
\ifdim  \shortenproofright<\wd\proofrulename
\then   \shortenproofright=\wd\proofrulename
\fi
\dimen2=\shortenproofleft \advance\dimen2 by\dimen1
\dimen3=\shortenproofright\advance\dimen3 by\dimen4
\ifproofdots
\then
        \dimen6=\shortenproofleft \advance\dimen6 .5\dimen0
        \setbox1=\vbox to\proofdotseparation{\vss\hbox{$\cdot$}\vss}
        \setbox0=\hbox{%
                \kern\dimen6
                $\vcenter to\proofdotnumber\proofdotseparation
                        {\leaders\box1\vfill}$%
                \unhbox\proofrulename}%
\else   \dimen6=\fontdimen22\the\textfont2 
        \dimen7=\dimen6
        \advance\dimen6by.5\proofrulebreadth
        \advance\dimen7by-.5\proofrulebreadth
        \setbox0=\hbox{%
                \kern\shortenproofleft
                \ifdoubleproof
                \then   \hbox to\dimen0{%
                        $\mathsurround0pt\mathord=\mkern-6mu%
                        \cleaders\hbox{$\mkern-2mu=\mkern-2mu$}\hfill
                        \mkern-6mu\mathord=$}%
                \else   \vrule height\dimen6 depth-\dimen7 width\dimen0
                \fi
                \unhbox\proofrulename}%
        \ht0=\dimen6 \dp0=-\dimen7
\fi
\let\doll\relax
\ifwasinsideprooftree
\then   \let\VBOX\vbox
\else   \ifmmode\else$\let\doll=$\fi
        \let\VBOX\vcenter
\fi
\VBOX   {\baselineskip\proofrulebaseline \lineskip.2ex
        \expandafter\lineskiplimit\ifproofdots0ex\else-0.6ex\fi
        \hbox   spread\dimen5   {\hfi\unhbox\proofabove\hfi}%
        \hbox{\box0}%
        \hbox   {\kern\dimen2 \box\proofbelow}}\doll%
\global\dimen2=\dimen2
\global\dimen3=\dimen3
\egroup 
\ifonleftofproofrule
\then   \shortenproofleft=\dimen2
\fi
\shortenproofright=\dimen3
\onleftofproofrulefalse
\ifinsideprooftree
\then   \hskip.5em plus 1fil \penalty2
\fi
}

\ifx\xyloaded\undefined\else \fi
\let\xyloaded=\relax
{\catcode96 12\catcode`\#6\catcode`\.12\catcode`\:12\catcode`\'12\catcode`\@11
\ifx\xywarnifdefined\undefined\else \immediate\write16{}%
 \immediate\write16{Xy-pic Warning: \string\xywarnifdefined\space redefined.}%
 \immediate\write16{}\fi
\gdef\xywarnifdefined#1{\ifx#1\undefined\else \immediate\write16{}%
 \immediate\write16{Xy-pic Warning: `\string#1' redefined.}%
 \immediate\write16{}\fi}
\xywarnifdefined\xydef@ \gdef\xydef@#1{\xywarnifdefined#1\def#1}
\xywarnifdefined\xylet@ \gdef\xylet@#1{\xywarnifdefined#1\let#1}
\xywarnifdefined\xynew@
 \gdef\xynew@#1#2{\xywarnifdefined#2\csname new#1\endcsname#2}}
\xywarnifdefined\xyuncatcodes
\xywarnifdefined\xyreuncatcodes \def\xyreuncatcodes{\edef\xyuncatcodes{%
 \catcode92 0 \catcode123 1 \catcode125 2 \catcode37 14
 \catcode 9 \the\catcode 9 \catcode10 \the\catcode10 \catcode12 \the\catcode12
 \catcode35 \the\catcode35 \catcode36 \the\catcode36 \catcode38 \the\catcode38 
 \catcode43 \the\catcode43 \catcode45 \the\catcode45 \catcode46 \the\catcode46 
 \catcode47 \the\catcode47 
 \catcode60 \the\catcode60 \catcode61 \the\catcode61 \catcode62 \the\catcode62 
 \catcode64 \the\catcode64 \catcode96 \the\catcode96
 \newlinechar \the\newlinechar \endlinechar \the\endlinechar }}
\xyreuncatcodes
\xywarnifdefined\xycatcodes \def\xycatcodes{%
 \catcode 9 10
 \catcode35 6 \catcode 36 3 \catcode 38 4
 \catcode43 12 \catcode 45 12 \catcode 46 12 \catcode 47 12
 \catcode60 12 \catcode 61 12 \catcode 62 12
 \catcode64 11 \catcode 96 12 }
\xycatcodes
{\catcode`\|0 \xywarnifdefined|DOCMODE
\gdef|DOCMODE#1{\ifx(#1\relax \xycatcodes \expandafter\ignorespaces
 \else \skipspecials@ \expandafter\docm@\fi}%
\xywarnifdefined\skipspecials@
\gdef\skipspecials@{%
 \catcode`\\12 \catcode`\{12 \catcode`\}12 \catcode`\#12 \catcode`\%12
 \catcode`\^^L12 \endlinechar`\^^J }%
\catcode`\/=12 \lccode`\/`\\%
\lccode`\D`\D \lccode`\O`\O \lccode`\C`\C \lccode`\M`\M \lccode`\E`\E
\lowercase{%
\xywarnifdefined\docm@ \gdef\docm@{\docm@i}%
\xywarnifdefined\docm@i \gdef\docm@i#1^^J{\docm@ii#1/DOCMODE\docm@iii}%
\xywarnifdefined\docm@ii
 \gdef\docm@ii#1/DOCMODE{\def\next@{#1}\futurelet\next\docm@iii}%
\xywarnifdefined\docm@iii \gdef\docm@iii#1\docm@iii{%
 \ifx\next\docm@iii \let\next\next@ \docecho@ \let\next@\docm@
 \else\ifx\next@\empty \let\next@\docfinish@
 \else \edef\next@{\noexpand\docm@iv\next@/DOCMODE#1\noexpand\docm@iv}%
 \fi\fi \next@}%
\xywarnifdefined\docm@iv
 \gdef\docm@iv#1/DOCMODE\docm@iv{\def\next{#1}\docecho@ \docm@}}%
\xywarnifdefined\docecho@ \global\let\docecho@\relax
\xywarnifdefined\docfinish@ \gdef\docfinish@{\xyuncatcodes|DOCMODE\next}}
\xydef@\xydefcsname@#1{\DN@{#1}\DNii@##1{%
 \ifx ##1\relax\else \xywarning@{`\string##1\string' redefined}\fi \def##1}%
 \expandafter\nextii@\csname\codeof\next@\endcsname}
\xydef@\xyletcsnamecsname@#1#2{\def\1{#1}\def\2{#2}\DN@##1##2{%
 \ifx ##1\relax\else \xywarning@{`\string##1\string' redefined}\fi
 \let##1=##2}%
 \expandafter\expandafter\expandafter\next@
 \expandafter\csname\expandafter\codeof\expandafter\1\expandafter\endcsname
 \csname\codeof\2\endcsname}
\xywarnifdefined\codeof
\xywarnifdefined\codeof@
{\catcode`\:=12
 \gdef\codeof#1{\expandafter\codeof@\meaning#1<-:}
 \gdef\codeof@#1:->#2<-:{#2}}
\xywarnifdefined\addAT@
\xywarnifdefined\addHASH@
\xywarnifdefined\addDOLL@
\xywarnifdefined\addAND@
\xywarnifdefined\addRQ@
\xywarnifdefined\addPLUS@
\xywarnifdefined\addDASH@
\xywarnifdefined\addDOT@
\xywarnifdefined\addLT@
\xywarnifdefined\addEQ@
\xywarnifdefined\addGT@
\xywarnifdefined\addLQ@
\xydef@\xymakeADD@#1#2 #3 {\ifnum\catcode#3=6 \def#1##1{##1#2#2}%
 \else \def#1##1{##1#2}\fi}
\xydef@\xyrecat@{\xymakeADD@\addAT@}
\xydef@\xyrecat{\xyrecat@}
\xydef@\xyresetcatcodes{\def\xyrecat{\xyrecat@}\xyreuncatcodes
\xyuncatcodes
\xyrecat @ 64 \catcode 64 11
\xymakeADD@\addHASH@ # 35
\xymakeADD@\addDOLL@ $ 36
\xymakeADD@\addAND@ & 38
\xymakeADD@\addRQ@ ' 39
\xymakeADD@\addPLUS@ + 43
\xymakeADD@\addDASH@ - 45
\xymakeADD@\addDOT@ . 46
\xymakeADD@\addLT@ < 60
\xymakeADD@\addEQ@ = 61
\xymakeADD@\addGT@ > 62
\xymakeADD@\addLQ@ ` 96
 
 \relax \xyuncatcodes}
\xyuncatcodes \xyresetcatcodes \xycatcodes
\ifx\xyidiomsloaded\empty   \fi
\let\xyidiomsloaded=\empty
\xywarnifdefined\A@ \dimendef\A@=4
\xywarnifdefined\B@ \dimendef\B@=6
\xywarnifdefined\R@ \dimendef\R@=8
\ifx\undefined\AveryUNLIKELYc@ntr@lSEQUENCE@@\else
\errmessage{Xy-pic Error: \string\undefined\space defined.}\fi
\ifx\undefined\literal@ \def\literal@#1{#1}\fi
\ifx\undefined\eat@ \def\eat@#1{}\fi
\xydef@\xyFN@{\futurelet\next}
\ifx\undefined\DN@ \def\DN@{\def\next@}\fi
\ifx\undefined\DNii@ \def\DNii@{\def\nextii@}\fi
\ifx\undefined\setboxz@h\def\setboxz@h{\setbox\z@\hbox}\fi
\ifx\undefined\wdz@ \def\wdz@{\wd\z@}\fi
\ifx\undefined\boxz@ \def\boxz@{\box\z@}\fi
\ifx\undefined\W@ \def\W@{\immediate\write16 }\fi
\ifx\undefined\space@ \def\space@.{\futurelet\space@\relax}\space@. \fi
\ifx\undefined\notempty \def\notempty#1{T\if @#1@F\else T\fi}\fi
\xydef@\xysetup@dummy#1{\xyuncatcodes#1}
\xywarnifdefined\xysetup@@
\ifx\AtEndDocument\undefined
 \expandafter\ifx\csname amsppt.sty\endcsname\relax
 \let\xysetup@@=\xysetup@dummy
 \else
 \def\xysetup@@#1{%
 \expandafter\def\expandafter\topmatter\expandafter{\topmatter
 #1\xyuncatcodes}}\fi
\else
 \def\xysetup@@#1{\AtBeginDocument{#1\xyuncatcodes}}
\fi
\ifx\xysetup@@\xysetup@dummy\else
 \xysetup@@{\let\xysetup@@=\xysetup@dummy \xyuncatcodes}\fi
\xywarnifdefined\xyclosedown@@
\ifx\AtEndDocument\undefined \let\xyclosedown@@=\eat@
\else \def\xyclosedown@@#1{\AtEndDocument{#1}}\fi
\ifx\amstexloaded@\relax
\xylet@\toks@ii=\toks@@   \fi
\ifx\@tempcnta\undefined
\xynew@{count}\count@@
\xynew@{count}\count@@@
\else
\xylet@\count@@=\@tempcnta
\xylet@\count@@@=\@tempcntb
\fi
\ifx\undefined\toks@ii \toksdef\toks@ii=2 \fi

\xydef@\stripRCS$#1${\stripRCS@#1: @@ @@@}
\xydef@\stripRCS@#1: #2@ #3@@@{%
 \ifx @#2\string?\else\ifx :#2\else\stripRCS@@#2\fi\fi}
\xydef@\stripRCS@@#1 #2: @{#1}
 \edef\next{\stripRCS$Revision: 3.2 $}
 \edef\next@{\stripRCS$Locker: $}
\xylet@\xyversion=\next
\def\next{ @}\ifx\next\next@
 \edef\next{\stripRCS$Date: 1995/09/19 18:22:27 $}
\else\edef\xyversion{\xyversion+}
 \edef\next{\number\year/\ifnum\month<10 0\fi\number\month
 /\ifnum\day<10 0\fi\number\day}\fi
\xylet@\xydate=\next
\xydef@\Xygreet@{%
 \W@{}%
 \W@{ Xy-pic version \xyversion\space<\xydate>}%
 \W@{ Copyright (c) 1991-1995 by Kristoffer H. Rose <kris@diku.dk>}%
 \W@{ Xy-pic is free software: see the User\string's Guide for details.}%
 \W@{}}
\Xygreet@
\expandafter\everyjob\expandafter{\the\everyjob\Xygreet@}
\xydef@\Xy{\leavevmode
 \hbox{\kern-.1em X\kern-.3em\lower.4ex\hbox{Y\kern-.15em}}}
\xylet@\XY=\Xy
\xywarnifdefined\thelineno@
\ifx\inputlineno\undefined \edef\thelineno@{\string?}
\else \def\thelineno@{\the\inputlineno}\fi
\xydef@\xytracelineno@{ \string[\jobname:\thelineno@\string]}
\xydef@\xywarning@#1{{\newlinechar=`\^^J%
 \W@{}\W@{Xy-pic Warning: #1\xytracelineno@.}\W@{}}}
\xydef@\xyerror@#1#2{\if\inxy@\xy@{ERROR #1}{}\fi
 {\def\2{#2}\newlinechar=`\^^J%
 \ifx\2\empty \errhelp{See the Xy-pic manual for further information.}%
 \else \errhelp{#2}\fi
 \errmessage{Xy-pic error: #1}}}
\xydef@\xybug@#1{{\newlinechar=`\^^J%
 \errhelp{This is a bug in Xy-pic and should not happen!^^J%
If it did then please send a bug report with the offending Xy-pic code^^J%
to the author of Xy-pic, kris@diku.dk.}%
 \errmessage{Xy-pic BUG: #1 -- notify kris@diku.dk.}}}
\xydef@\xyfont@#1{\ifx#1\undefined \DN@{\global\font#1}\expandafter\next@
 \else \xywarning@{Using previously loaded \string#1\space font}\fi}
\xyfont@\xydashfont=xydash10
\xydef@\xydashl@{\fontdimen6\xydashfont}
\xydef@\xydashh@{\fontdimen5\xydashfont}
\xydef@\xydashw@{\fontdimen8\xydashfont}
\xyfont@\xyatipfont=xyatip10
\xyfont@\xybtipfont=xybtip10
\xyfont@\xybsqlfont=xybsql10
\xydef@\xybsqll@{\fontdimen6\xybsqlfont}
\xydef@\xybsqlh@{\fontdimen5\xybsqlfont}
\xydef@\xybsqlw@{\fontdimen8\xybsqlfont}
\xyfont@\xycircfont=xycirc10
\xynew@{dimen}\X@c
\xynew@{dimen}\Y@c
\xynew@{dimen}\U@c
\xynew@{dimen}\D@c
\xynew@{dimen}\L@c
\xynew@{dimen}\R@c
\xynew@{toks}\Edge@c
\xynew@{dimen}\X@p
\xynew@{dimen}\Y@p
\xynew@{dimen}\U@p
\xynew@{dimen}\D@p
\xynew@{dimen}\L@p
\xynew@{dimen}\R@p
\xynew@{toks}\Edge@p
\xynew@{dimen}\X@origin \X@origin=\z@
\xynew@{dimen}\Y@origin \X@origin=\z@
\xynew@{dimen}\X@xbase \X@xbase=1mm
\xynew@{dimen}\Y@xbase \Y@xbase=\z@
\xynew@{dimen}\X@ybase \X@ybase=\z@
\xynew@{dimen}\Y@ybase \Y@ybase=1mm
\xynew@{dimen}\X@min
\xynew@{dimen}\Y@min
\xynew@{dimen}\X@max
\xynew@{dimen}\Y@max
\xynew@{box}\lastobjectbox@
\xynew@{box}\zerodotbox@
\setbox\zerodotbox@=\hbox{\dimen@=.5\xydashw@
 \kern-\dimen@ \vrule width\xydashw@ height\dimen@ depth\dimen@}
\wd\zerodotbox@=\z@ \ht\zerodotbox@=\z@ \dp\zerodotbox@=\z@
\xynew@{dimen}\almostz@ \almostz@=50sp
\xydef@\zz@#1{\ifdim#1<\z@-\fi#1<\almostz@\relax}
\xynew@{if}\iftmp@
\xynew@{dimen}\d@X
\xynew@{dimen}\d@Y
\xydef@\sd@X{}
\xydef@\sd@Y{}
\xynew@{count}\K@ \K@=1024
\xynew@{count}\KK@ \KK@=32
\xynew@{count}\Direction
\xynew@{dimen}\K@dXdY
\xynew@{dimen}\K@dYdX
\xydef@\cosDirection{}
\xydef@\sinDirection{}
\xywarnifdefined\DirectionChar
\xywarnifdefined\SemiDirectionChar
\xynew@{read}\xyread@
\xynew@{write}\xywrite@
\xynew@{count}\csp@
\xynew@{dimen}\quotPTK@
\xydef@\addtotoks@#1{\toks@=\expandafter{\the\toks@#1}}
\xydef@\xyinputorelse@#1#2{%
 \expandafter\let\expandafter\next@\csname#1loaded\endcsname
 \ifx\next@\empty \let\next@=\relax \else
 \openin\xyread@=#1 \ifeof\xyread@ \DN@{\openin\xyread@=#1.doc
 \ifeof\xyread@ \DN@{#2}\else \DN@{\closein\xyread@\input#1.doc }\fi \next@}%
 \else \DN@{\closein\xyread@\input#1 }\fi\fi \next@}
\global\csp@=\z@
\xydef@\enter@#1{\global\advance\csp@\@ne
 \expandafter\xdef\csname cs@\number\csp@\endcsname{#1}\ignorespaces}
\xydef@\nter@#1{\global\advance\csp@\@ne
 \expandafter\gdef\csname cs@\number\csp@\endcsname{#1}\ignorespaces}
\xydef@\dontleave@{\csname cs@\number\csp@\endcsname}
\xydef@\unenter@{\global\advance\csp@\m@ne}
\xydef@\leave@{\expandafter\unenter@\csname cs@\number\csp@\endcsname}
\quotPTK@=\p@ \divide\quotPTK@\K@
\xylet@\quotsign@@=\empty
\xywarnifdefined\removePT@
{\catcode`p=12 \catcode`t=12 \gdef\removePT@#1pt{#1}}
\xydef@\quotient@#1#2#3{\A@=#2\relax \B@=#3\relax
 \ifdim\A@<\z@\def\quotsign@@{-}\else\def\quotsign@@{+}\fi
 \ifdim\quotsign@@\A@<15pt \multiply\A@\K@
 \else\ifdim\quotsign@@\A@<511pt \multiply\A@\KK@ \divide\B@\KK@
 \else \divide\B@\K@ \fi\fi
 \ifdim\ifdim\B@<\z@-\fi\B@<\quotPTK@ \xywarning@{division overflow}%
 \else \advance\A@.5\B@ \divide\A@\B@ \fi
 \multiply\A@\quotPTK@ \edef#1{\expandafter\removePT@\the\A@}}
\xydef@\quotient@@#1#2#3{\A@=#2\relax \B@=#3\relax
 \multiply\A@\KK@ \divide\B@\KK@ \divide\B@ 8
 \ifdim\B@=\z@\else \advance\A@.5\B@ \divide\A@\B@ \fi
 \B@=.125\quotPTK@ \multiply\A@\B@ \edef#1{\expandafter\removePT@\the\A@}}
\xydef@\loop@#1\repeat@{\def\body@{#1}\iterate@}\xylet@\repeat@=\fi
\xydef@\iterate@{\body@\expandafter\iterate@\else\fi}
\xydef@\xyinitial@#1#2{\DN@{#1}%
 \xyerror@{command used out of context: \codeof\next@}{}}
\xylet@\xy@=\xyinitial@
\xylet@\oxy@=\xy@
\xydef@\inxy@{T\ifx\xy@\xyinitial@ F\else T\fi}
\xydef@\xyxy@@ix@{\begingroup
 \xyuncatcodes\afterassignment\endgroup\global\toks9=}
\xydef@\xy@@{\xy@{}}
\xydef@\plainxy@{\let\xy@=\xyxy@ \let\oxy@=\xy@ \let\xy@@ix@=\xyxy@@ix@}
\xydef@\xy{\ifmmode\expandafter\xymath@\else\expandafter\xynomath@\fi}
\xydef@\xymath@{\hbox\bgroup \dimen@=\the\fontdimen22\textfont\tw@ \xyinside@}
\xydef@\xynomath@{\hbox\bgroup \dimen@=\z@ \xyinside@}
\xydef@\xyinside@{%
 \aftergroup\xycheck@end
 \setboxz@h\bgroup
 \plainxy@
 \X@c=\z@ \Y@c=\z@ \czeroEdge@
 \X@p=\z@ \Y@p=\z@ \U@p=\z@ \D@p=\z@ \L@p=\z@ \R@p=\z@ \Edge@p={\zeroEdge}%
 \X@min=\hsize \X@max=-\hsize \Y@min=\hsize \Y@max=-\hsize
 \expandafter\POS\everyxy@@}
\xydef@\czeroEdge@{\U@c=\z@ \D@c=\U@c \L@c=\U@c \R@c=\U@c \Edge@c={\zeroEdge}}
\xydef@\xyxy@#1#2{#2}
\xywarnifdefined\everyxy
\expandafter\def\addEQ@\everyxy#1{\def\everyxy@@{#1}\ignorespaces}
\xylet@\everyxy@@=\empty
\xydef@\endxy{\if\inxy@\else\xyerror@{Unexpected \string\endxy}{}\fi
 \relax
 \dimen@=\Y@max \advance\dimen@-\Y@min
 \ifdim\dimen@<\z@ \dimen@=\z@ \Y@min=\z@ \Y@max=\z@ \fi
 \dimen@=\X@max \advance\dimen@-\X@min
 \ifdim\dimen@<\z@ \dimen@=\z@ \X@min=\z@ \X@max=\z@ \fi
 \edef\tmp@{\egroup
 \setboxz@h{\kern-\the\X@min\boxz@}%
 \ht\z@=\the\Y@max \dp\z@=-\the\Y@min \wdz@=\the\dimen@
 \noexpand\maybeunraise@ \raise\dimen@\boxz@
 \egroup \noexpand\xy@end
 \U@c=\the\Y@max \D@c=-\the\Y@min \L@c=-\the\X@min \R@c=\the\X@max}\tmp@}
\xydef@\maybeunraise@{\if\inxy@\else \dimen@ii=\dp\z@
 \ifdim\dimen@ii<\z@ \advance\dimen@\dimen@ii \fi\fi}
\xydef@\xycheck@end{\xyFN@\xycheck@end@}
\xydef@\xycheck@end@{\ifx\next\xy@end\DN@\xy@end{}\else\DN@{\xy@end}\fi\next@}
\xydef@\xy@end{%
 \xyerror@{An \string\xy\space environment is not closed correctly.}%
 {I expected \string\endxy. You probably have an umatched {} grouping.}}
\xydef@\POS{\afterPOS{}}
\xydef@\afterPOS#1{%
 \DN@##1{\def\afterPOS@{\def\afterPOS@{##1}#1}}%
 \expandafter\next@\expandafter{\afterPOS@}%
 \afterCOORD{\xyFN@\POS@}}
\xylet@\afterPOS@=\empty
\xydef@\afterCOORD#1{%
 \DN@##1{\def\afterCOORD@{\def\afterCOORD@{##1}#1}}%
 \expandafter\next@\expandafter{\afterCOORD@}%
 \afterVECTORorEMPTY{\xy@@\czeroEdge@ \afterCOORD@}{\xyFN@\COORD@}}
\xylet@\afterCOORD@=\empty
\xydef@\afterVECTORorEMPTY#1#2{%
 \DN@##1{\def\afterVECTOR@{\def\afterVECTOR@{##1}%
 \ifVECTORempty@\DN@{#2}\else\DN@{#1}\fi \next@}}%
 \expandafter\next@\expandafter{\afterVECTOR@}%
 \xyFN@\VECTOR@}
\xynew@{if}\ifVECTORempty@
\xylet@\afterVECTOR@=\empty
\xydef@\xyVECTOR@{%
 \ifx \space@\next \expandafter\DN@\space{\xyFN@\VECTOR@}%
 \else \ifcat A\noexpand\next \let\next@=\VECTOR@letter
 \else \let\next@=\VECTOR@other \fi\fi \next@}
\xylet@\VECTOR@=\xyVECTOR@
\def\notrelaxorelse@#1#2{\ifx#1\relax \expandafter#2\else\expandafter#1\fi}
\xydef@\VECTOR@letter{%
 \ifx a\next \expandafter\VECTOR@a \else \expandafter\CORNER@ \fi}
\xydef@\VECTOR@a a(#1){\xy@{a(#1)}{\vfromcartesianangle@{#1}}%
 \VECTORempty@false \afterVECTOR@}
\xydef@\CORNER@{%
 \xy@{}{\A@=-.5\L@c \advance\A@.5\R@c \B@=-.5\D@c \advance\B@.5\U@c
 \let\nextii@=\zeroit@}%
 \VECTORempty@true\CORNER@i}
\xydef@\zeroit@#1{#1=\z@}
\xydef@\CORNER@i{%
 \ifx D\next \DN@ D{\xy@{D}{\Y@c=-\D@c \nextii@\X@c \B@=\Y@c}\CORNER@ii}%
 \else\ifx U\next \DN@ U{\xy@{U}{\Y@c= \U@c \nextii@\X@c \B@=\Y@c}\CORNER@ii}%
 \else\ifx L\next \DN@ L{\xy@{L}{\X@c=-\L@c \nextii@\Y@c \A@=\X@c}\CORNER@ii}%
 \else\ifx R\next \DN@ R{\xy@{R}{\X@c= \R@c \nextii@\Y@c \A@=\X@c}\CORNER@ii}%
 \else\ifx C\next \DN@ C{\xy@{C}{\X@c= \A@ \Y@c= \B@}\CORNER@ii}%
 \else\ifx E\next \DN@ E{\xy@{E}{%
 \A@=\X@c \B@=\Y@c \the\Edge@c\z@ \advance\X@c-\A@ \advance\Y@c-\B@}%
 \CORNER@ii}%
 \else\ifx P\next \DN@ P{\xy@{P}{%
 \A@=\X@c \B@=\Y@c \the\Edge@c\thr@@ \advance\X@c-\A@ \advance\Y@c-\B@}%
 \CORNER@ii}%
 \else\ifx (\next
 \DN@(##1){\xy@{(##1)}{\X@c=##1\X@c \Y@c=##1\Y@c}\afterVECTOR@}%
 \else \let\next@=\afterVECTOR@
 \fi\fi\fi\fi\fi\fi\fi\fi \next@}
\xydef@\CORNER@ii{\xy@@{\let\nextii@=\eat@}%
 \VECTORempty@false \xyFN@\CORNER@i}
\xydef@\VECTOR@other{%
 \addLT@\ifx \next
 \addGT@{\addLT@\DN@##1}{\xy@{<##1>}{\vfromabsolute@{##1}}%
 \VECTORempty@false\afterVECTOR@}%
 \else\ifx (\next
 \DN@({\xyFN@\VECTOR@other@open}%
 \else\ifx /\next
 \DN@/##1/{\xy@@ix@{{##1}}%
 \xy@{/##1/}{\expandafter\vfromslide@\the\toks9}%
 \VECTORempty@false\afterVECTOR@}%
 \else\ifx 0\next
 \DN@ 0{\xy@{0}{\X@c=\z@ \Y@c=\z@}\VECTORempty@false\afterVECTOR@}%
 \else
 \DN@{\VECTORempty@true\afterVECTOR@}%
 \fi\fi\fi\fi \next@}
\xydef@\VECTOR@other@open{%
 \ifx *\next \DN@{\VECTORempty@true \xyFN@\afterVECTOR@(}%
 \else
 \DN@##1){\xy@{(##1)}{\vfromcartesian@{##1}}\VECTORempty@false\afterVECTOR@}%
 \fi \next@}
\xydef@\xyCOORD@{%
 \ifx \space@\next \expandafter\DN@\space{\xyFN@\COORD@}%
 \else \ifcat A\noexpand\next \let\next@=\xyCOORD@letter
 \else \let\next@=\xyCOORD@other \fi\fi \next@}
\xylet@\COORD@=\xyCOORD@
\xydef@\xyCOORD@letter{%
 \ifx c\next
 \DN@ c{\xy@{c}{}\afterCOORD@}%
 \else\ifx p\next
 \DN@ p{\xy@{p}\cfromp@ \afterCOORD@}%
 \else\ifx x\next
 \DN@ x{\xy@{x}{\R@c=\X@xbase \U@c=\Y@xbase \intersect@}\afterCOORD@}%
 \else\ifx y\next
 \DN@ y{\xy@{y}{\R@c=\X@ybase \U@c=\Y@ybase \intersect@}\afterCOORD@}%
 \else\ifx s\next
 \DN@ s##1{\xy@{s{##1}}{\cfroms@{##1}}\afterCOORD@}%
 \else \let\next@=\afterCOORD@ \fi\fi\fi\fi\fi \next@}
\xydef@\xyCOORD@other{%
 \ifx "\next
 \DN@"##1"{\xy@{"##1"}{\cfromid@{##1}}\afterCOORD@}%
 \else\ifx \bgroup\next
 \DN@##1{\xy@{{##1}}{\enter@{\pfromthep@\basefromthebase@}}%
 \silencexy@ \POS##1\relax \unsilencexy@ \xy@@\leave@ \afterCOORD@}%
 \else\ifx (\next
 \DN@({\xyFN@\xyCOORD@other@open}%
 \else \let\next@=\afterCOORD@ \fi\fi\fi \next@}
\xynew@{if}\ifsilentxy@
\xydef@\silencexy@{%
 \ifsilentxy@ \nter@{}%
 \else \nter@{\silentxy@false \let\xy@=\unsilent@@xy@}
 \silentxy@true \let\unsilent@@xy@=\xy@ \def\xy@##1##2{\unsilent@@xy@{}{##2}}%
 \fi}
\xydef@\unsilencexy@{\leave@}
\xydef@\xyCOORD@other@open{%
 \ifx *\next
 \DN@*##1*){\xy@{(*}{\enter@{\pfromthep@\basefromthebase@}}%
 \POS##1\relax \xy@{*)}\leave@ \afterCOORD@}%
 \else \DN@{\xyFN@\afterCOORD@(}%
 \fi \next@}
\xydef@\xyPOS@{%
 \ifx \space@\next \expandafter\DN@\space{\xyFN@\POS@}%
 \else\addPLUS@\ifx \next
 \addPLUS@\DN@{\xy@+{\enter@\cplusthec@}%
 \afterCOORD{\xy@@\leave@ \xyFN@\POS@}}%
 \else\addDASH@\ifx \next
 \addDASH@\DN@{\xy@-{\enter@\cplusthec@}%
 \afterCOORD{\xy@@{\X@c=-\X@c \Y@c=-\Y@c\leave@}\xyFN@\POS@}}%
 \else\ifx !\next
 \DN@ !{\xy@!{\enter@\cskewthec@}\afterCOORD{\xy@@\leave@ \xyFN@\POS@}}%
 \else\addDOT@\ifx \next
 \addDOT@\DN@{\xy@.{\enter@\cmergethec@}%
 \afterCOORD{\xy@@\leave@ \xyFN@\POS@}}%
 \else\ifx ,\next
 \DN@ ,{\xy@,{\comma@@}\afterCOORD{\xyFN@\POS@}}%
 \else\ifx ;\next
 \DN@ ;{\xy@;{\swap@}\afterCOORD{\xyFN@\POS@}}%
 \else\ifx :\next
 \DN@ :{\xyFN@\POS@colon}%
 \else\addEQ@\ifx \next
 \addEQ@\DN@{\xyFN@\saveid@}%
 \else\ifx *\next
 \DN@ *{\xyFN@\POS@star}%
 \else \ifx ?\next
 \DN@?{\xy@?{}\afterPLACE{\xyFN@\POS@}}%
 \else \addAT@\ifx \next
 \addAT@\DN@{\xyFN@\STACK@}%
 \else
 \let\next@=\afterPOS@
 \fi\fi\fi\fi\fi\fi\fi\fi\fi\fi\fi\fi \next@}
\xylet@\comma@@=\relax
\xylet@\POS@=\xyPOS@
\xydef@\POS@colon{\DNii@{\afterCOORD{\xyFN@\POS@}}%
 \ifx :\next \xy@{::}{\setbase@@\X@c\Y@c}\DN@:{\nextii@}%
 \else \xy@:{\setbase@\X@p\Y@p\X@c\Y@c}\let\next@=\nextii@ \fi
 \next@}
\xydef@\POS@star{%
 \ifx *\next
 \DN@*##1##{\nextii@{##1}}%
 \DNii@##1##2{\xy@@ix@{{##1}{##2}}%
 \xy@{**##1{##2}}{\expandafter\connect@\the\toks9}\xyFN@\POS@}%
 \else
 \DN@##1##{\nextii@{##1}}%
 \DNii@##1##2{\xy@@ix@{{##1}{##2}}%
 \xy@{*##1{##2}}{\expandafter\drop@\the\toks9}\xyFN@\POS@}%
 \fi
 \next@}
\xydef@\cfromp@{\X@c=\X@p \Y@c=\Y@p \U@c=\U@p \D@c=\D@p \L@c=\L@p \R@c=\R@p
 \Edge@c=\expandafter{\the\Edge@p}}
\xydef@\pfromc@{\X@p=\X@c \Y@p=\Y@c \U@p=\U@c \D@p=\D@c \L@p=\L@c \R@p=\R@c
 \Edge@p=\expandafter{\the\Edge@c}}
\xydef@\swapdimen@#1#2{\dimen@=#1\relax #1=#2\relax #2=\dimen@}
\xynew@{toks}\swaptoks@@
\xydef@\swap@{\swapdimen@\X@c\X@p \swapdimen@\Y@c\Y@p
 \swapdimen@\U@c\U@p\swapdimen@\D@c\D@p \swapdimen@\L@c\L@p\swapdimen@\R@c\R@p
 \swaptoks@@=\Edge@c \Edge@c=\Edge@p \Edge@p=\swaptoks@@}
\xydef@\vfromabsolute@#1{\vfromabsolute@@#1,@}
\xydef@\vfromabsolute@@#1,#2@{\X@c=#1\relax
 \DN@{#2}\ifx\next@\empty \Y@c=\X@c
 \else \DN@##1,{\Y@c=##1}\next@#2\relax \fi
 \advance\X@c 1sp \advance\Y@c 1sp
}
\xydef@\cfromthec@{\X@c=\the\X@c \Y@c=\the\Y@c
 \U@c=\the\U@c \D@c=\the\D@c \L@c=\the\L@c \R@c=\the\R@c
 \Edge@c={\expandafter\noexpand\the\Edge@c}}
\xydef@\cfromthep@{\X@c=\the\X@p \Y@c=\the\Y@p
 \U@c=\the\U@p \D@c=\the\D@p \L@c=\the\L@p \R@c=\the\R@p
 \Edge@c={\expandafter\noexpand\the\Edge@p}}
\xydef@\pfromthep@{\X@p=\the\X@p \Y@p=\the\Y@p
 \U@p=\the\U@p \D@p=\the\D@p \L@p=\the\L@p \R@p=\the\R@p
 \Edge@p={\expandafter\noexpand\the\Edge@p}}
\xydef@\pfromthec@{\X@p=\the\X@c \Y@p=\the\Y@c
 \U@p=\the\U@c \D@p=\the\D@c \L@p=\the\L@c \R@p=\the\R@c
 \Edge@p={\expandafter\noexpand\the\Edge@c}}
\xydef@\cplusthec@{\advance\X@c\the\X@c \advance\Y@c\the\Y@c}
\xydef@\cskewthec@{%
 \noexpand\cskew@{\the\Y@c}{\the\X@c}{\the\D@c}{\the\U@c}{\the\L@c}{\the\R@c}}
\xydef@\cskew@#1#2#3#4#5#6{%
 \D@c=#3\advance\D@c \Y@c \ifdim\D@c<\z@ \D@c=\z@ \fi
 \U@c=#4\advance\U@c-\Y@c \ifdim\U@c<\z@ \U@c=\z@ \fi
 \advance\Y@c#1%
 \L@c=#5\advance\L@c \X@c \ifdim\L@c<\z@ \L@c=\z@ \fi
 \R@c=#6\advance\R@c-\X@c \ifdim\R@c<\z@ \R@c=\z@ \fi
 \advance\X@c#2%
 \Edge@c={\rectangleEdge}}
\xydef@\cmergethec@{%
 \noexpand\cmerge@{\the\Y@c}{\the\X@c}{\the\D@c}{\the\U@c}{\the\L@c}{\the\R@c}}
\xydef@\cmerge@#1#2#3#4#5#6{\the\Edge@c4%
 \A@=#2\advance\A@-\X@c \B@=#1\advance\B@-\Y@c
 \dimen@=#5\advance\L@c \A@ \ifdim\L@c<\dimen@ \L@c=\dimen@ \fi
 \dimen@=#6\advance\R@c-\A@ \ifdim\R@c<\dimen@ \R@c=\dimen@ \fi
 \dimen@=#3\advance\D@c \B@ \ifdim\D@c<\dimen@ \D@c=\dimen@ \fi
 \dimen@=#4\advance\U@c-\B@ \ifdim\U@c<\dimen@ \U@c=\dimen@ \fi
 \advance\X@c\A@ \advance\Y@c\B@}
\xydef@\halfroottwo{.70710678}
\xydef@\partroottwo{.29289322}
\xydef@\halfrootthree{.8660254}
\xydef@\vfromcartesian@#1{\vfromcartesian@@#1@}
\xydef@\vfromcartesian@@#1,#2@{%
 \X@c=\X@origin \advance\X@c#1\X@xbase \advance\X@c#2\X@ybase
 \Y@c=\Y@origin \advance\Y@c#1\Y@xbase \advance\Y@c#2\Y@ybase}
\xydef@\setbase@#1#2#3#4{%
 \X@origin=#1\relax \Y@origin=#2\relax
 \X@xbase=#3\relax \advance\X@xbase-\X@origin
 \Y@xbase=#4\relax \advance\Y@xbase-\Y@origin
 \X@ybase=-\Y@xbase \Y@ybase=\X@xbase}
\xydef@\setbase@@#1#2{%
 \X@ybase=#1\relax \advance\X@ybase-\X@origin
 \Y@ybase=#2\relax \advance\Y@ybase-\Y@origin}
\xydef@\basefromthebase@{\X@origin=\the\X@origin \Y@origin=\the\Y@origin
 \X@xbase=\the\X@xbase \Y@xbase=\the\Y@xbase
 \X@ybase=\the\X@ybase \Y@ybase=\the\Y@ybase}
\xydef@\vfromcartesianangle@#1{\enter@\basefromthebase@ \R@=#1\p@
 \B@=360\p@
 \loop@ \ifdim\R@<\z@ \advance\R@\B@ \repeat@
 \loop@ \ifdim\R@>\B@ \advance\R@-\B@ \repeat@
 \ifdim\R@<.5\B@\else \R@=-\R@ \advance\R@\B@
 \X@ybase=-\X@ybase \Y@ybase=-\Y@ybase \fi
 \B@=180\p@
 \ifdim\R@<.5\B@\else \R@=-\R@ \advance\R@\B@
 \X@xbase=-\X@xbase \Y@xbase=-\Y@xbase \fi
 \B@=90\p@
 \ifdim\R@<.5\B@ \let\nextiii@=\literal@
 \else \R@=-\R@ \advance\R@\B@ \def\nextiii@##1,##2@{##2,##1@}\fi
 \dimen@=\z@ \DN@{1,0@}%
 \dimen@ii=45\p@ \DNii@{.70710678,.70710678@}%
 \chooseangleinterval@
 {\chooseangleinterval@
 {\chooseangleinterval@
 {\chooseangleinterval@
 {\chooseangleinterval@
 {}%
 {4.090909}{.99677570,.08023846@}%
 {}}%
 {6}{.99452190,.10452846@}%
 {\chooseangleinterval@
 {}%
 {8.181818}{.98982144,.14231484@}%
 {}}}%
 {10}{.98480775,.17364818@}%
 {\chooseangleinterval@
 {}%
 {12.857143}{.97492791,.22252093@}%
 {}}}%
 {15}{.96592583,.25881905@}%
 {\chooseangleinterval@
 {\chooseangleinterval@
 {}%
 {16.363636}{.95949297,.28173256@}%
 {}}%
 {18}{.95105652,.30901699@}%
 {\chooseangleinterval@
 {}%
 {20}{.93969262,.34202014@}%
 {}}}}%
 {22.5}{.92387953,.38268343@}%
 {\chooseangleinterval@
 {\chooseangleinterval@
 {\chooseangleinterval@
 {}%
 {24.545455}{.90963200,.41541501@}%
 {}}%
 {25.714286}{.90096887,.43388374@}%
 {}}%
 {30}{.86602540,.5@}%
 {\chooseangleinterval@
 {\chooseangleinterval@
 {}%
 {32.727273}{.84125353,.54064082@}%
 {}}%
 {36}{.80901699,.58778525@}%
 {\chooseangleinterval@
 {\chooseangleinterval@
 {}%
 {38.571429}{.78183148,.62348980@}%
 {}}%
 {40.909091}{.75574957,.65486073@}%
 {\chooseangleinterval@
 {}%
 {40}{.76604444,.64278761@}%
 {}}}}}%
 \A@=\R@ \advance\A@-\dimen@
 \ifdim\ifdim\A@<\z@-\fi\A@<.01\p@ \edef\next@{\expandafter\nextiii@\next@}%
 \else \B@=\dimen@ii \advance\B@-\R@ 
 \ifdim\A@<\B@ \dimen@=\toradians@\A@
 \edef\next@{\next@ \expandafter\removePT@\the\dimen@ @}%
 \else \dimen@=-\toradians@\B@
 \edef\next@{\nextii@ \expandafter\removePT@\the\dimen@ @}%
 \fi
 \expandafter\interpolatepoint@\next@
 \edef\next@{\expandafter\nextiii@\next@}%
 \fi 
 \expandafter\vfromcartesian@@\next@
 \leave@}
\xydef@\chooseangleinterval@#1#2#3#4{%
 \B@=#2\p@ \def\next{#3}%
 \ifdim\R@<\B@ \dimen@ii=\B@ \let\nextii@=\next #1%
 \else \dimen@=\B@ \let\next@=\next \ifdim\B@<\R@ #4\fi\fi}
\xydef@\interpolateinterval@#1,#2@#3,#4@{%
 \A@=#1\p@ \dimen@=#3\p@ \advance\dimen@-\A@ \advance\A@\next\dimen@
 \B@=#2\p@ \dimen@=#4\p@ \advance\dimen@-\B@ \advance\B@\next\dimen@
 \edef\next@{\expandafter\removePT@\the\A@,\expandafter\removePT@\the\B@ @}}
\xydef@\toradians@{0.01745329}
\xydef@\interpolatepoint@#1,#2@#3@{%
 \A@=#1\p@ \dimen@ii=#3\A@ \dimen@ii=-.5\dimen@ii \advance\A@#3\dimen@ii
 \dimen@=-#2\p@ \advance\A@#3\dimen@
 \B@=#2\p@ \dimen@ii=#3\B@ \dimen@ii=-.5\dimen@ii \advance\B@#3\dimen@ii
 \dimen@=#1\p@ \advance\B@#3\dimen@
 \edef\next@{\expandafter\removePT@\the\A@,\expandafter\removePT@\the\B@ @}}
\xydef@\drop@#1#2{%
 \global\setbox\lastobjectbox@=\object#1{#2}%
 \ifHidden@ \dimen@=\X@c \advance\dimen@-\L@c \else
 \dimen@=\Y@c \advance\dimen@ \U@c \ifdim\Y@max<\dimen@ \Y@max=\dimen@ \fi
 \dimen@=\Y@c \advance\dimen@-\D@c \ifdim\dimen@<\Y@min \Y@min=\dimen@ \fi
 \dimen@=\X@c \advance\dimen@ \R@c \ifdim\X@max<\dimen@ \X@max=\dimen@ \fi
 \dimen@=\X@c \advance\dimen@-\L@c \ifdim\dimen@<\X@min \X@min=\dimen@ \fi \fi
 \ifInvisible@\else
 \setboxz@h{\kern\dimen@ \raise\Y@c\box\lastobjectbox@}%
 \ht\z@=\z@ \dp\z@=\z@ \wd\z@=\z@ {\Drop@@}\fi}
\xydef@\connect@#1#2{\setupDirection@ \enter@{\cfromthec@}%
 \global\setbox\lastobjectbox@=\object#1{#2}\leave@
 \Connect@@}
\xydef@\afterPLACE#1{%
 \DN@##1{\def\afterPLACE@{\xy@@\leave@ \def\afterPLACE@{##1}#1}}%
 \expandafter\next@\expandafter{\afterPLACE@}%
 \def\PLACEf@{{.5}}%
 \xy@@{\enter@{\pfromthep@}%
 \Creset@@
 \def\PLACEf@{{.5}}%
 \let\PLACEedgep@@=\PLACEedgep@ \let\PLACEedgec@@=\PLACEedgec@}%
 \xyFN@\PLACE@}
\xydef@\PLACEf@{}
\xydef@\PLACEedgep@@{}
\xydef@\PLACEedgec@@{}
\xydef@\PLACEedgep@{\Cshavep@@ \def\PLACEedgep@@{\Cslidep@@\jot}}
\xydef@\PLACEedgec@{\Cshavec@@ \def\PLACEedgec@@{\Cslidec@@{-\jot}}}
\xylet@\afterPLACE@=\empty
\xydef@\PLACE@{%
 \ifx \space@\next \expandafter\DN@\space{\xyFN@\PLACE@}%
 \else\addLT@\ifx \next
 \addLT@\DN@{\addLT@\xy@{\def\PLACEf@{{0}}\PLACEedgep@@}\xyFN@\PLACE@}%
 \else\addGT@\ifx \next
 \addGT@\DN@{\addGT@\xy@{\def\PLACEf@{{1}}\PLACEedgec@@}\xyFN@\PLACE@}%
 \else\ifx (\next
 \DN@(##1){\def\PLACEf@{{##1}}\xy@{(##1)}{\def\PLACEf@{{##1}}}\xyFN@\PLACE@}%
 \else\ifx !\next
 \DN@!{\xyFN@\PLACE@intercept}%
 \else
 \DN@{\xy@@{\expandafter\Calong@@\PLACEf@ \czeroEdge@}\PLACE@@}%
 \fi\fi\fi\fi\fi \next@}
\xydef@\PLACE@intercept{%
 \ifx \space@\next \expandafter\DN@\space{\xyFN@\PLACE@intercept}%
 \else\ifx \bgroup\next
 \DN@##1{\xy@{!{##1}}{}\PLACE@intercept@{##1}}%
 \else\ifx (\next
 \DN@(*##1*){\xy@{!(*##1*)}{}\PLACE@intercept@{##1}}%
 \else \DN@{\xyerror@{{<pos>} expected after ! in <place>}{}}%
 \fi\fi\fi \next@}
\xydef@\PLACE@intercept@#1{%
 \xy@@{\enter@{\pfromthep@\basefromthebase@}\begingroup}%
 \xy@@ix@{#1}\xy@@{\plainxy@ \expandafter\POS\the\toks9\relax
 \edef\next@{\endgroup
 \X@c =\the\X@c \Y@c=\the\Y@c \X@p=\the\X@p \Y@p=\the\Y@p}%
 \next@ \Cintercept@@ \leave@}%
 \PLACE@@}
\xydef@\PLACE@@{%
 \ifx \space@\next \expandafter\DN@\space{\xyFN@\PLACE@@}%
 \else\ifx /\next \DN@/##1/{\xy@{/##1/}{\Cslidec@@{##1}}\afterPLACE@}%
 \else \let\next@=\afterPLACE@
 \fi\fi \next@}
\xydef@\intersect@{%
 \d@X=\X@c \advance\d@X-\X@p \d@Y=\Y@c \advance\d@Y-\Y@p
 \A@=\X@c \advance\A@-\X@origin \B@=\Y@c \advance\B@-\Y@origin
 \edef\next@{\expandafter\removePT@\the\R@c}%
 \edef\nextii@{\expandafter\removePT@\the\U@c}%
 \D@c=\next@\d@Y \advance\D@c-\nextii@\d@X \divide\D@c\KK@
 \L@c=\next@\B@ \advance\L@c-\nextii@\A@ \divide\L@c\KK@
 \ifdim\D@c=\z@\zeroDivide@\else \quotient@\next@\L@c\D@c \fi
 \advance\X@c-\next@\d@X \advance\Y@c-\next@\d@Y
 \czeroEdge@}
\xydef@\zeroDivide@@{\zeroDivide@message{\intersect@}{treated as 0}\DN@{0}}
\xydef@\zeroDivide@message#1#2{\xywarning@{division by 0 in \string#1, #2}}
\xylet@\zeroDivide@=\zeroDivide@@
\xydef@\zeroDivideLimit@@{\ifdim\L@c=\z@ \DN@{0}%
 \else\ifdim\L@c<\z@\DN@{-\zeroDivide@Limit}%
 \else\DN@{\zeroDivide@Limit}\fi\fi
 \zeroDivide@message{\intersect@}{replaced by \zeroDivide@Limit}}
\xydef@\zeroDivideLimit@#1{\edef\zeroDivide@Limit{#1}%
 \let\zeroDivide@=\zeroDivideLimit@@}
\xylet@\zeroDivideLimit=\zeroDivideLimit@
\xydef@\vfromslide@#1{\enter@\DirectionfromtheDirection@ \begingroup
 \plainxy@\afterDIRECTIONorEMPTY\vfromslide@i\vfromslide@i#1@}
\xydef@\vfromslide@i#1@{\DN@{#1}%
 \edef\next{\endgroup
 \ifx\next@\empty \dimen@=.5pc \else \dimen@=#1\relax \fi
 \X@c=\cosDirection\dimen@ \Y@c=\sinDirection\dimen@}\next
 \leave@}
\xydef@\s@bot{-1}
\xydef@\s@top{-1}
\xydef@\cfroms@#1{\tests@{#1}\runs@\outofranges@}
\xydef@\tests@#1#2#3{\DN@{#3}%
 \count@=\s@top \advance\count@-#1\relax
 \ifnum\count@>\s@bot\relax \ifnum\count@>\s@top\else\DN@{#2}\fi\fi
 \next@}
\xydef@\runs@{\csname S@\the\count@\endcsname}
\xydef@\outofranges@{\count@=\s@top \advance\count@-\s@bot
 \xyerror@{stack index out of range (should be 0..\the\count@)}{}}
\xydef@\STACK@{%
 \addPLUS@\ifx\next
 \addPLUS@\DN@{\xy@{@+}{}\afterCOORD{\xy@@\spushc@ \xyFN@\POS@}}%
 \else\addDASH@\ifx\next
 \addDASH@\DN@{\xy@{@-}{}\afterCOORD{\xy@@\spop@ \xyFN@\POS@}}%
 \else \ifx i\next \DN@ i{\xy@{@i}\sinit@ \xyFN@\POS@}%
 \else \ifx (\next \DN@ ({\xy@{@(}\senter@ \xyFN@\POS@}%
 \else \ifx )\next \DN@ ){\xy@{@)}\sleave@ \xyFN@\POS@}%
 \else\addEQ@\ifx\next \addEQ@\DN@{\STACK@load}%
 \else\addAT@\ifx\next \addAT@\DN@{\xy@{@@}{}\smap@}%
 \else \DN@##1{\xyerror@{illegal stack command ##1}{}\afterCOORD{\xyFN@\POS@}}%
 \fi\fi\fi\fi\fi\fi\fi \next@}
\xydef@\STACK@load{\xy@{@=}{%
 \if\sempty@\else \xywarning@{loading on top of non-empty stack}\sinit@ \fi
 \let\comma@@=\spushc@}%
 \afterCOORD{\xy@@{\spushc@ \let\comma@@=\relax}\xyFN@\POS@}}
\xydef@\spushc@{%
 \count@=\s@top \advance\count@\@ne \edef\s@top{\the\count@}%
 \expandafter\edef\csname S@\s@top\endcsname{\cfromthec@}}
\xydef@\spushid@#1{\DNii@{#1}\edef\nextii@{\codeof\nextii@}%
 \expandafter\let\expandafter\next@\csname Q@\nextii@\endcsname
 \ifx\next@\relax \xyerror@{<pos> \string"\nextii@\string" not defined}{}%
 \else
 \count@=\s@top \advance\count@\@ne \edef\s@top{\the\count@}%
 \DNii@##1{\expandafter\def\csname S@\s@top\endcsname{##1}}%
 \expandafter\nextii@\expandafter{\next@}%
 \fi}
\xydef@\idfroms@#1#2{%
 \tests@{#2}{\DN@{\idfromxy@{#1}}%
 \expandafter\expandafter\expandafter\next@
 \expandafter\expandafter\expandafter{\csname S@\the\count@\endcsname}%
 }\outofranges@}
\xydef@\spop@{\count@=\s@top
 \ifnum\count@>\s@bot \advance\count@\m@ne \edef\s@top{\the\count@}%
 \else \xyerror@{nothing to pop from stack}{}\fi}
\xydef@\sinit@{\edef\s@top{\s@bot}}
\xydef@\senter@{%
 \count@=\s@top \advance\count@\@ne
 \expandafter\edef\csname S@\the\count@\endcsname{\s@bot}%
 \edef\s@bot{\the\count@}\edef\s@top{\the\count@}}
\xydef@\sleave@{%
 \ifnum\s@bot=\s@top\else \xywarning@{leaving non-empty stack}\sinit@ \fi
 \ifnum\s@bot>\m@ne \edef\s@bot{\csname S@\s@top\endcsname}%
 \count@=\s@top \advance\count@\m@ne \edef\s@top{\the\count@}%
 \edef\sbot{\the\count@}\fi}
\xydef@\sempty@{\ifnum\s@top=\s@bot TT\else TF\fi}
\xydef@\xytotoks@#1#2{\addtotoks@{#2}}
\xydef@\xytotoks@@toksix@#1{\addtotoks@{\toks9={#1}}}
\xydef@\smap@{%
 \begingroup \toks@={}\let\xy@=\xytotoks@ \let\oxy@=\xy@
 \let\xy@@ix@=\xytotoks@@toksix@
 \afterCOORD{\expandafter\endgroup
 \expandafter\smapxy@@\expandafter{\the\toks@}\xyFN@\POS@}}
\xydef@\smapxy@@#1{\xy@@{\edef\smapp@@{\s@bot}\smapxy@i{#1}}}
\xylet@\smapp@@=\empty
\xydef@\smapxy@i#1{%
 \ifnum\smapp@@<\s@top
 \count@=\smapp@@ \advance\count@\@ne \edef\smapp@@{\the\count@}%
 \DN@{\csname S@\smapp@@\endcsname #1\relax \smapxy@i{#1}}%
 \else \let\next@=\relax
 \fi \next@}
\xydef@\saveid@{%
 \ifx \space@\next \expandafter\DN@\space{\xyFN@\saveid@}%
 \else \ifx "\next\DN@"##1"{\xy@{="##1"}{\idfromc@{##1}}\xyFN@\POS@}%
 \else \ifx :\next\DN@:##1"##2"{\xy@{=:"##2"}{\idfrombase@{##2}}\xyFN@\POS@}%
 \else\addAT@\ifx\next
 \addAT@\DN@"##1"{\xy@{=@"##1"}{\idfromstack@{##1}}\xyFN@\POS@}%
 \else \ifx s\next
 \DN@ s##1"##2"{\xy@{=s##1"##2"}{\idfroms@{##2}{##1}}\xyFN@\POS@}%
 \else\addEQ@\ifx\next \let\saveid@COORD@@=\saveid@COORDii
 \addEQ@\DN@{\xyFN@\saveid@COORD}%
 \else \let\saveid@COORD@@=\saveid@COORDi \let\next@=\saveid@COORD
 \fi\fi\fi\fi\fi\fi \next@}
\xylet@\saveid@COORD@@=\relax
\xydef@\idfromc@#1{\DN@{#1}%
 \expandafter\edef\csname Q@\codeof\next@\endcsname{\cfromthec@}}
\xydef@\idfrombase@#1{\DN@{#1}%
 \expandafter\edef\csname Q@\codeof\next@\endcsname{\basefromthebase@}}
\xydef@\idfromstack@#1{%
 \toks@={\if\sempty@\else
 \xywarning@{loading on top of non-empty stack}\sinit@ \fi}%
 \count@=\s@bot \advance\count@\@ne
 \ifnum\count@>\s@top\else
 \loop@
 \expandafter\let\expandafter\next@\csname S@\the\count@\endcsname
 \expandafter\addtotoks@\expandafter{\next@}%
 \ifnum\count@<\s@top \advance\count@\@ne \addtotoks@{\spushc@}\repeat@
 \fi
 \DN@{#1}\edef\next@##1{%
 \def\expandafter\noexpand\csname Q@\codeof\next@\endcsname{##1}}%
 \expandafter\next@\expandafter{\the\toks@}}
\xydef@\saveid@COORD{%
 \begingroup \toks@={}\let\xy@=\xytotoks@ \let\oxy@=\xy@
 \let\xy@@ix@=\xytotoks@@toksix@
 \afterCOORD{\expandafter\saveid@COORDi\expandafter{\the\toks@}}}
\xydef@\saveid@COORDi#1#2"#3"{\endgroup \xy@@{\idfromxy@{#3}{#1}}\xyFN@\POS@}
\xydef@\saveid@COORDii#1#2"#3"{\endgroup \xy@@{\idfromcxy@{#3}{#1}}\xyFN@\POS@}
\xydef@\idfromxy@#1#2{\DN@{#1}%
 \expandafter\def\csname Q@\codeof\next@\endcsname{#2}}
\xydef@\idfromcxy@#1#2{\DN@{#1}%
 \expandafter\edef\csname Q@\codeof\next@\endcsname{\cfromthec@#2}}
\xydef@\cfromid@#1{\DNii@{#1}\edef\nextii@{\codeof\nextii@}%
 \expandafter\let\expandafter\next@\csname Q@\nextii@\endcsname
 \ifx\next@\relax \xyerror@{<pos> \string"\nextii@\string" not defined}{}%
 \else \expandafter\next@\fi}
\xydef@\OBJECT@{%
 \ifx \space@\next \expandafter\DN@\space{\xyFN@\OBJECT@}%
 \else\ifcat A\noexpand\next \let\next@=\OBJECT@letter
 \else \let\next@=\OBJECT@other \fi\fi \next@}
\xydef@\OBJECT@letter{%
 \ifx i\next \DN@ i{\addtotoks@\Invisible@true \xyFN@\OBJECT@}%
 \else\ifx h\next\DN@ h{\addtotoks@\Hidden@true \xyFN@\OBJECT@}%
 \else\ifx o\next\DN@ o{\xywarning@{Obsolete o modifier used}\OBJECT@shape{o}}%
 \else\ifx x\next\DN@ x{\xywarning@{Obsolete x modifier used}\OBJECT@shape{}}%
 \else \let\next@=\OBJECT@direction
 \fi\fi\fi\fi \next@}
\xydef@\OBJECT@other{%
 \ifx !\next \DN@!{\OBJECT@shift}%
 \else\addPLUS@\ifx \next \DN@{\OBJECT@change}%
 \else\addDASH@\ifx \next \DN@{\OBJECT@change}%
 \else\addEQ@\ifx \next \DN@{\OBJECT@set}%
 \else\ifx [\next
 \DN@[##1]{\xy@{[##1]}{\OBJECT@shape{##1}}}%
 \else\ifx ^\next \let\next@=\OBJECT@direction
 \else\ifx _\next \let\next@=\OBJECT@direction
 \else\ifx :\next \let\next@=\OBJECT@direction
 \else\ifx ?\next
 \DN@ ?{\xywarning@{\string? modifier used}\xyFN@\OBJECT@direction}%
 \else\ifx (\next
 \let\next@=\OBJECT@direction
 \else\addAT@\ifx\next \addAT@\DN@##1##{\OBJECT@@{\dir##1}}%
 \else \DN@##1##{\OBJECT@@{##1}}%
 \fi\fi\fi\fi\fi\fi\fi\fi\fi\fi\fi \next@}
\xydef@\prevEdge@@{\zeroEdge}
\xydef@\OBJECT@@#1#2{%
 \expandafter\def\expandafter\prevEdge@@\expandafter{\the\Edge@c}%
 \expandafter\Edge@c\expandafter{\objectEdge}%
 \Invisible@false\Hidden@false \def\Leftness@{.5}\def\Upness@{.5}%
 \def\Drop@@{\boxz@}\def\Connect@@{\straight@\relax}%
 \DN@{#1}\ifx\next@\empty \DNii@{#2}%
 \ifx\nextii@\empty \DN@{\hbox\bgroup\no@}\else \let\next@=\objectbox \fi\fi
 \setbox\z@=\next@{#2}\L@c=\Leftness@\wdz@ \R@c=\wdz@ \advance\R@c-\L@c
 \D@c=\dp\z@ \advance\D@c\ht\z@ \U@c=\Upness@\D@c \advance\D@c-\U@c
 \R@p=\z@ \L@p=\L@c \U@p=\U@c \advance\U@p-\ht\z@ \D@p=-\U@p
 \the\toks@\toks@={}\setboxz@h{\kern\R@p \raise\U@p\boxz@}%
 \checkZeroEdge@
 \dimen@=\L@c \advance\dimen@\R@c \wdz@=\dimen@ \ht\z@=\U@c \dp\z@=\D@c \boxz@
 \OBJECT@x}
\xydef@\adjustLR@{%
 \ifdim\zz@\wdz@ \L@c=\z@ \R@c=\z@ \dimen@=\Leftness@\p@
 \ifdim\dimen@<\z@ \L@c=\dimen@ \R@c=-\L@c
 \else\ifdim\dimen@>\p@ \L@c=\dimen@ \advance\L@c-\p@ \R@c=-\L@c \fi\fi
 \else \L@c=\Leftness@\wdz@ \R@c=\wdz@ \advance\R@c-\L@c \fi}
\xydef@\adjustUD@{\dimen@=\ht\z@ \advance\dimen@\dp\z@
 \ifdim\zz@\dimen@ \U@c=\z@ \D@c=\z@ \dimen@=\Upness@\p@
 \ifdim\dimen@<\z@ \U@c=\dimen@ \D@c=-\L@c
 \else\ifdim\dimen@>\p@ \U@c=\dimen@ \advance\U@c-\p@ \D@c=-\L@c \fi\fi
 \else \D@c=\dimen@ \U@c=\Upness@\dimen@ \advance\D@c-\U@c \fi}
\xydef@\checkZeroEdge@{%
 \expandafter\DN@\expandafter{\expandafter\noexpand\the\Edge@c}%
 \DNii@{\noexpand\zeroEdge}%
 \ifx\next@\nextii@ \DN@{\expandafter\Edge@c\expandafter{\objectEdge}}%
 \ifdim\zz@\L@c \ifdim\zz@\R@c \ifdim\zz@\U@c \ifdim\zz@\D@c
 \DN@{}\fi\fi\fi\fi
 \else
 \ifdim\zz@\L@c \ifdim\zz@\R@c \ifdim\zz@\U@c \ifdim\zz@\D@c
 \DN@{\Edge@c={\zeroEdge}}\fi\fi\fi\fi
 \fi \next@}
\xydef@\OBJECT@x{\toks@={\egroup\def\Drop@@}%
 \expandafter\addtotoks@\expandafter{\expandafter{\Drop@@}\def\Connect@@}%
 \expandafter\addtotoks@\expandafter{\expandafter{\Connect@@}}%
 \edef\tmp@{\D@c=\the\D@c \U@c=\the\U@c \L@c=\the\L@c \R@c=\the\R@c
 \Edge@c={\expandafter\noexpand\the\Edge@c}%
 \ifInvisible@\noexpand\Invisible@true\else\noexpand\Invisible@false\fi
 \ifHidden@\noexpand\Hidden@true\else\noexpand\Hidden@false\fi
 \def\noexpand\Leftness@{\Leftness@}\def\noexpand\Upness@{\Upness@}}%
 \expandafter\addtotoks@\expandafter{\tmp@}\the\toks@}
\xynew@{if}\ifInvisible@
\xynew@{if}\ifHidden@
\xydef@\Leftness@{}
\xydef@\Upness@{}
\xydef@\Drop@@{\boxz@}
\xydef@\Connect@@{}
\xydef@\objectbox#1{\hbox{$\m@th\objectstyle{#1}$}}
\xylet@\objectstyle=\textstyle
\xydef@\object{\hbox\bgroup\object@}
\xydef@\object@{%
 \edef\next@{={\DirectionfromtheDirection@}}\expandafter\toks@\next@
 \plainxy@ \xyFN@\OBJECT@}
\xydef@\composite#1#{\hbox\bgroup\composite@{#1}}
\xydef@\composite@#1#2{%
 \DN@{#1}\ifx\next@\empty\else\xywarning@{no variants of
 \string\composite\space allowed}\fi
 \global\setbox9=\hbox\bgroup
 \D@p=-\maxdimen \U@p=-\maxdimen \L@p=-\maxdimen \R@p=-\maxdimen
 \xyFN@\composite@i#2@}
\xydef@\composite@i{%
 \ifx \space@\next \expandafter\DN@\space{\xyFN@\composite@i}%
 \else\ifx *\next \DN@ *{\xyFN@\composite@i}%
 \else\ifx @\next \DN@ @{\composite@x}%
 \xyerror@{<composite> object expected}{}\czeroEdge@
 \else \DN@{\composite@ii}\fi\fi\fi \next@}
\xydef@\composite@ii#1#{\composite@iii{#1}}
\xydef@\composite@iii#1#2{%
 \setbox\z@=\object#1{#2}%
 \ifInvisible@ \setboxz@h{}%
 \else \setboxz@h{\kern-\L@c \boxz@}\ht\z@=\z@ \dp\z@=\z@ \wd\z@=\z@ {\Drop@@}\fi
 \ifHidden@\else
 \ifdim\U@p<\U@c \U@p=\U@c \fi \ifdim\D@p<\D@c \D@p=\D@c \fi
 \ifdim\R@p<\R@c \R@p=\R@c \fi \ifdim\L@p<\L@c \L@p=\L@c \fi
 \fi
 \xyFN@\composite@iv}
\xydef@\composite@iv{%
 \ifx \space@\next \expandafter\DN@\space{\xyFN@\composite@iv}%
 \else \ifx @\next \DN@ @{\composite@x}%
 \else \let\next@=\composite@i \fi\fi \next@}
\xydef@\composite@x{%
 \edef\tmp@{\egroup \D@c=\the\D@p \U@c=\the\U@p \L@c=\the\L@p \R@c=\the\R@p}\tmp@
 \setboxz@h{\kern\L@c\box9}\ht\z@=\U@c \dp\z@=\D@c
 \dimen@=\L@c \advance\dimen@\R@c \wdz@=\dimen@
 \Edge@c={\rectangleEdge}\computeLeftUpness@ \boxz@
 \OBJECT@x}
\xydef@\computeLeftUpness@{%
 \dimen@=\L@c \advance\dimen@\R@c
 \ifdim\zz@\dimen@ \def\Connect@@{\straight@{\dottedSpread@\jot}}%
 \ifdim\zz@\L@c\else
 \DN@{\zeroEdge}\expandafter\DNii@\expandafter{\the\Edge@c}%
 \ifx\next@\nextii@\Edge@c={\rectangleEdge}\fi\fi
 \else \quotient@\Leftness@\L@c\dimen@ \fi
 \dimen@=\U@c \advance\dimen@\D@c 
 \ifdim\zz@\dimen@ \def\Connect@@{\straight@{\dottedSpread@\jot}}%
 \ifdim\zz@\U@c\else
 \DN@{\zeroEdge}\expandafter\DNii@\expandafter{\the\Edge@c}%
 \ifx\next@\nextii@\Edge@c={\rectangleEdge}\fi\fi
 \else \quotient@\Upness@\U@c\dimen@ \fi}
\xydef@\xybox#1{\xy#1\endxy \Edge@c={\rectangleEdge}\computeLeftUpness@}
\xydef@\OBJECT@shift{%
 \let\xy@=\xytotoks@ \afterVECTORorEMPTY
 {\OBJECT@shift@}%
 {\addtotoks@{\X@c=-\L@c \advance\X@c\R@p \advance\X@c\L@p \Y@c=\U@p}\OBJECT@shift@}}
\xydef@\OBJECT@shift@{%
 \addtotoks@{\advance\U@p-\Y@c
 \advance\L@c\X@c \advance\R@c-\X@c \advance\D@c\Y@c \advance\U@c-\Y@c
 \computeLeftUpness@}%
 \let\xy@=\oxy@
 \xyFN@\OBJECT@}
\xylet@\objectmargin@=\jot
\xylet@\objectwidth@=\z@
\xylet@\objectheight@=\z@
\xydef@\objectmargin{\afterADDOP{\Addop@@\objectmargin@}}
\xydef@\objectwidth{\afterADDOP{\Addop@@\objectwidth@}}
\xydef@\objectheight{\afterADDOP{\Addop@@\objectheight@}}
\xydef@\OBJECT@change{%
 \afterADDOP{%
 \addEQ@\ifx \next
 \addtotoks@{\X@c=\D@c \advance\X@c\U@c \Y@c=\L@c \advance\Y@c\R@c}%
 \else
 \addtotoks@{\X@c=\objectmargin@ \advance\X@c\X@c \Y@c=\X@c}%
 \fi
 \let\xy@=\xytotoks@ 
 \afterVECTORorEMPTY\OBJECT@change@\OBJECT@change@}}
\xydef@\OBJECT@set{%
 \afterADDOP{%
 \let\xy@=\xytotoks@ \afterVECTORorEMPTY\OBJECT@change@
 {\addtotoks@{\X@c=\objectwidth@ \Y@c=\objectheight@}\OBJECT@change@}}}
\xydef@\OBJECT@change@{%
 \addtotoks@{\advance\R@c\L@c \advance\R@p-\L@c \let\tmp@=\R@c}%
 \expandafter\addtotoks@\expandafter{\Addop@@\tmp@\X@c\R@c=\tmp@
 \L@c=\Leftness@\R@c \advance\R@p\L@c \advance\R@c-\L@c}%
 \addtotoks@{\advance\D@c\U@c \let\tmp@=\D@c}%
 \expandafter\addtotoks@\expandafter{\Addop@@\tmp@\Y@c\D@c=\tmp@
 \U@c=\Upness@\D@c \advance\D@c-\U@c}%
 \let\xy@=\oxy@ \xyFN@\OBJECT@}
\xydef@\afterADDOP#1{\def\afterADDOP@{#1}\xyFN@\ADDOP@}
\xylet@\afterADDOP@=\empty
\xydef@\ADDOP@{%
 \ifx \space@\next \expandafter\DN@\space{\xyFN@\ADDOP@}%
 \else\addPLUS@\ifx \next \addPLUS@\DN@{\xyFN@\ADDOP@plus}%
 \else\addDASH@\ifx \next \addDASH@\DN@{\xyFN@\ADDOP@minus}%
 \else\addEQ@\ifx \next
 \addEQ@\DN@{\def\Addop@@{\Addop@0+=}\afterADDOP@}%
 \else
 \DN@{\def\Addop@@{\Addop@0+=}\afterADDOP@}%
 \fi\fi\fi\fi \next@}
\xydef@\ADDOP@plus{%
 \addEQ@\ifx \next
 \addEQ@\DN@{\def\Addop@@{\Addop@0+<}\afterADDOP@}%
 \else
 \DN@{\def\Addop@@{\Addop@1+=}\afterADDOP@}%
 \fi \next@}
\xydef@\ADDOP@minus{%
 \addEQ@\ifx \next
 \addEQ@\DN@{\def\Addop@@{\Addop@0+>}\afterADDOP@}%
 \else
 \DN@{\def\Addop@@{\Addop@1-=}\afterADDOP@}%
 \fi \next@}
\xydef@\Addop@#1#2#3#4#5{%
 \dimen@=#4\relax \edef#4{\the\dimen@}%
 \dimen@=#1\dimen@ \advance\dimen@#2#5\relax \advance\dimen@ 1sp
 \ifdim\dimen@#3#4\else \edef#4{\the\dimen@}\fi
 \ifx\xy@\xyinitial@\else \DN@##1{\xy@@{\edef#4{##1}\checkZeroEdge@}}%
 \expandafter\next@\expandafter{#4}\fi}
\xydef@\objectEdge{\rectangleEdge}
\xydefcsname@{shape [r]}{%
 \advance\R@c\L@c \L@c=.5\U@c \advance\L@c.5\D@c \advance\R@c-\L@c}
\xydefcsname@{shape [l]}{%
 \advance\L@c\R@c \R@c=.5\U@c \advance\R@c.5\D@c \advance\L@c-\R@c}
\xydefcsname@{shape [u]}{%
 \advance\U@c\D@c \D@c=.5\L@c \advance\D@c.5\R@c \advance\U@c-\D@c}
\xydefcsname@{shape [d]}{%
 \advance\D@c\U@c \U@c=.5\L@c \advance\U@c.5\R@c \advance\D@c-\U@c}
\xydefcsname@{shape [c]}{%
 \advance\D@c\U@c \D@c=.5\D@c \U@c=\D@c \advance\L@c\R@c \L@c=.5\L@c \R@c=\L@c}
\xydef@\OBJECT@shape#1{\DN@{shape [#1]}%
 \expandafter\let\expandafter\nextii@\csname\codeof\next@\endcsname
 \ifx\nextii@\relax\DN@{style [#1]}%
 \expandafter\let\expandafter\nextii@\csname\codeof\next@\endcsname
 \ifx\nextii@\relax \DN@{\OBJECT@shapei[#1]}%
 \else\DN@{\nextii@\xyFN@\OBJECT@}\fi
 \else \expandafter\addtotoks@\expandafter{\nextii@}%
 \DN@{\xyFN@\OBJECT@}%
 \fi \next@}
\xydefcsname@{shape []}{\the\Edge@c5\relax}%
\xydefcsname@{shape [o]}{\Edge@c={\circleEdge}\the\Edge@c5\relax 
 \Edge@c={\circleEdge}\def\prevEdge@@{\circleEdge}}
\xydefcsname@{shape [.]}{\czeroEdge@}
\xydef@\OBJECT@shapei[#1#2]{\DN@{shape [#1...]}%
 \expandafter\let\expandafter\next\csname\codeof\next@\endcsname
 \ifx\next\relax\DN@{*stylechar@#1@}%
 \expandafter\let\expandafter\next\csname\codeof\next@\endcsname
 \ifx\next\relax\DNii@{shape [#1#2]}%
 \xywarning@{illegal [<shape>] ignored: \codeof\nextii@\space not defined}%
 \DN@{\xyFN@\OBJECT@}\else \DN@{\next{#2}\xyFN@\OBJECT@}\fi
 \else
 \expandafter\addtotoks@\expandafter{\next{#2}}\DN@{\xyFN@\OBJECT@}%
 \fi \next@}
\xywarnifdefined\preXY@style@
\xywarnifdefined\postXY@style@
 \gdef\preXY@style@{}
 \gdef\postXY@style@{}
\xydef@\xy@style@{}
\xydef@\preStyle@@{}
\xydef@\postStyle@@{}
\xydef@\xypre@Style@{\let\xypost@Style@=\xypost@Style@@}
\xydef@\xypost@Style@{\let\xypre@Style@=\xypre@Style@@}
\xydef@\xypre@skipStyle@#1\xypost@Style@@{#1}
\xydef@\xypre@Style@@{\styletoks@={}\preXY@style@
 \expandafter\DN@\expandafter{\the\styletoks@}%
 \ifx\next@\empty \DN@{\let\xypre@Style@=\relax \xypre@skipStyle@}%
 \else
 \let\xypre@Style@=\relax \let\xypost@Style@=\xypost@Style@@
 \DN@{\expandafter\xydoprestyles@\expandafter{\the\styletoks@}}%
 \fi \next@ }
\xydef@\xypost@Style@@{\styletoks@={}\postXY@style@
 \expandafter\xydopoststyles@\expandafter{\the\styletoks@}%
 \let\xypost@Style@=\relax \let\xypre@Style@=\xypre@Style@@ }
\xydef@\xydoprestyles@@{\literal@}
\xydef@\xydopoststyles@@{\literal@}
\xylet@\xydoprestyles@=\xydoprestyles@@
\xylet@\xydopoststyles@=\xydopoststyles@@
\xydef@\loadxystyle@{\let\xypre@Style@=\xypre@Style@@}\loadxystyle@
\xydef@\preXYstyle@@{}
\xydef@\postXYstyle@@{}
\xydef@\Unloadstyle@{%
 \let\preXYstyle@=\preXYstyle@@
 \let\postXYstyle@=\postXYstyle@@}
\xydef@\checkXyStyle@{\ifx\xy@style@\empty\resetStyle@
 \ifInvisible@\else\ifHidden@\else\DN@{\no@@}\ifx\next@\Connect@@
 \else\styleDrop@\styleConnect@\fi\fi\fi\fi}
\xydef@\resetStyle@{\gdef\preXY@style@{}\gdef\postXY@style@{}}
\xydef@\styleDrop@{\let\xy@style@=\relax
 \expandafter\def\expandafter\Drop@@\expandafter{%
 \expandafter\xypre@Style@@\Drop@@\xypost@Style@@}}
\xydef@\styleConnect@{\let\xy@style@=\relax
 \expandafter\def\expandafter\Connect@@\expandafter{%
 \expandafter\xypre@Style@@\Connect@@\xypost@Style@@}}
\xydefcsname@{shape [=...]}#1{\checkXyStyle@ \addtotoks@{\xynamestyle@{#1}}}
\xydef@\xynamestyle@#1{\checkXyStyle@
 \expandafter\DNii@\expandafter{\csname shape [#1]\endcsname}%
 \expandafter\ifx\nextii@\relax\xywarning@{Defining new style [#1]}%
 \else\xywarning@{Redefining style [#1]}\fi
 \expandafter\xynamestyle@@\csname shape [#1]\endcsname }
\xydef@\xynamestyle@@#1{%
 \expandafter\def\expandafter\tmp@\expandafter{\preXY@style@}%
 \DN@##1{\def\tmp@{\checkXyStyle@\gdef\preXY@style@{##1}}}%
 \expandafter\next@\expandafter{\preXY@style@}%
 \DN@##1{\expandafter\gdef\expandafter#1\expandafter{\tmp@
 \gdef\postXY@style@{##1}}}%
 \expandafter\next@\expandafter{\postXY@style@}\DN@{}}
\xydef@\newxystyle#1#2#3{%
 \DN@{#3}\ifx\next@\empty 
 \xydefcsname@{shape [#1]}{\csname xyshape@#1@\endcsname}%
 \else \expandafter\def\csname shape [#1]\endcsname{%
 \csname xyshape@#1@\endcsname}\fi
 \DN@{#2}\ifx\next@\empty
 \expandafter\def\csname xyshape@#1@\endcsname{%
 \xyundefinedStyle@{#1}{}@@}%
 \else \expandafter\def\csname xyshape@#1@\endcsname{#2}\fi}
\xydef@\xyundefinedStyle@#1#2@@{%
 \xywarning@{style #1 not defined, nothing to apply}}
\xynew@{toks}{\styletoks@}
\xydef@\addtostyletoks@#1{%
 \expandafter\styletoks@\expandafter{\the\styletoks@#1}}
\xydef@\applyFIFOstyle@#1#2#3#4{\bgroup
 \styletoks@={\egroup\gdef\preXY@style@}%
 \expandafter\toks@\expandafter{\preXY@style@}%
 \expandafter\addtotoks@\expandafter{\expandafter#1\expandafter{#2}}%
 \expandafter\addtostyletoks@\expandafter{\expandafter{\the\toks@}%
 \gdef\postXY@style@}%
 \expandafter\toks@\expandafter{\expandafter#3\expandafter{#4}}%
 \expandafter\addtotoks@\expandafter{\postXY@style@}%
 \expandafter\addtostyletoks@\expandafter{\expandafter{\the\toks@}}%
 \the\styletoks@ }
\xydef@\applyLIFOstyle@#1#2#3#4{\bgroup
 \styletoks@={\egroup\gdef\preXY@style@}%
 \expandafter\toks@\expandafter{\expandafter#1\expandafter{#2}}%
 \expandafter\addtotoks@\expandafter{\preXY@style@}%
 \expandafter\addtostyletoks@\expandafter{\expandafter{\the\toks@}%
 \gdef\postXY@style@}%
 \expandafter\toks@\expandafter{\postXY@style@}%
 \expandafter\addtotoks@\expandafter{\expandafter#3\expandafter{#4}}%
 \expandafter\addtostyletoks@\expandafter{\expandafter{\the\toks@}}%
 \the\styletoks@ }
\xydef@\OBJECT@direction{\afterDIRECTIONorEMPTY{%
 \edef\next@{{\DirectionfromtheDirection@}}\expandafter\addtotoks@\next@
 \xyFN@\OBJECT@}%
 {\xyFN@\OBJECT@}}
\xydef@\afterDIRECTIONorEMPTY#1#2{%
 \DN@##1{\def\afterDIRECTION@{\def\afterDIRECTION@{##1}%
 \ifDIRECTIONempty@\DN@{#2}\else\DN@{#1}\fi \next@}}%
 \expandafter\next@\expandafter{\afterDIRECTION@}%
 \xyFN@\DIRECTION@}
\xylet@\afterDIRECTION@=\empty
\xynew@{if}\ifDIRECTIONempty@
\xydef@\DIRECTION@{%
 \ifx \space@\next \expandafter\DN@\space{\xyFN@\DIRECTION@}%
 \else\ifx v\next \DN@ v{\DIRECTION@v}%
 \else\ifx \bgroup\next \let\next@=\DIRECTION@group
 \else\ifx (\next \DN@({\xyFN@\DIRECTION@open}%
 \else
 \DN@{\count@=8
 \afterDIAG{\ifnum\count@=8 \DN@{\DIRECTIONempty@true \xyFN@\DIRECTION@i}%
 \else \DN@{\xy@@{\dimen@=\xydashl@}\Directionfromdiag@}\fi \next@}}%
 \fi\fi\fi\fi \next@}
\xydef@\DIRECTION@open{%
 \ifx *\next \DN@*##1*){\DIRECTION@group{##1}}%
 \else \DN@{\xyerror@{(* <pos> *) expected}{} \xyFN@\DIRECTION@i}%
 \fi \next@}
\def\afterDIAG#1{\def\afterDIAG@{#1}\xyFN@\DIAG@}
\xydef@\DIAG@{%
 \ifx d\next \DN@ d{\count@=1 \xyFN@\DIAG@@}%
 \else\ifx r\next \DN@ r{\count@=3 \xyFN@\DIAG@@}%
 \else\ifx u\next \DN@ u{\count@=5 \xyFN@\DIAG@@}%
 \else\ifx l\next \DN@ l{\count@=7 \xyFN@\DIAG@@}%
 \else \let\next@=\afterDIAG@
 \fi\fi\fi\fi \next@}
\xydef@\DIAG@@{\ifcase\count@ \or
 \DIAG@@@ l0r2\or\or \DIAG@@@ d2u4\or\or \DIAG@@@ r4l6\or\or \DIAG@@@ u6d0%
 \else\xybug@{impossible <diag> number}\fi
 \next@}
\xydef@\DIAG@@@#1#2#3#4{%
 \ifx #1\next \count@=#2\DN@#1{\afterDIAG@}%
 \else \ifx #3\next \count@=#4\DN@#3{\afterDIAG@}%
 \else \let\next@=\afterDIAG@ \fi\fi}
\xydef@\Directionfromdiag@{\ifcase\count@
 \xy@@{\dlDirection@\dimen@}%
 \or \xy@@{\dDirection@\dimen@}%
 \or \xy@@{\drDirection@\dimen@}%
 \or \xy@@{\rDirection@\dimen@}%
 \or \xy@@{\urDirection@\dimen@}%
 \or \xy@@{\uDirection@\dimen@}%
 \or \xy@@{\ulDirection@\dimen@}%
 \or \xy@@{\lDirection@\dimen@}%
 \or
 \else\xybug@{impossible <diag>}\fi
 \DIRECTIONempty@false\xyFN@\DIRECTION@i}
\xydef@\DIRECTION@v{%
 \xy@{v}{\enter@{\cfromthec@ \X@origin=\the\X@origin \Y@origin=\the\Y@origin
 \X@p=\the\X@p \Y@p=\the\Y@p}%
 \X@origin=\z@ \Y@origin=\z@}%
 \afterVECTORorEMPTY
 {\xy@@{\X@p=\z@ \Y@p=\z@ \setupDirection@ \leave@}%
 \DIRECTIONempty@false \xyFN@\DIRECTION@i}%
 {\xy@@\leave@ \xyerror@{<vector> expected after v}{}%
 \DIRECTIONempty@false \xyFN@\DIRECTION@i}}
\xydef@\DIRECTION@group#1{%
 \xy@@{\begingroup}\xy@@ix@{#1}\xy@@{\plainxy@\expandafter\POS\the\toks9\relax
 \setupDirection@\edef\next@{\endgroup \DirectionfromtheDirection@}\next@}%
 \DIRECTIONempty@false \xyFN@\DIRECTION@i}
\xydef@\DIRECTION@i{%
 \ifx ^\next \DN@ ^{\xy@^{\aboveDirection@\xydashl@}%
 \DIRECTIONempty@false \xyFN@\DIRECTION@i}%
 \else\ifx _\next \DN@ _{\xy@_{\belowDirection@\xydashl@}%
 \DIRECTIONempty@false \xyFN@\DIRECTION@i}%
 \else\ifx :\next \DN@ :{%
 \xy@{:}{\enter@{\cfromthec@ \basefromthebase@ \X@p=\the\X@p \Y@p=\the\Y@p}%
 \X@origin=\z@ \Y@origin=\z@
 \X@xbase=\cosDirection\xydashl@ \Y@xbase=\sinDirection\xydashl@
 \X@ybase=-\Y@xbase \Y@ybase=\X@xbase}%
 \afterVECTORorEMPTY
 {\xy@@{\X@p=\z@ \Y@p=\z@ \setupDirection@ \leave@}%
 \DIRECTIONempty@false \xyFN@\DIRECTION@i}%
 {\xy@@\leave@ \xyerror@{<vector> expected after :}{}%
 \DIRECTIONempty@false \xyFN@\DIRECTION@i}}%
 \else
 \let\next@=\afterDIRECTION@
 \fi\fi\fi \next@}
\xydef@\drop#1#{\DN@##1{\xy@@ix@{{#1}{##1}}%
 \xy@{\drop#1{##1}}{\expandafter\drop@\the\toks9}\ignorespaces}\next@}
\xydef@\connect#1#{\DN@##1{\xy@@ix@{{#1}{##1}}%
 \xy@{\connect#1{##1}}{\expandafter\connect@\the\toks9}\ignorespaces}\next@}
\xydef@\save{\relax\saveC}
\xydef@\saveC{\xy@\save\save@ \POS}
\xydef@\save@{\enter@{\cfromthec@ \pfromthep@ \basefromthebase@}}
\xydef@\restore{\xy@\restore\leave@ \ignorespaces}
\xydef@\xyecho{%
 \xy@\xyecho{\let\xy@=\xyecho@ \message{\string\xyecho}}\POS}
\xydef@\xyecho@#1#2{{\def\1{#1}\ifx\1\empty\else\message{\codeof\1}\fi}%
 \oxy@{#1}{#2}}
\xydef@\xyverbose{%
 \xy@\xyverbose{\let\xy@=\xyverbose@
 \W@{Xy: \string\xyverbose\xytracelineno@}}\POS}
\xydef@\xyverbose@#1#2{%
 {\def\1{#1}\ifx\1\empty\else\W@{Xy: \codeof\1\xytracelineno@}\fi}%
 \oxy@{#1}{#2}}
\xydef@\xytracing{%
 \xy@\xytracing{\let\xy@=\xytracing@
 \W@{Xy TRACE: \string\xytracing\xytracelineno@}\xystatus@:}\POS}
\xydef@\xytracing@#1#2{{\def\1{#1}\def\2{#2}%
 \W@{Xy TRACE: \codeof\1 {\codeof\2}\xytracelineno@}}\oxy@{#1}{#2}\xystatus@:}
\xydef@\xystatus@#1{%
 \W@{#1 c=<\the\X@c,\the\Y@c> \expandafter\string\the\Edge@c
 \string[\the\L@c+\the\R@c,\the\D@c+\the\U@c\string]}%
 \W@{#1 p=<\the\X@p,\the\Y@p> \expandafter\string\the\Edge@p
 \string[\the\L@p+\the\R@p,\the\D@p+\the\U@p\string]}%
 \W@{#1 [d=<\the\d@X,\the\d@Y>
 Direction=\the\Direction=\string(\cosDirection,\sinDirection\string)]
 S=\the\csp@}%
 \W@{#1 base = <\the\X@origin,\the\Y@origin> +
 x\string*<\the\X@xbase,\the\Y@xbase> +
 y\string*<\the\X@ybase,\the\Y@ybase>}%
 \W@{#1 min/max = <\the\X@min,\the\Y@min> / <\the\X@max,\the\Y@max>}}
\xydef@\xyquiet{\xy@\xyquiet{\let\xy@=\oxy@}}
\xydef@\xyignore#1{\xy@\xyignore{\xyignore@{#1}}\ignorespaces}
\xydef@\xyignore@#1{{\let\xy@=\xyeat@ \let\oxy@=\xy@ \POS#1\relax}}
\xydef@\xyeat@#1#2{}
\xydef@\xycompile@@{\jobname-}
\xydef@\xycompileno@@{0}
\xydef@\CompilePrefix#1{%
 \def\xycompile@@{#1}\xdef\xycompile@@{\codeof\xycompile@@}%
 \xdef\xycompileno@@{0}}
\xydef@\xycompile{%
 \count@=\xycompileno@@ \advance\count@\@ne
 \xdef\xycompileno@@{\ifnum10>\count@ 0\fi \the\count@}%
 \edef\next{\noexpand\xycompileto{\xycompile@@\xycompileno@@}}\next}
\xylet@\compilename@@=\empty
\xylet@\xyrecompile@@=\relax
\xydef@\xycompileto#1#2{%
 \if\inxy@ \DN@{\xy@@{\nter@{}}}%
 \else \DN@{\xy \xy@@{\nter@{\endxy}}}\fi \next@
 \ifxysaving@ \xyerror@{Compilations can not be nested}{}\fi
 \DN@{#1}\edef\compilename@@{\codeof\next@}\DNii@{#2}%
 \def\xyrecompile@@{recompiling TRUNCATED}%
 \expandafter\xyinputorelse@\expandafter{\compilename@@.xyc}%
 {\def\xyrecompile@@{compiling to}}%
 \ifx\xyrecompile@@\relax\else \expandafter\xyrecompile@ \fi \xy@@\leave@
 \ignorespaces}
\xydef@\xyrecompile@{%
 \message{(\xyrecompile@@\space\string`\compilename@@.xyc\string'}%
 \DN@{\immediate\openout\xywrite@=}\expandafter\next@\compilename@@.xyc
 \immediate\write\xywrite@{%
 \string\xycompiled{\compilename@@}%
 {\the\year/\the\month/\the\day\string:\the\time\xytracelineno@}%
 {Xy-pic \xyversion}\xycomment@}%
 \immediate\write\xywrite@{{\codeof\nextii@}\relax}%
 {\xysaving@ \expandafter\POS\nextii@ \relax}%
 \immediate\write\xywrite@{\string\xyendcompiled}%
 \immediate\closeout\xywrite@ \message{done)}%
 \expandafter\input\compilename@@.xyc }
\xydef@\xysaving@{\let\xy@=\xysave@ \let\oxy@=\xy@
 \let\xy@@ix@=\xysave@@toksix@ \xysaving@true}
\xynew@{if}\ifxysaving@ \xysaving@false
\xydef@\xysave@#1#2{{\DN@{{#1}{#2}}%
 \immediate\write\xywrite@{\string\xy@\codeof\next@\xycomment@}}}
\xydef@\xysave@@toksix@#1{{\DN@{{#1}}%
 \immediate\write\xywrite@{\string\xy@@ix@\codeof\next@\relax}}}
\xywarnifdefined\xycomment@
{\catcode`\%=12 \catcode`\(=1 \catcode`\)=2 \gdef\xycomment@(
\xydef@\xycompiled#1#2#3#4{\DN@{#1}\edef\next@{\codeof\next@}%
 \ifx\next@\compilename@@\else
 \xywarning@{This file does not contain the result of
 \string\xycompileto{\compilename@@}{...}^^J%
 but of \string\xycompileto{\next@}}\fi
 \edef\next{Xy-pic \xyversion}\DN@{#3}\ifx\next\next@
 \DN@{#4}\ifx\next@\nextii@ \xycatcodes
 \else \def\xyrecompile@@{recompiling to}  \fi
 \else \def\xyrecompile@@{Xy-pic version change - recompiling}  \fi}
\xydef@\xyendcompiled{\let\xyrecompile@@=\relax \xyuncatcodes }
\message{kernel objects:}
\message{directionals,}
\xydef@\dir{\hbox\bgroup\xyFN@\dir@i}
\xydef@\dir@i{\ifx *\next \DN@*{\object@}\else \let\next@=\dir@ii \fi \next@}
\xydef@\dir@ii#1#{\dir@{#1}}
\xydef@\dir@#1#2{\DN@{dir#1{#2}}%
 \expandafter\let\expandafter\next\csname\codeof\next@\endcsname
 \ifx\next\relax \DN@{dir{#2}}%
 \expandafter\let\expandafter\next\csname\codeof\next@\endcsname
 \ifx\next\relax \DN@{\dir#1{#2}}%
 \xyerror@{illegal <dir>: \codeof\next@\space not defined}{}%
 \let\next=\no@ \fi\fi \next}
\xydefcsname@{dir{}}{\no@}
\xyletcsnamecsname@{dir0{}}{dir{}}
\xyletcsnamecsname@{dir1{}}{dir{}}
\xyletcsnamecsname@{dir^{}}{dir{}}
\xyletcsnamecsname@{dir_{}}{dir{}}
\xyletcsnamecsname@{dir2{}}{dir{}}
\xyletcsnamecsname@{dir3{}}{dir{}}
\xyletcsnamecsname@{dir{ }}{dir{}}
\xydef@\no@{\egroup \czeroEdge@ \Invisible@false \Hidden@false
 \def\Leftness@{.5}\def\Upness@{.5}%
 \def\Drop@@{\setbox\z@=\copy\voidb@x}\def\Connect@@{\no@@}}
\xydefcsname@{dir1{-}}{\line@}
\xydefcsname@{dir2{-}}{\line@ \double@\xydashh@}
\xydefcsname@{dir3{-}}{\line@ \triple@\xydashh@}
\xyletcsnamecsname@{dir0{-}}{dir{}}
\xyletcsnamecsname@{dir{-}}{dir1{-}}
\xyletcsnamecsname@{dir{=}}{dir2{-}}
\xydef@\line@{\dimen@=\sd@Y\sinDirection\xydashl@
 \ifnum\SemiDirectionChar<31 \D@c=\z@ \U@c=\dimen@ \DN@{\d@Y<\z@}%
 \else\ifnum\SemiDirectionChar<64 \D@c=\dimen@ \U@c=\z@ \DN@{\z@<\d@Y}%
 \else\ifnum\SemiDirectionChar<96 \D@c=\dimen@ \U@c=\z@ \DN@{\d@X<\z@}%
 \else \D@c=\z@ \U@c=\dimen@ \DN@{\d@X<\z@}\fi\fi\fi
 \setboxz@h{\line@@}\ht\z@=\U@c \dp\z@=\D@c
 \L@c=\z@ \R@c=\wdz@
 \ifdim\next@ \dimen@=\R@c \R@c=\L@c \L@c=\dimen@
 \dimen@=\U@c \U@c=\D@c \D@c=\dimen@ \advance\dimen@-\U@c
 \lower\dimen@\boxz@
 \else \boxz@ \fi
 \edef\tmp@{\egroup \U@c=\the\U@c \D@c=\the\D@c \L@c=\the\L@c \R@c=\the\R@c}%
 \tmp@
 \Edge@c={\rectangleEdge}\Invisible@false\Hidden@false
 \ifdim\z@<\U@c \def\Upness@{1}\else \def\Upness@{0}\fi
 \ifdim\z@<\L@c \def\Leftness@{1}\else \def\Leftness@{0}\fi
 \def\Drop@@{\boxz@}\def\Connect@@{\solid@}}
\xydef@\line@@{{\xydashfont\SemiDirectionChar\/}}
\xydef@\solid@{%
 \ifInvisible@ \DN@{\no@@}%
 \else \dimen@=\Y@c \advance\dimen@-\Y@p
 \ifjusthvtest@.05pt>\ifdim\dimen@<\z@-\fi\dimen@ \DN@{\solidhrule@}%
 \else \dimen@=\X@c \advance\dimen@-\X@p
 \ifjusthvtest@.05pt>\ifdim\dimen@<\z@-\fi\dimen@ \DN@{\solidvrule@}%
 \else \DN@{\straight@\solidSpread@}\fi\fi\fi
 \next@}
\xydef@\solidSpread@{\ifnum\z@<\count@@ \advance\count@@\@ne \fi}
\xylet@\ifjusthvtest@=\ifdim
\xydef@\NoRules{\let\ifjusthvtest@=\iffalse}
\xydef@\UseRules{\let\ifjusthvtest@=\ifdim}
\xydef@\solidvrule@{\no@@{%
 \ifdim\Y@c<\Y@p \dimen@=\Y@c \Y@c=\Y@p \Y@p=\dimen@ \advance\Y@c-\D@p \advance\Y@p\U@c
 \else \advance\Y@c-\D@c \advance\Y@p\U@p \fi
 \advance\X@c-.5\xydashw@
 \setboxz@h{\kern\X@c \vrule width\xydashw@ height\Y@c depth-\Y@p}%
 \ht\z@=\z@ \wd\z@=\z@ \dp\z@=\z@ {\Drop@@}}}
\xydef@\solidhrule@{\no@@{%
 \ifdim\X@c<\X@p \advance\X@c\R@c \advance\X@p-\L@p
 \else \dimen@=\X@c \X@c=\X@p \X@p=\dimen@ \advance\X@c\R@p \advance\X@p-\L@c \fi
 \advance\X@p-\X@c \advance\Y@c.5\xydashw@ \advance\Y@p-.5\xydashw@
 \setboxz@h{\kern\X@c \vrule width\X@p height\Y@c depth-\Y@p}%
 \ht\z@=\z@ \wd\z@=\z@ \dp\z@=\z@ {\Drop@@}}}
\xydef@\zerodot{\copy\zerodotbox@}
\xydefcsname@{dir1{.}}{\point@}
\xydefcsname@{dir2{.}}{\point@ \double@\xydashh@}
\xydefcsname@{dir3{.}}{\point@ \triple@\xydashh@}
\xyletcsnamecsname@{dir0{.}}{dir{}}
\xyletcsnamecsname@{dir{.}}{dir1{.}}
\xyletcsnamecsname@{dir{:}}{dir2{.}}
\xydef@\point@{\pointlike@\zerodot\p@}
\xydef@\pointlike@#1#2{%
 \setboxz@h{#1}\wdz@=\z@ \ht\z@=\z@ \dp\z@=\z@ \boxz@\egroup
 \Invisible@false \Hidden@false \def\Leftness@{.5}\def\Upness@{.5}\ctipEdge@
 \def\Drop@@{\boxz@}\def\Connect@@{\straight@{\dottedSpread@{#2}}}}
\xydef@\dottedSpread@#1{\setupDirection@
 \dimen@=#1\relax \dimen@=2\dimen@
 \A@=\sd@X\cosDirection\dimen@ \B@=\sd@Y\sinDirection\dimen@
 \global\setbox\lastobjectbox@=\hbox to\A@{\hss
 \kern.5\A@\box\lastobjectbox@\kern.5\A@\hss}%
 \dp\lastobjectbox@=.5\B@ \ht\lastobjectbox@=.5\B@
 \advance\d@X\sd@X\A@ \advance\d@Y\sd@Y\B@
 \advance\X@c\sd@X.5\A@ \advance\Y@c\sd@Y.5\B@
 \ifdim\sd@Y\d@Y<\sd@X\d@X \dimen@=\sd@X\d@X
 \ifdim\zz@\A@\else \divide\dimen@\A@ \fi \count@@=\dimen@
 \else\ifdim\z@=\sd@Y\d@Y\else
 \dimen@=\sd@Y\d@Y \ifdim\zz@\B@\else \divide\dimen@\B@ \fi \count@@=\dimen@
 \fi\fi \advance\count@@\@ne}
\xydefcsname@{dir1{~}}{\squiggle@}
\xydefcsname@{dir2{~}}{\squiggle@ \double@\xybsqlh@}
\xydefcsname@{dir3{~}}{\squiggle@ \triple@\xybsqlh@}
\xyletcsnamecsname@{dir0{~}}{dir{}}
\xyletcsnamecsname@{dir{~}}{dir1{~}}
\xydef@\squiggle@{\xybsqlfont
 \dimen@=\sd@X\cosDirection\xybsqll@ \advance\dimen@.1\p@
 \dimen@ii=\sd@Y\sinDirection\xybsqll@
 \kern\dimen@\squiggle@@
 \edef\tmp@{\egroup \U@c=\the\dimen@ii \L@c=\the\dimen@}\tmp@
 \wdz@=2\L@c \R@c=\L@c \ht\z@=\U@c \D@c=\U@c \dp\z@=\U@c \Edge@c={\rectangleEdge}%
 \Invisible@false \Hidden@false \def\Leftness@{.5}\def\Upness@{.5}%
 \def\Drop@@{\boxz@}\def\Connect@@{\straight@\squiggledSpread@}}
\xydef@\squiggle@@{\DirectionChar \count@=\DirectionChar
 \advance\count@-64 \ifnum\count@<\z@ \advance\count@128 \fi \char\count@}
\xydef@\squiggledSpread@{%
 \dimen@=\d@X \advance\dimen@-\sd@X\count@@\A@ \advance\dimen@\sd@X.3\p@
 \advance\X@c-.5\dimen@ \advance\d@X-\dimen@
 \dimen@=\d@Y \advance\dimen@-\sd@Y\count@@\B@ \advance\dimen@\sd@Y.3\p@
 \advance\Y@c-.5\dimen@ \advance\d@Y-\dimen@}
\xydef@\double@#1{\edef\Drop@@{\dimen@=#1\relax
 \dimen@=.5\dimen@ \A@=-\sinDirection\dimen@ \B@=\cosDirection\dimen@
 \setbox2=\hbox{\kern\A@\raise\B@\copy\z@}\dp2=\z@ \ht2=\z@ \wd2=\z@ \box2
 \setbox2=\hbox{\kern-\A@\raise-\B@\boxz@}\dp2=\z@ \ht2=\z@ \wd2=\z@ \box2 }}
\xydef@\triple@#1{\edef\Drop@@{\dimen@=#1\relax
 \A@=-\sinDirection\dimen@ \B@=\cosDirection\dimen@
 \setbox2=\hbox{\kern\A@\raise\B@\copy\z@}\dp2=\z@ \ht2=\z@ \wd2=\z@ \box2
 \setbox2=\hbox{\kern-\A@\raise-\B@\copy\z@}\dp2=\z@ \ht2=\z@ \wd2=\z@ \box2
 \dp\z@=\z@ \ht\z@=\z@ \wdz@=\z@ \boxz@}}
\xydefcsname@{dir1{--}}{\dash@}
\xydefcsname@{dir2{--}}{\dash@ \double@\xydashh@}
\xydefcsname@{dir3{--}}{\dash@ \triple@\xydashh@}
\xyletcsnamecsname@{dir0{--}}{dir{}}
\xyletcsnamecsname@{dir{--}}{dir1{--}}
\xyletcsnamecsname@{dir{==}}{dir2{--}}
\xydef@\dash@{\line@ \wdz@=2\wdz@ \ht\z@=2\ht\z@ \dp\z@=2\dp\z@
 \multiply\D@c\tw@ \multiply\U@c\tw@ \multiply\L@c\tw@ \multiply\R@c\tw@
 \def\Connect@@{\straight@\dashedSpread@}}
\xydef@\dashedSpread@{\ifnum\z@<\count@@ \advance\count@@\@ne \fi
 \advance\d@X\sd@X.5\A@ \advance\d@Y\sd@Y.5\B@
 \ifdim\z@<\d@X \advance\X@c.5\A@ \fi \advance\Y@c\sd@Y.5\B@}
\xydefcsname@{dir1{~~}}{\dashsquiggle@}
\xydefcsname@{dir2{~~}}{\dashsquiggle@ \double@\xybsqlh@}
\xydefcsname@{dir3{~~}}{\dashsquiggle@ \triple@\xybsqlh@}
\xyletcsnamecsname@{dir0{~~}}{dir{}}
\xyletcsnamecsname@{dir{~~}}{dir1{~~}}
\xydef@\dashsquiggle@{\squiggle@
 \multiply\D@c\tw@ \multiply\U@c\tw@ \multiply\L@c\tw@ \multiply\R@c\tw@
 \dimen@=\L@c \advance\dimen@\R@c \wdz@=\dimen@ \ht\z@=\U@c \dp\z@=\D@c
 \def\Connect@@{\straight@\dashsquiggledSpread@}}
\xydef@\dashsquiggledSpread@{\ifnum\z@<\count@@ \advance\count@@\@ne \fi
 \advance\X@c.5\A@ \advance\d@X.5\A@ \advance\Y@c.25\B@ \advance\d@Y.5\B@}
\xyletcsnamecsname@{dir1{..}}{dir{.}}
\xyletcsnamecsname@{dir2{..}}{dir2{.}}
\xyletcsnamecsname@{dir3{..}}{dir3{.}}
\xyletcsnamecsname@{dir{..}}{dir1{.}}
\xyletcsnamecsname@{dir{::}}{dir2{.}}
\xylet@\ctipEdge@=\czeroEdge@
\xydefcsname@{dir1{>}}{\tip@}
\xydefcsname@{dir^{>}}{\atip@}
\xydefcsname@{dir_{>}}{\btip@}
\xyletcsnamecsname@{dir0{>}}{dir{}}
\xyletcsnamecsname@{dir{>}}{dir1{>}}
\xydefcsname@{dir1{<}}{\reverseDirection@\tip@}
\xydefcsname@{dir^{<}}{\reverseDirection@\btip@}
\xydefcsname@{dir_{<}}{\reverseDirection@\atip@}
\xyletcsnamecsname@{dir0{<}}{dir{}}
\xyletcsnamecsname@{dir{<}}{dir1{<}}
\xydef@\tip@{\tip@x\tip@@}
\xydef@\atip@{\tip@x\atip@@}
\xydef@\btip@{\tip@x\btip@@}
\xydef@\tip@x#1{#1\egroup
 \ctipEdge@ \Invisible@false \Hidden@false \def\Leftness@{.5}\def\Upness@{.5}%
 \def\Drop@@{\boxz@}\def\Connect@@{\straight@{\dottedSpread@\jot}}}
\xydef@\tip@@{\atip@@\btip@@}
\xydef@\atip@@{\xyatipfont\DirectionChar}
\xydef@\btip@@{\xybtipfont\DirectionChar}
\xydefcsname@{dir2{>}}{\Tip@}
\xydefcsname@{dir2{<}}{\reverseDirection@\Tip@}
\xydef@\Tip@{\kern2.5pt \vrule height2.5pt depth2.5pt width\z@
 \Tip@@ \kern2.5pt \egroup
 \U@c=2.5pt \D@c=2.5pt \L@c=2.5pt \R@c=2.5pt \Edge@c={\circleEdge}%
 \Invisible@false \Hidden@false \def\Leftness@{.5}\def\Upness@{.5}%
 \def\Drop@@{\boxz@}\def\Connect@@{\straight@{\dottedSpread@\jot}}}
\xydef@\Tip@@{\count@=\DirectionChar
 \advance\count@-4 \ifnum\count@<\z@ \advance\count@128 \fi
 \xyatipfont\char\count@
 \advance\count@ 8 \ifnum127<\count@ \advance\count@-128 \fi
 \xybtipfont\char\count@}
\xydefcsname@{dir3{>}}{\Ttip@}
\xydefcsname@{dir3{<}}{\reverseDirection@\Ttip@}
\xydef@\Ttip@{\kern3.2pt \vrule height3.2pt depth3.2pt width\z@
 \Ttip@@ \kern3.2pt \egroup
 \U@c=3.2pt \D@c=3.2pt \L@c=3.2pt \R@c=3.2pt \Edge@c={\circleEdge}%
 \Invisible@false \Hidden@false \def\Leftness@{.5}\def\Upness@{.5}%
 \def\Drop@@{\boxz@}\def\Connect@@{\straight@{\dottedSpread@\jot}}}
\xydef@\Ttip@@{%
 \setboxz@h\bgroup\reverseDirection@\line@ \wdz@=\z@ \ht\z@=\z@ \dp\z@=\z@
 \kern-\L@c \boxz@ \kern\L@c
 {\vDirection@(1,-.31)\xydashl@ \xyatipfont\char\DirectionChar}%
 {\vDirection@(1,+.31)\xydashl@ \xybtipfont\char\DirectionChar}}
\xydefcsname@{dir1{|}}{\stopper@}
\xydefcsname@{dir^{|}}{\aboveDirection@\xydashl@
 \shiftdir@\line@\z@ \pointlike@{}\xydashh@}
\xydefcsname@{dir_{|}}{\belowDirection@\xydashl@
 \shiftdir@\line@\z@ \pointlike@{}\xydashh@}
\xydefcsname@{dir2{|}}{\stopper@ \double@\xydashh@}
\xydefcsname@{dir3{|}}{\stopper@ \double@{2\xydashh@}}
\xyletcsnamecsname@{dir0{|}}{dir{}}
\xyletcsnamecsname@{dir{|}}{dir1{|}}
\xydef@\stopper@{\tip@x\stopper@@}
\xydef@\stopper@@{\setboxz@h{\count@=\SemiDirectionChar \advance\count@64
 \ifnum127<\count@ \advance\count@-128 \fi \xydashfont\char\count@\/}%
 \setboxz@h{\kern-.5\wdz@ \dimen@=\sd@Y\cosDirection\xydashl@ 
 \ifnum\SemiDirectionChar=95 \dimen@=\sd@X\sd@Y\dimen@ \fi
 \raise.5\dimen@\boxz@}%
 \wdz@=\z@ \ht\z@=\z@ \dp\z@=\z@ \boxz@}
\xydefcsname@{dir1{(}}{\hook@}
\xydefcsname@{dir^{(}}{\ahook@}
\xydefcsname@{dir_{(}}{\bhook@}
\xyletcsnamecsname@{dir0{(}}{dir{}}
\xyletcsnamecsname@{dir{(}}{dir1{(}}
\xydefcsname@{dir1{)}}{\reverseDirection@\hook@}
\xydefcsname@{dir^{)}}{\reverseDirection@\bhook@}
\xydefcsname@{dir_{)}}{\reverseDirection@\ahook@}
\xyletcsnamecsname@{dir0{)}}{dir{}}
\xyletcsnamecsname@{dir{)}}{dir1{)}}
\xydef@\hook@{\tip@x\hook@@}
\xydef@\hook@@{\setboxz@h{\xybsqlfont
 \vDirection@(1,-1){.707107\xybsqll@}%
 \hbox{\DirectionChar
 \kern-\d@Y\raise\d@X\hbox{\count@=\DirectionChar \advance\count@-32
 \ifnum\count@<\z@ \advance\count@128 \fi \char\count@}}}%
 \wdz@=\z@ \ht\z@=\z@ \dp\z@=\z@ \boxz@}
\xydef@\ahook@{\tip@x\ahook@@}
\xydef@\ahook@@{\setboxz@h{\xybsqlfont
 \vDirection@(1,-1){.707107\xybsqll@}\kern-\d@X
 \lower\d@Y\hbox{\DirectionChar
 \kern-\d@Y\raise\d@X\hbox{\count@=\DirectionChar \advance\count@-32
 \ifnum\count@<\z@ \advance\count@128 \fi \char\count@}}}%
 \wdz@=\z@ \ht\z@=\z@ \dp\z@=\z@ \boxz@}
\xydef@\bhook@{\tip@x\bhook@@}
\xydef@\bhook@@{\setboxz@h{\xybsqlfont
 \vDirection@(-1,-1){.707107\xybsqll@}\DirectionChar
 \kern\d@X\raise\d@Y\hbox{\count@=\DirectionChar \advance\count@-96
 \ifnum\count@<\z@ \advance\count@128 \fi \char\count@}}%
 \wdz@=\z@ \ht\z@=\z@ \dp\z@=\z@ \boxz@}
\xydefcsname@{dir^{'}}{\reverseDirection@\bturn@}
\xydefcsname@{dir_{'}}{\reverseDirection@\aturn@}
\xydefcsname@{dir^{`}}{\aturn@}
\xydefcsname@{dir_{`}}{\bturn@}
\xydef@\aturn@{\tip@x\aturn@@}
\xydef@\aturn@@{\setboxz@h{\xybsqlfont
 \vDirection@(1,-1){.707107\xybsqll@}\kern-\d@X
 \lower\d@Y\hbox{\DirectionChar}}%
 \wdz@=\z@ \ht\z@=\z@ \dp\z@=\z@ \boxz@}
\xydef@\bturn@{\tip@x\bturn@@}
\xydef@\bturn@@{\setboxz@h{\xybsqlfont
 \vDirection@(-1,-1){.707107\xybsqll@}\DirectionChar}%
 \wdz@=\z@ \ht\z@=\z@ \dp\z@=\z@ \boxz@}
\xydef@\newdir#1#{\newdir@{#1}}
\xydef@\newdir@#1#2#3{\xydefcsname@{dir#1{#2}}{\composite@{}{#3}}}
\xydef@\shiftdir@#1#2{%
 \setbox\z@=\hbox\bgroup#1\relax
 \setboxz@h{\dimen@ii=#2\relax
 \dimen@=-\cosDirection\dimen@ii \advance\dimen@-\L@c
 \kern\dimen@ \lower\sinDirection\dimen@ii\boxz@}%
 \wdz@\z@ \ht\z@=\z@ \dp\z@=\z@ \boxz@}
\xylet@\tipjot@=\jot
\xydefcsname@{dir1{>>}}{\shiftdir@\tip@\tipjot@ \tip@}
\xydefcsname@{dir^{>>}}{\shiftdir@\atip@\tipjot@ \atip@}
\xydefcsname@{dir_{>>}}{\shiftdir@\btip@\tipjot@ \btip@}
\xydefcsname@{dir2{>>}}{\composite@{}{h!/\tipjot@/\dir2{>}*\dir2{>}}}
\xydefcsname@{dir3{>>}}{\composite@{}{h!/\tipjot@/\dir3{>}*\dir3{>}}}
\xyletcsnamecsname@{dir0{>>}}{dir{}}
\xyletcsnamecsname@{dir{>>}}{dir1{>>}}
\xydefcsname@{dir1{<<}}{\reverseDirection@ \shiftdir@\tip@\tipjot@ \tip@}
\xydefcsname@{dir^{<<}}{\reverseDirection@ \shiftdir@\btip@\tipjot@ \btip@}
\xydefcsname@{dir_{<<}}{\reverseDirection@ \shiftdir@\atip@\tipjot@ \atip@}
\xydefcsname@{dir2{<<}}{\composite@{}{h!/-\tipjot@/\dir2{<}*\dir2{<}}}
\xydefcsname@{dir3{<<}}{\composite@{}{h!/-\tipjot@/\dir3{<}*\dir3{<}}}
\xyletcsnamecsname@{dir0{<<}}{dir{}}
\xyletcsnamecsname@{dir{<<}}{dir1{<<}}
\xydefcsname@{dir{||}}{\shiftdir@\stopper@\xydashh@ \shiftdir@\stopper@\z@
 \pointlike@{}\jot}
\xydefcsname@{dir^{||}}{\shiftdir@{\aboveDirection@\xydashl@\line@}\xydashh@
 \shiftdir@{\aboveDirection@\xydashl@\line@}\z@ \pointlike@{}\jot}
\xydefcsname@{dir_{||}}{\shiftdir@{\belowDirection@\xydashl@\line@}\xydashh@
 \shiftdir@{\belowDirection@\xydashl@\line@}\z@ \pointlike@{}\jot}
\xydefcsname@{dir2{||}}{\shiftdir@\stopper@\xydashh@ \shiftdir@\stopper@\z@
 \pointlike@{}\jot \double@\xydashh@}
\xydefcsname@{dir3{||}}{\shiftdir@\stopper@\xydashh@ \shiftdir@\stopper@\z@
 \pointlike@{}\jot \double@{2\xydashh@}}
\xydefcsname@{dir{>|}}{\shiftdir@\stopper@\z@ \tip@}
\xydefcsname@{dir2{>|}}{\composite@{}{\dir2{>}*\dir2{|}}}
\xydefcsname@{dir3{>|}}{\composite@{}{\dir3{>}*\dir3{|}}}
\xydefcsname@{dir{>>|}}{\shiftdir@\stopper@\z@ \shiftdir@\tip@\tipjot@ \tip@}
\xydefcsname@{dir2{>>|}}{\composite@{}{h!/\tipjot@/\dir2{>}*\dir2{>}*\dir2{|}}}
\xydefcsname@{dir3{>>|}}{\composite@{}{h!/\tipjot@/\dir3{>}*\dir3{>}*\dir3{|}}}
\xydefcsname@{dir{|<}}{\reverseDirection@ \shiftdir@\stopper@\z@ \tip@}
\xydefcsname@{dir2{|<}}{\reverseDirection@ \shiftdir@\stopper@\z@ \Tip@}
\xydefcsname@{dir3{|<}}{\reverseDirection@ \shiftdir@\stopper@\z@ \Ttip@}
\xydefcsname@{dir{|<<}}{\reverseDirection@
 \shiftdir@\stopper@\z@ \shiftdir@\tip@\tipjot@ \tip@}
\xydefcsname@{dir2{|<<}}{%
 \composite@{}{h!/-\tipjot@/\dir2{<}*\dir2{<}*\dir2{|}}}
\xydefcsname@{dir3{|<<}}{%
 \composite@{}{h!/-\tipjot@/\dir3{<}*\dir3{<}*\dir3{|}}}
\xydefcsname@{dir{|-}}{\shiftdir@\stopper@\z@
 \shiftdir@\line@\z@ \pointlike@{}\jot}
\xydefcsname@{dir^{|-}}{\shiftdir@{\aboveDirection@\xydashl@ \line@}\z@
 \shiftdir@\line@\z@ \pointlike@{}\jot}
\xydefcsname@{dir_{|-}}{\shiftdir@{\belowDirection@\xydashl@ \line@}\z@
 \shiftdir@\line@\z@ \pointlike@{}\jot}
\xydefcsname@{dir2{|-}}{\shiftdir@\stopper@\z@
 \shiftdir@\line@\z@ \pointlike@{}\jot \double@\xydashh@}
\xydefcsname@{dir3{|-}}{\shiftdir@\stopper@\z@
 \shiftdir@\line@\z@ \pointlike@{}\jot \triple@\xydashh@}
\xyletcsnamecsname@{dir0{|=}}{dir{}}
\xyletcsnamecsname@{dir{|=}}{dir2{|-}}
\xydefcsname@{dir{+}}{%
 \DN@##1{\composite@{}{##10\dir{|}*!C##10\dir{-}}}\addEQ@\next@}
\xydefcsname@{dir{x}}{\vDirection@(1,1)\jot
 \DN@##1{\composite@{}{##10\dir{|}*!C##10\dir{-}}}\addEQ@\next@}
\xydefcsname@{dir{/}}{\vDirection@(1,-.3)\jot \stopper@}
\xydefcsname@{dir{//}}{\vDirection@(1,-.3)\jot
 \shiftdir@\stopper@\xydashh@ \stopper@}
\xydefcsname@{dir{*}}{\solidpoint@}
\xydef@\solidpoint@{%
 \pointlike@{\kern-1.8pt\lower1.8pt\hbox{$\scriptstyle\bullet$}}\jot}
\xydefcsname@{dir{o}}{\hollowpoint@}
\xydef@\hollowpoint@{%
 \pointlike@{\kern-1.8pt\lower1.8pt\hbox{$\scriptstyle\circ$}}\jot}
\message{circles,}
\xydef@\cir#1#{\hbox\bgroup
 \afterVECTORorEMPTY{\xy@@{\R@=\X@c}\cir@}{\xy@@{\R@=\R@c}\cir@}#1@}
\xydef@\cir@#1@#2{%
 \DN@{#1}\ifx\next@\empty\else \xyerror@{illegal circle <radius>: must be
 <vector> or <empty>}{}\fi
 \afterCIRorDIAG{\xyFN@\cir@cir}{\xyFN@\cir@diag}#2@}
\xydef@\cir@cir{%
 \ifx \space@\next \expandafter\DN@\space{\xyFN@\cir@cir}%
 \else \ifx @\next \DN@ @{\cir@i}%
 \else \xyerror@{illegal <cir>: must have form <diag><orient><diag> or
 <empty>}{}%
 \fi\fi \next@}
\xydef@\cir@diag{%
 \DN@{\xyerror@{illegal <cir>: must have form <diag><orient><diag> or
 <empty>}{}}%
 \ifx @\next \ifnum\count@=8
 \DN@ @{\def\CIRin@@{0}\def\CIRorient@@{\CIRfull@}\def\CIRout@@{7}\cir@i}%
 \fi\fi \next@}
\xydef@\cir@i{%
 \ifnum\CIRin@@=8 \xyerror@{incomplete <cir> specification}{%
The <cir> you specified as <diag><orient><diag> is not sufficiently specific.}%
 \def\CIRin@@{0}\fi
 \ifdim\R@<.5\p@ \R@=\z@ \zerodot
 \else \CIRorient@@ \cirbuild@ \fi
 \L@c=\R@ \R@c=\R@ \D@c=\R@ \U@c=\R@ \def\Leftness@{.5}\def\Upness@{.5}%
 \def\Drop@@{\boxz@}\def\Connect@@{\straight@\relax}\Edge@c={\circleEdge}%
 \OBJECT@x}
\xydef@\CIRin@@{3}
\xydef@\CIRout@@{3}
\xylet@\CIRorient@@=\empty
\xydef@\afterCIRorDIAG#1#2{\def\afterCIR@{#1}\def\afterCIRDIAG@{#2}\xyFN@\CIR@}
\xylet@\afterCIR@=\empty
\xylet@\afterCIRDIAG@=\empty
\xydef@\CIR@{\count@=8 \afterDIAG{\edef\CIRin@@{\the\count@}\xyFN@\CIR@@}}
\xydef@\CIR@@{%
 \ifx \space@\next \expandafter\DN@\space{\xyFN@\CIR@@}%
 \else\ifx ^\next
 \DN@ ^{\def\CIRorient@@{\CIRacw@}%
 \afterDIAG{\edef\CIRout@@{\the\count@}\afterCIR@}}%
 \else\ifx _\next
 \DN@_{\def\CIRorient@@{\CIRcw@}%
 \afterDIAG{\edef\CIRout@@{\the\count@}\afterCIR@}}%
 \else
 \DN@{\def\CIRorient@@{\relax}\afterCIRDIAG@}%
 \fi\fi\fi \next@}
\xylet@\CIRtest@@=\relax
\xydef@\CIRlo@@{0}
\xydef@\CIRhi@@{0}
\xydef@\CIRfull@{\def\CIRtest@@##1##2{##2}}
\xydef@\cirbuild@{\cirrestrict@@ \multiply\count@8
 \circhar@0\circhar@7\kern\dimen@
 \circhar@1\circhar@6\kern\dimen@
 \circhar@2\circhar@5\kern\dimen@
 \circhar@3\circhar@4\kern\dimen@}
\xydef@\circhar@#1{%
 \setboxz@h{\circhar@@{#1}}\dimen@=\wdz@ \wdz@=\z@ \ht\z@=\R@ \dp\z@=\R@
 \CIRtest@@#1{\boxz@}\setbox\z@=\copy\voidb@x}
\xydef@\circhar@@#1{{\xycircfont \advance\count@#1\relax \char\count@}}
\xydef@\cirrestrict@@{\begingroup \dimen@=\R@
 \setboxz@h{\xycircfont\char\z@\char\@ne}\A@=\wdz@
 \ifdim\R@<8\A@ \count@=\dimen@ \divide\count@\A@ \advance\count@\m@ne
 \else\ifdim\R@<16\A@ \count@=\dimen@
 \dimen@=2\A@ \divide\count@\dimen@ \advance\count@3
 \else\ifdim\R@<32\A@ \count@=\dimen@
 \dimen@=4\A@ \divide\count@\dimen@ \advance\count@7
 \else \count@=15 \fi\fi\fi
 \R@=\A@
 \ifnum\count@<8 \multiply\R@\count@ \advance\R@\A@
 \else\ifnum\count@<12 \multiply\R@\count@ \multiply\R@\tw@ \advance\R@-6\A@
 \else\ifnum\count@<16 \multiply\R@\count@ \multiply\R@ 4 \advance\R@-28\A@
 \else \multiply\R@ 32 \fi\fi\fi
 \edef\@tmp{\endgroup \R@=\the\R@ \count@=\the\count@}\@tmp}
\xydef@\CIRacw@{\count@@=\CIRin@@ \count@=\CIRout@@
 \ifnum\count@=8 \count@=\count@@
 \ifnum\count@<6 \advance\count@\tw@ \else \advance\count@-6 \fi \fi
 \ifnum\count@@<\@ne \advance\count@@7 \else \advance\count@@\m@ne \fi
 \ifnum\count@<\@ne \advance\count@7 \else \advance\count@\m@ne \fi
 \ifnum\count@@>\count@ \let\CIRtest@@=\CIRtest@outside
 \edef\CIRlo@@{\the\count@}\edef\CIRhi@@{\the\count@@}%
 \else \let\CIRtest@@=\CIRtest@inside
 \edef\CIRlo@@{\the\count@@}\edef\CIRhi@@{\the\count@}%
 \fi}
\xydef@\CIRcw@{\count@@=\CIRin@@ \count@=\CIRout@@
 \ifnum\count@=8 \count@=\count@@
 \ifnum\count@>\@ne \advance\count@-\tw@ \else \advance\count@6 \fi \fi
 \ifnum\count@@<5 \advance\count@@\thr@@ \else \advance\count@@-5 \fi
 \ifnum\count@<5 \advance\count@\thr@@ \else \advance\count@-5 \fi
 \ifnum\count@@<\count@ \let\CIRtest@@=\CIRtest@outside
 \edef\CIRlo@@{\the\count@@}\edef\CIRhi@@{\the\count@}%
 \else \let\CIRtest@@=\CIRtest@inside
 \edef\CIRlo@@{\the\count@}\edef\CIRhi@@{\the\count@@}%
 \fi}
\xydef@\CIRtest@inside#1#2{\let\next@=\relax
 \ifnum\CIRlo@@>#1\else \ifnum#1<\CIRhi@@\DN@{#2}\fi\fi \next@}
\xydef@\CIRtest@outside#1#2{\let\next@=\relax
 \ifnum\CIRlo@@>#1\DN@{#2}\else \ifnum#1<\CIRhi@@\else\DN@{#2}\fi\fi \next@}
\message{text;}
\xydef@\txt{\hbox\bgroup \xyFN@\txt@}
\xydef@\txt@{%
 \addLT@\ifx\next \addGT@{\addLT@\DN@##1}{\A@=##1\txt@i}%
 \else \DN@{\A@=\maxdimen \txt@i}\fi \next@}
\xydef@\txt@i#1#{%
 \setboxz@h{#1\mathstrut}\dimen@=\ht\z@ \advance\dimen@\dp\z@
 \baselineskip=1.1\dimen@ \lineskip=.2\dimen@ \lineskiplimit=\lineskip
 \def\txtline@@##1{\txtline@{#1}{##1}}\object@\txt@ii}
\xylet@\txtline@@=\eat@
\xydef@\txtline@#1#2{\relax\setboxz@h{#1\ignorespaces #2\unskip}%
 \ifdim\A@<\wdz@ \setboxz@h{\hsize=\A@
 \leftskip=0pt plus4em \rightskip=\leftskip
 \parfillskip=0pt \parindent=0pt
 \spaceskip=.3333em \xspaceskip=.5em
 \pretolerance=9999 \tolerance=9999
 \hyphenpenalty=9999 \exhyphenpenalty=9999
 \vbox{#1\noindent\ignorespaces #2\unskip}}%
 \else\ifdim\A@<\maxdimen\setboxz@h to\A@{\hfil\boxz@\hfil}\fi\fi
 \boxz@}
\xydef@\txt@ii#1{\vbox{%
 \let\\=\cr
 \tabskip=\z@skip \halign{\relax\hfil\txtline@@{##}\hfil\cr#1\crcr}}}
\message{options;}
\xylet@\xyoption@@=\relax
\xydef@\xyoption#1{\DN@{#1}\edef\next@{\codeof\next@}%
 \csname xyeveryrequest@\next@ @\endcsname
 \xyinputorelse@{xy#1}{\xyinputorelse@{xy#1.doc}{\xyoption@truncated#1@}}%
 \def\xyoption@@{#1}\edef\xyoption@@{\codeof\xyoption@@}
 \expandafter\let\expandafter\next@\csname xy\xyoption@@ version\endcsname
 \expandafter\let\csname xy\xyoption@@ loaded\endcsname=\next@
 \runxywith@ \ignorespaces}
\xydef@\xyoption@truncated#1#2#3#4#5#6#7@{%
 \DN@{#7}\ifx\next@\empty \DN@##1##2{##2}\else\let\next@=\xyinputorelse@ \fi
 \next@{xy#1#2#3#4#5#6}%
 {\DN@{#1}\edef\next@{\codeof\next@}\xyerror@{No `\next@' option}{%
Your \string\xyoption\string{\next@\string}%
 request could not be granted: the required^^J%
file `xy\next@.tex' could not be located. Please make sure that it is^^J%
properly installed before continuing.}}}
\xydef@\xyrequire#1{\DN@{#1}\edef\next@{\codeof\next@}%
 \csname xyeveryrequest@\next@ @\endcsname
 \expandafter\let\expandafter\next@\csname xy\codeof\next@ loaded\endcsname
 \ifx \next@\relax \DN@{\xyoption{#1}}\else \DN@{\ignorespaces}\fi \next@}
\xylet@\xywith@@=\empty
\xydef@\runxywith@{\let\xywithdo@@=\xywithtest@ \xywith@@}
\xydef@\xywithoption#1#2{\DN@{#1}\edef\next@{\codeof\next@}%
 \expandafter\let\expandafter\nextii@\csname xy\next@ loaded\endcsname
 \ifx \nextii@\relax
 \expandafter\def\expandafter\xywith@@\expandafter{\xywith@@ 
 \xywithdo@@{#1}{#2}}%
 \else \expandafter\xywithrun@\expandafter{\next@}{#2}\fi}
\xydef@\xywithtest@#1#2{\DN@{#1}\edef\next@{\codeof\next@}%
 \ifx\next@\xyoption@@ \expandafter\xywithrun@\expandafter{\next@}{#2}\fi}
\xydef@\xywithrun@#1#2{\csname xyeverywithoption@#1@\endcsname #2}
\xydef@\xyevery@#1#2#3{\DN@{#2}\edef\next@{\codeof\next@}%
 \expandafter\ifx\csname xyevery#1@\next@ @\endcsname\relax
 \expandafter\let\csname xyevery#1@\next@ @\endcsname=\empty \fi
 \DNii@##1{\expandafter\def\expandafter##1\expandafter{##1#3}}%
 \expandafter\nextii@\csname xyevery#1@\next@ @\endcsname}
\xydef@\xyeveryrequest{\xyevery@{request}}
\xydef@\xyeverywithoption{\xyevery@{withoption}}
\xydef@\xyprovide#1#2#3#4#5#6{%
 \def\xyoption@@{#1}\edef\xyoption@@{\codeof\xyoption@@}\edef\next@{#3}%
 \message{ Xy-pic option: #2 v.\next@}%
 \expandafter\let\expandafter\nextii@\csname xy\xyoption@@ loaded\endcsname
 \ifx \next@\nextii@ \message{not reloaded} 
 \else
 \ifx \nextii@\relax\else \xyerror@{Option `\xyoption@@' version mismatch}{%
You previously loaded, or the format has preloaded, a different version^^J%
of this option. Just continue to try to load this version instead (and^^J%
be prepared for a lot of warnings about redefinitions).}%
 \fi
 \expandafter\let\csname xy\xyoption@@ version\endcsname=\next@
 \expandafter\let\expandafter\xyenddocmode@\csname DOCMODE\endcsname
 \expandafter\let\csname DOCMODE\endcsname\xyprovidedocmode@
 \xycatcodes
 \fi \ignorespaces}
\xydef@\xyendinput{\expandafter\let\csname DOCMODE\endcsname=\xyenddocmode@
 \message{loaded}\xyuncatcodes }
\expandafter\xylet@\expandafter\xyprovidedocmode@\csname DOCMODE\endcsname
\xylet@\xyenddocmode@=\relax
\xydef@\xydriversloaded@@{\do{unload}}
\xylet@\xydriversselected@@=\empty
\xylet@\xydriver@unload@support@@=\empty
\xydef@\selectdriver@#1{\DN@{#1}\edef\next@{\codeof\next@}%
 \expandafter\selectdriver@@\expandafter{\next@}}
\xydef@\selectdriver@single#1{\xysetup@@{\def\xydriversselected@@{\do{#1}}%
 \xyLoadDrivers@}\let\selectdriver@@=\changedriver@single}
\xydef@\changedriver@single#1{\xysetup@@{\def\xydriversselected@@{\do{#1}}%
 \xyReloadDrivers@}}
\xydef@\selectdriver@multiple#1{\expandafter\xysetup@@\expandafter{%
 \expandafter\def\expandafter\xydriversselected@@\expandafter{%
 \xydriversselected@@ \do{#1}}%
 \let\do=\activatedriver@ \xydriversselected@@}}
\xydef@\UseSingleDriver{\let\selectdriver@@=\selectdriver@single}
\xydef@\MultipleDrivers{\let\selectdriver@@=\selectdriver@multiple}
\xydef@\xyLoadDrivers@{\let\do=\activatedriver@ \xydriversselected@@}
\xydef@\xyReloadDrivers@{\activatedriver@{unload}\xyLoadDrivers@}
\xylet@\xyReloadDrivers=\xyReloadDrivers@
\xydef@\activatedriver@#1{%
 \let\doii=\activatedriversupport@ \csname xydriver@#1@support@@\endcsname}
\xydef@\activatedriversupport@#1#2{%
 \expandafter\ifx\csname xy#1loaded\endcsname\relax\DN@{}%
 \else\DN@{#2}\fi \next@}
\UseSingleDriver
\xydef@\xyselectoptionsupport@{\let\do=\activatedriver@ \xydriversselected@@}
\xydef@\xyShowDrivers{{\W@{Loaded:}\let\do=\doShow \xydriversloaded@@
 \let\next@=\empty
 \def\do##1{\ifx\next@\empty \DN@{##1}%
 \else \expandafter\DN@\expandafter{\next@, ##1}\fi}%
 \xydriversselected@@\W@{Selected: \next@.}}}
\def\doShow#1{\let\next@=\empty
 \def\doii##1##2{\ifx\next@\empty \DN@{##1}%
 \else \expandafter\DN@\expandafter{\next@, ##1}\fi}%
 \csname xydriver@#1@support@@\endcsname
 \W@{ <driver> #1 supports \next@.}}
\xydef@\newdriver#1{%
 \def\nextiii@##1{%
 \expandafter\def\expandafter\xydriversloaded@@
 \expandafter{\xydriversloaded@@\do{##1}}%
 \expandafter\let\csname xydriver@\xyoption@@ @support@@\endcsname=\empty}%
 \def\do##1{\DNii@{##1}\ifx\xyoption@@\nextii@ \let\nextiii@=\eat@ \fi}%
 \xydriversloaded@@ \expandafter\nextiii@\expandafter{\xyoption@@}%
 #1\relax
 \DN@##1{\xywithoption{##1}{%
 \selectdriver@{##1}\xyeveryrequest{##1}{\selectdriver@{##1}}}}%
 \expandafter\next@\expandafter{\xyoption@@}\ignorespaces}
\xydef@\xyaddsupport{\expandafter\xyadddriversupport@\expandafter{\xyoption@@}}
\xydef@\xyaddunsupport{\xyadddriversupport@{unload}}
\xydef@\xyadddriversupport@#1#2#3{%
 \DNii@{#1}\edef\nextii@{\codeof\nextii@}%
 \def\nextiii@{#2}\edef\nextiii@{\codeof\nextiii@}%
\def\next{\xybug@{<driver> \nextii@\space not loaded}}%
\def\do##1{\DN@{##1}\edef\next@{\codeof\next@}%
 \ifx\next@\nextii@ \let\next=\relax \fi}\xydriversloaded@@ \next
\def\next{\xybug@{<driver> \nextii@\space already supports \nextiii@}}%
\def\doii##1##2{\DN@{##1}\edef\next@{\codeof\next@}%
 \ifx\next@\nextiii@ \let\next=\relax \fi}\xydriversloaded@@ \next
 \DN@##1##2{\ifx##1\relax \let##1=\empty \fi
 \expandafter\def\expandafter##1\expandafter{##1\doii{##2}{#3}}}%
 \expandafter\expandafter\expandafter\next@
 \expandafter\expandafter\csname xydriver@\nextii@ @support@@\endcsname
 \expandafter{\nextiii@}%
 \DN@##1{\expandafter\xyeverywithoption\expandafter{\nextiii@}{%
 \xytestforsupport@{##1}}}%
 \expandafter\next@\expandafter{\nextii@}\xydriversloaded@@
 \ignorespaces}
\xydef@\xytestforsupport@#1{\def\do##1{\DN@{##1}\DNii@{#1}%
 \ifx\next@\nextii@ \expandafter\xyselectoptionsupport@ \fi}}
\xynew@{if}\ifunsupportwarnings@
\xydef@\xyunsupportwarning@#1#2{{%
 \DN@{#1}\edef\next@{\codeof\next@}
 \expandafter\ifx\csname xywarn@unload@\next@ @\endcsname\relax
 \expandafter\gdef\csname xywarn@unload@\next@ @\endcsname{}%
 \ifunsupportwarnings@ \xyclosedown@@\xyunsupportwarnings@@ \fi
 \global\unsupportwarnings@true
 \expandafter\gdef\expandafter\xyunsupportwarnings@@\expandafter{%
 \xyunsupportwarnings@@ \W@{ #2.}}%
 \W@{}%
 \W@{Xy-pic Warning: `\next@' reproduction is NOT EXACT\xytracelineno@:}%
 \W@{ #2.}%
 \W@{}%
 \fi}}
\xydef@\xyunsupportwarnings@@{\W@{}%
 \W@{Xy-pic Warning: The produced DVI file is NOT EXACT:}}
\xynew@{if}\ifsupportwarnings@
\xydef@\xysupportwarning@#1#2{{%
 \DN@{#1}\edef\next@{\codeof\next@}\DNii@{#2}\edef\nextii@{\codeof\nextii@}%
 \expandafter\ifx\csname xywarn@\next@ @\nextii@ @\endcsname\relax
 \expandafter\gdef\csname xywarn@\next@ @\nextii@ @\endcsname{}%
 \ifsupportwarnings@ \xyclosedown@@\xysupportwarnings@@ \fi
 \global\supportwarnings@true
 \expandafter\gdef\expandafter\xysupportwarnings@@\expandafter{%
 \xysupportwarnings@@ \driverextensioncomplain@{#1}{#2}}%
 \W@{}%
 \W@{Xy-pic Warning: The produced DVI file is NOT PORTABLE\xytracelineno@:}%
 \driverextensioncomplain@{#1}{#2}%
 \W@{}%
 \fi}}
\xydef@\xysupportwarnings@@{%
 \W@{Xy-pic Warning: The produced DVI file is NOT PORTABLE:}}
\xydef@\driverextensioncomplain@#1#2{%
 \DN@{#1}\edef\next@{\codeof\next@}\DNii@{#2}\edef\nextii@{\codeof\nextii@}%
 \W@{ It contains <driver> `\next@' \string\special s for the `\nextii@'
 extension.}}
\xydef@\dvimessage@#1#2{\xysupportwarning@{#1}{#2}}
\xynew@{if}\ifxydrivernoloads@
\xydef@\xydrivernoload@#1{\xyerror@{Could not load xy#1.tex}{}%
 \xydrivernoloads@true}
\xydef@\xydriverfail@#1{\xyerror@{Sorry, #1 not supported.}{}}
\xydef@\xyunload@#1{\xywarning@{Unloading #1.}}
\message{algorithms: directions,}
\xydef@\DirectionfromtheDirection@{\noexpand\Direction=\the\Direction
 \noexpand\d@X=\the\d@X \noexpand\d@Y=\the\d@Y
 \def\noexpand\sd@X{\sd@X}\def\noexpand\sd@Y{\sd@Y}%
 \noexpand\K@dXdY=\the\K@dXdY \noexpand\K@dYdX=\the\K@dYdX
 \chardef\noexpand\DirectionChar=\the\DirectionChar
 \chardef\noexpand\SemiDirectionChar=\the\SemiDirectionChar
 \def\noexpand\cosDirection{\cosDirection}%
 \def\noexpand\sinDirection{\sinDirection}%
 \noexpand\resetupDirection@}
\xydef@\Directiontest@@#1#2{#2}
\xydef@\setupDirection@{%
 \d@X=\X@c\advance\d@X-\X@p \d@Y=\Y@c\advance\d@Y-\Y@p
 \Directiontest@@\relax\setupDirection@i}
\xydef@\setupDirection@i{\DN@{\setupDirection@ii}%
 \ifdim\d@X=\d@Y
 \ifdim\zz@\d@Y \DN@{}%
 \else\ifdim\d@X<\z@ \DN@{\dlDirection@{-1.4142\d@X}}%
 \else \DN@{\urDirection@{1.4142\d@X}}\fi\fi
 \else\ifdim\d@X<\d@Y
 \ifdim\zz@\d@X \DN@{\uDirection@\d@Y}%
 \else\ifdim\zz@\d@Y \DN@{\lDirection@{-\d@X}}%
 \else\ifdim-\d@X=\d@Y \DN@{\ulDirection@{-1.4142\d@X}}\fi\fi\fi
 \else
 \ifdim\zz@\d@X \DN@{\dDirection@{-\d@Y}}%
 \else\ifdim\zz@\d@Y \DN@{\rDirection@\d@X}%
 \else\ifdim\d@X=-\d@Y \DN@{\drDirection@{1.4142\d@X}}\fi\fi\fi
 \fi\fi \next@}
\xydef@\dlDirection@{\Direction=\z@
 \def\cosDirection{-.7071}\def\sinDirection{-.7071}\def\sd@X{-}\def\sd@Y{-}%
 \chardef\DirectionChar=127\relax\chardef\SemiDirectionChar=127\relax
 \K@dXdY=1\K@ \K@dYdX=1\K@ \fixedDirection@}
\xydef@\dDirection@{\Direction=\K@
 \def\cosDirection{0}\def\sinDirection{-1}\def\sd@X{+}\def\sd@Y{-}%
 \chardef\DirectionChar=15\relax\chardef\SemiDirectionChar=31\relax
 \K@dXdY=\z@ \K@dYdX=\KK@\K@ \fixedDirection@}
\xydef@\drDirection@{\dimen@ii=2\K@ \Direction=\dimen@ii
 \def\cosDirection{+.7071}\def\sinDirection{-.7071}\def\sd@X{+}\def\sd@Y{-}%
 \chardef\DirectionChar=31\relax\chardef\SemiDirectionChar=63\relax
 \K@dXdY=-1\K@ \K@dYdX=-1\K@ \fixedDirection@}
\xydef@\rDirection@{\dimen@ii=3\K@ \Direction=\dimen@ii
 \def\cosDirection{+1}\def\sinDirection{0}\def\sd@X{+}\def\sd@Y{+}%
 \chardef\DirectionChar=47\relax\chardef\SemiDirectionChar=95\relax
 \K@dXdY=\KK@\K@ \K@dYdX=\z@ \fixedDirection@}
\xydef@\urDirection@{\dimen@ii=4\K@ \Direction=\dimen@ii
 \def\cosDirection{+.7071}\def\sinDirection{+.7071}\def\sd@X{+}\def\sd@Y{+}%
 \chardef\DirectionChar=63\relax\chardef\SemiDirectionChar=127\relax
 \K@dXdY=1\K@ \K@dYdX=1\K@ \fixedDirection@}
\xydef@\uDirection@{\dimen@ii=-3\K@ \Direction=\dimen@ii
 \def\cosDirection{0}\def\sinDirection{+1}\def\sd@X{+}\def\sd@Y{+}%
 \chardef\DirectionChar=79\relax\chardef\SemiDirectionChar=31\relax
 \K@dXdY=\z@ \K@dYdX=\KK@\K@ \fixedDirection@}
\xydef@\ulDirection@{\dimen@ii=-2\K@ \Direction=\dimen@ii
 \def\cosDirection{-.7071}\def\sinDirection{+.7071}\def\sd@X{-}\def\sd@Y{+}%
 \chardef\DirectionChar=95\relax\chardef\SemiDirectionChar=63\relax
 \K@dXdY=-1\K@ \K@dYdX=-1\K@ \fixedDirection@}
\xydef@\lDirection@{\Direction=-\K@
 \def\cosDirection{-1}\def\sinDirection{0}\def\sd@X{-}\def\sd@Y{+}%
 \chardef\DirectionChar=111\relax\chardef\SemiDirectionChar=95\relax
 \K@dXdY=\KK@\K@ \K@dYdX=\z@ \fixedDirection@}
\xydef@\fixedDirection@#1{\dimen@ii=#1\relax
 \d@X=\cosDirection\dimen@ii \d@Y=\sinDirection\dimen@ii
 \resetupDirection@}
\xydef@\setupDirection@ii{%
 \ifdim\d@X<\z@ \def\sd@X{-}\else \def\sd@X{+}\fi
 \ifdim\d@Y<\z@ \def\sd@Y{-}\else \def\sd@Y{+}\fi
 \K@dXdY=\sd@X\d@X \ifdim\K@dXdY<500pt \multiply\K@dXdY\KK@ \fi \dimen@=\sd@Y\d@Y
 \advance\dimen@.5\KK@ \divide\dimen@\KK@
 \ifdim\dimen@=\z@\else
 \advance\K@dXdY by.5\dimen@\relax \divide\K@dXdY\dimen@
 \fi \K@dXdY=\sd@X\sd@Y\K@dXdY
 \K@dYdX=\sd@Y\d@Y \ifdim\K@dYdX<500pt \multiply\K@dYdX\KK@ \fi \dimen@=\sd@X\d@X
 \advance\dimen@.5\KK@ \divide\dimen@\KK@
 \ifdim\dimen@=\z@\else
 \advance\K@dYdX by.5\dimen@\relax \divide\K@dYdX\dimen@
 \fi \K@dYdX=\sd@X\sd@Y\K@dYdX
 \Direction=\maxdimen
 \ifnum\K@dXdY<-\K@ \else \ifnum\K@<\K@dXdY \else
 \ifdim \d@Y<\z@
 \Direction=\K@ \advance\Direction-\K@dXdY
 \else
 \Direction=\K@ \multiply\Direction-\thr@@ \advance\Direction-\K@dXdY
 \fi\fi\fi
 \ifnum-\K@<\K@dYdX \ifnum\K@dYdX<\K@
 \ifdim \d@X<\z@
 \Direction=-\K@ \advance\Direction\K@dYdX
 \else
 \Direction=\K@ \multiply\Direction\thr@@ \advance\Direction\K@dYdX
 \fi\fi\fi
 \ifnum\Direction=\maxdimen
 \Direction=\K@dYdX \advance\Direction-\K@dXdY \divide\Direction\tw@ 
 \ifnum\K@dXdY<\z@ \advance\Direction\K@ \advance\Direction\K@
 \else \advance\Direction-\K@ \advance\Direction-\K@ \fi
 \fi
 \imposeDirection@i
 \resetupDirection@
 }
\xydef@\imposeDirection@{\count@@=\Direction 
 \loop@\ifnum\count@@>4096 \advance\count@@-8192 \repeat@
 \loop@\ifnum\count@@<-4096 \advance\count@@8192 \repeat@
 \def\sd@X{+}\ifnum\count@@<\K@ \relax
 \ifnum\count@@>-3072 \def\sd@X{-}\fi\fi
 \def\sd@Y{+}\ifnum\count@@<3072 \relax
 \ifnum\count@@>-\K@ \relax\def\sd@Y{-}\fi\fi
 \imposeDirection@i
 \d@X=\cosDirection\xydashl@ \d@Y=\sinDirection\xydashl@
 \resetupDirection@}
\xydef@\imposeDirection@i{%
 \count@@=\K@ \multiply\count@@ by8 \advance\count@@\Direction
 \count@=\count@@ \advance\count@\KK@ \divide\count@64 \advance\count@\m@ne
 \loop@\ifnum127<\count@ \advance\count@-128 \repeat@
 \chardef\DirectionChar\count@
 \advance\count@@16 \divide\count@@\KK@ \advance\count@@\m@ne
 \loop@\ifnum127<\count@@ \advance\count@@-128 \repeat@
 \chardef\SemiDirectionChar\count@@
 \setbox8=\hbox{\xydashfont\SemiDirectionChar\/}%
 \quotient@@\cosDirection{\sd@X\wd8}\xydashl@
 \setbox8=\hbox{\xydashfont\count@=\SemiDirectionChar\advance\count@-64
 \ifnum\count@<\z@ \advance\count@128 \fi \char\count@\/}%
 \quotient@@\sinDirection{\sd@Y\wd8}\xydashl@}
\xydef@\getxyDirection@#1{\xy@@\bgroup\xy@@ix@{#1}%
 \xy@@{\expandafter\POS\the\toks9\relax\setupDirection@
 \edef\next@{\egroup \Direction=\the\Direction}\next@ \imposeDirection@ }}
\xydef@\reverseDirection@{%
 \d@X=-\d@X \d@Y=-\d@Y
 \ifdim\d@X<\z@ \def\sd@X{-}\else \def\sd@X{+}\fi
 \ifdim\d@Y<\z@ \def\sd@Y{-}\else \def\sd@Y{+}\fi
 \dimen@=4\K@ \ifnum\Direction<\z@ \advance\Direction\dimen@
 \else \advance\Direction-\dimen@ \fi
 \count@=\DirectionChar \ifnum\count@<64 \advance\count@64
 \else \advance\count@-64 \fi \chardef\DirectionChar=\count@
 \edef\cosDirection{\if-\cosDirection\else-\cosDirection\fi}%
 \edef\sinDirection{\if-\sinDirection\else-\sinDirection\fi}%
 \resetupDirection@}
\xydef@\aboveDirection@#1{%
 \dimen@=\d@X \d@X=-\d@Y \d@Y=\dimen@
 \dimen@=\K@dXdY \K@dXdY=-\K@dYdX \K@dYdX=-\dimen@
 \ifdim\d@X<\z@ \def\sd@X{-}\else \def\sd@X{+}\fi
 \ifdim\d@Y<\z@ \def\sd@Y{-}\else \def\sd@Y{+}\fi
 \dimen@=2\K@ \ifdim 1\Direction<\dimen@\else \dimen@=-6\K@ \fi
 \advance\Direction\dimen@
 \count@=\DirectionChar \ifnum\count@<96 \advance\count@32
 \else \advance\count@-96 \fi \chardef\DirectionChar=\count@
 \count@=\SemiDirectionChar \ifnum\count@<64 \advance\count@64
 \else \advance\count@-64 \fi \chardef\SemiDirectionChar=\count@
 \let\tmp@=\cosDirection
 \edef\cosDirection{\if-\sinDirection\else-\sinDirection\fi}%
 \let\sinDirection=\tmp@
 \dimen@=#1\relax \d@X=\cosDirection\dimen@ \d@Y=\sinDirection\dimen@
 \resetupDirection@}
\xydef@\belowDirection@#1{%
 \dimen@=\d@X \d@X=\d@Y \d@Y=-\dimen@
 \dimen@=\K@dXdY \K@dXdY=-\K@dYdX \K@dYdX=-\dimen@
 \ifdim\d@X<\z@ \def\sd@X{-}\else \def\sd@X{+}\fi
 \ifdim\d@Y<\z@ \def\sd@Y{-}\else \def\sd@Y{+}\fi
 \dimen@=-2\K@\ifdim 1\Direction<\dimen@\dimen@=6\K@\fi \advance\Direction\dimen@
 \count@=\DirectionChar \ifnum\count@<32 \advance\count@96
 \else \advance\count@-32 \fi \chardef\DirectionChar=\count@
 \count@=\SemiDirectionChar \ifnum\count@<64 \advance\count@64
 \else \advance\count@-64 \fi \chardef\SemiDirectionChar=\count@
 \let\tmp@=\sinDirection
 \edef\sinDirection{\if-\cosDirection\else-\cosDirection\fi}%
 \let\cosDirection=\tmp@
 \dimen@=#1\relax \d@X=\cosDirection\dimen@ \d@Y=\sinDirection\dimen@
 \resetupDirection@}
\xydef@\vDirection@(#1,#2)#3{\dimen@ii=#3\relax
 \dimen@=#1\dimen@ii \dimen@ii=#2\dimen@ii
 \d@X=\cosDirection\dimen@ \advance\d@X-\sinDirection\dimen@ii
 \d@Y=\sinDirection\dimen@ \advance\d@Y \cosDirection\dimen@ii
 \X@p=\X@c \advance\X@p-\d@X \Y@p=\Y@c \advance\Y@p-\d@Y
 \setupDirection@\ignorespaces}
\xydef@\resetDirection@{%
 \d@X=\X@c\advance\d@X-\X@p \d@Y=\Y@c\advance\d@Y-\Y@p \let\next@=\resetupDirection@
 \ifdim\sd@X\d@X<\z@ \let\next@=\setupDirection@i \fi
 \ifdim\sd@Y\d@Y<\z@ \let\next@=\setupDirection@i \fi
 \next@}
\xydef@\resetupDirection@{%
 \edef\Directiontest@@##1##2{\noexpand\DN@{##2}%
 \noexpand\ifdim\noexpand\d@X=\the\d@X\relax
 \noexpand\ifdim\noexpand\d@Y=\the\d@Y\relax \noexpand\DN@{##1}%
 \noexpand\fi\noexpand\fi \noexpand\next@}}
\xydef@\unsetupDirection@{\def\Directiontest@@##1##2{##2}}
\uDirection@\xydashl@
\message{edges,}
\xynew@{if}\ifInside@
\xydef@\zeroEdge#1{%
 \ifcase#1\relax \or \Inside@false \or \dimen@=\z@
 \or \else \Edge@c={\rectangleEdge}\fi}
\xydef@\circleEdge#1{\ifcase#1\expandafter\circleEdge@
 \or \expandafter\circleUnder@ \or \dimen@=\R@c
 \or \expandafter\circleProp@ \or \expandafter\circleInner@
 \else \expandafter\circleOuter@ \fi}
\xydef@\circleEdge@{\DN@{\ellipseEdge@}%
 \ifdim\R@c=\L@c\relax \ifdim\U@c=\D@c\relax
 \ifdim\R@c=\U@c\DN@{\circlecentredEdge@}%
 \else\DN@{\ellipsecentredEdge@}\fi
 \fi\fi \next@}
\xydef@\circleProp@{\DN@{\reverseDirection@\ellipseEdge@}%
 \ifdim\R@c=\L@c\relax \ifdim\U@c=\D@c\relax
 \ifdim\R@c=\U@c\DN@{\reverseDirection@\circlecentredEdge@}%
 \else\DN@{\reverseDirection@\ellipsecentredEdge@}\fi
 \fi\fi \next@}
\xydef@\circleUnder@{\Inside@false
 \ifdim\X@p=\X@c \relax \ifdim\Y@p=\Y@c \Inside@true \fi \fi 
 \ifInside@ \else \expandafter \circleCentre@ \fi}
\xydef@\circleCentre@{{%
 \ifdim\L@c=\R@c \relax\else
 \dimen@=\R@c\advance\dimen@-\L@c \divide\dimen@\tw@
 \advance\X@c\dimen@ \advance\R@c-\dimen@ \fi
 \d@X=\X@c \advance\d@X-\X@p \d@X=\ifdim\d@X<\z@-\fi\d@X
 \ifdim\U@c=\D@c\relax \else
 \dimen@=\U@c\advance\dimen@-\D@c \divide\dimen@\tw@
 \advance\Y@c\dimen@ \advance\U@c-\dimen@ \fi
 \d@Y=\Y@c \advance\d@Y-\Y@p \d@Y=\ifdim\d@Y<\z@-\fi\d@Y
 \DN@{}\ifdim\d@X>\R@c \relax \else \ifdim\d@Y>\U@c \relax 
 \else \ifdim\U@c=\R@c \DN@{\circlecentredUnder@}%
 \else \DN@{\ellipsecentredUnder@}\fi
 \fi\fi \next@}}
\xydef@\circleInner@{\DN@{\ellipseInner@}%
 \ifdim\R@c=\L@c\relax \ifdim\U@c=\D@c\relax
 \ifdim\R@c=\U@c\DN@{\circlecentredInner@}%
 \else\DN@{\ellipsecentredInner@}\fi
 \fi\fi \next@}
\xydef@\circleOuter@{\DN@{\ellipseOuter@}%
 \ifdim\R@c=\L@c\relax \ifdim\U@c=\D@c\relax
 \ifdim\R@c=\U@c\DN@{\circlecentredOuter@}%
 \else\DN@{\ellipsecentredOuter@}\fi
 \fi\fi \next@}
\xydef@\circlecentredEdge@{%
 \dimen@=-\cosDirection\R@c \advance\X@c\dimen@
 \dimen@=-\sinDirection\R@c \advance\Y@c\dimen@}
\xydef@\circlecentredUnder@{%
 \loop\ifdim\R@c>100\p@ \circlescale@ \repeat
 \edef\tmp@{\expandafter\removePT@\the\R@c}\dimen@=\tmp@\R@c 
 \edef\tmp@{\expandafter\removePT@\the\d@X}\advance\dimen@-\tmp@\d@X
 \edef\tmp@{\expandafter\removePT@\the\d@Y}\advance\dimen@-\tmp@\d@Y
 \ifdim\dimen@>\z@ \aftergroup\Inside@true \fi}
\xydef@\circlescale@{\divide\R@c\KK@ \divide\d@X\KK@ \divide\d@Y\KK@ }
\xydef@\circlecentredInner@{%
 \L@c=\sd@X\cosDirection\R@c \D@c=\sd@Y\sinDirection\R@c
 \R@c=\L@c \U@c=\D@c \Edge@c={\rectangleEdge}}
\xydef@\circlecentredOuter@{%
 \L@c=\R@c \D@c=\R@c \U@c=\D@c \Edge@c={\rectangleEdge}}
\xydef@\ellipsecentredEdge@{\bgroup \X@p=\X@c \Y@p=\Y@c 
 \ifdim\U@c>\R@c
 \X@c=\cosDirection\U@c \Y@c=\sinDirection\U@c
 \quotient@\tmp@\U@c\R@c \X@c=\tmp@\X@c \R@c=\U@c 
 \else
 \X@c=\cosDirection\R@c \Y@c=\sinDirection\R@c
 \quotient@\tmp@\R@c\U@c \Y@c=\tmp@\Y@c 
 \fi
 \advance\X@c\X@p \advance\Y@c\Y@p 
 \setupDirection@ \X@c=\X@p \Y@c=\Y@p \circlecentredEdge@
 \d@X=\X@c \advance\d@X-\X@p \d@Y=\Y@c \advance\d@Y-\Y@p 
 \ifdim\U@c>\L@c \quotient@\tmp@\L@c\U@c \d@X=\tmp@\d@X
 \else \quotient@\tmp@\U@c\R@c \d@Y =\tmp@\d@Y \fi
 \X@c=\X@p \advance\X@c\d@X \Y@c=\Y@p \advance\Y@c\d@Y 
 \edef\next@{\egroup \X@c=\the\X@c \Y@c=\the\Y@c}\next@ }%
\xydef@\ellipsecentredUnder@{%
 \ifdim\R@c>64\p@ \circlescale@ \divide\U@c\KK@ 
 \else \ifdim\U@c>64\p@ \circlescale@ \divide\U@c\KK@ \fi\fi
 \edef\tmp@{\expandafter\removePT@\the\R@c}\d@Y=\tmp@\d@Y 
 \edef\tmp@{\expandafter\removePT@\the\U@c}\d@X=\tmp@\d@X 
 \R@c=\tmp@\R@c \circlecentredUnder@ }
\xydef@\ellipsecentredOuter@{\Edge@c={\rectangleEdge}}
\xydef@\ellipsecentredInner@{%
 \bgroup \X@p=\X@c \Y@p=\Y@c \ellipsecentredEdge@
 \advance\X@c-\X@p \L@c=\ifdim\X@c<\z@-\fi\L@c
 \advance\Y@c-\Y@p \D@c=\ifdim\Y@c<\z@-\fi\D@c
 \edef\next@{\egroup \L@c=\the\L@c \D@c=\the\D@c}\next@
 \R@c=\L@c \U@c=\D@c \Edge@c={\rectangleEdge}}
\xydef@\ellipseEdge@{\bgroup
 \A@=\R@c \B@=\U@c
 \ifdim\R@c=\L@c \d@X=\z@ 
 \else \d@X=.5\R@c \advance\d@X-.5\L@c 
 \advance\A@\L@c \divide\A@\tw@ \fi
 \ifdim\U@c=\D@c \d@Y=\z@ 
 \else \d@Y=.5\U@c \advance\d@Y-.5\D@c 
 \advance\B@\D@c \divide\B@\tw@ \fi
 \bgroup
 \L@c=\A@ \U@c=\B@
 \R@p=\U@c \advance\R@p\L@c \multiply\R@p\tw@
 \ifdim\B@<\A@ \quotient@\tmp@\U@c\L@c \R@c=\tmp@\p@ \D@c=\p@
 \quotient@\tmp@\R@p\L@c \R@p=\tmp@\p@
 \else
 \ifdim\A@<\B@ \quotient@\tmp@\L@c\U@c \D@c=\tmp@\p@ \R@c=\p@
 \quotient@\tmp@\R@p\U@c \R@p=\tmp@\p@
 \else
 \R@c=\p@ \D@c=\p@ \quotient@\tmp@\R@p\U@c \R@p=\tmp@\p@
 \fi\fi
 \quotient@\sd@X\d@X\L@c \d@X=\sd@X\p@
 \quotient@\sd@Y\d@Y\U@c \d@Y=\sd@Y\p@
 \loop
 \bgroup \U@p=-\p@ \D@p=\z@
 \ifdim\R@c<\p@
 \edef\tmp@{\expandafter\removePT@\the\R@c}\dimen@=\tmp@\R@p 
 \advance\d@X\cosDirection\dimen@
 \else \advance\d@X\cosDirection\R@p \fi
 \edef\sd@X{\expandafter\removePT@\the\d@X}%
 \advance\U@p\sd@X\d@X
 \ifdim\R@c<\p@
 \edef\tmp@{\expandafter\removePT@\the\R@c}\dimen@=\tmp@\d@X 
 \advance\D@p\cosDirection\dimen@
 \else \advance\D@p\cosDirection\d@X \fi
 \ifdim\D@c<\p@
 \edef\tmp@{\expandafter\removePT@\the\D@c}\dimen@=\tmp@\R@p 
 \advance\d@Y\sinDirection\dimen@
 \else \advance\d@Y\sinDirection\R@p \fi
 \edef\sd@Y{\expandafter\removePT@\the\d@Y}%
 \advance\U@p\sd@Y\d@Y
 \ifdim\D@c<\p@
 \edef\tmp@{\expandafter\removePT@\the\D@c}\dimen@=\tmp@\d@Y 
 \advance\D@p\sinDirection\dimen@
 \else \advance\D@p\sinDirection\d@Y \fi
 \multiply\D@p\tw@ 
 \dimen@=\ifdim\U@p<\z@-\fi\U@p 
 \ifdim\dimen@<.01\p@ \U@p=\z@ 
 \else
 \quotient@\tmp@\U@p\D@p \U@p=\tmp@\p@
 \ifdim\U@p<\z@\xywarning@{poor convergence}\U@p=\z@
 \else \advance\R@p-\U@p \U@p=\ifdim\U@p<\z@-\fi\U@p 
 \fi \fi
 \edef\next@{\egroup \R@p=\the\R@p \U@p=\the\U@p \D@p=\the\D@p}\next@
 \ifdim\U@p>\z@ \repeat
 \edef\next@{\egroup \dimen@=\the\R@p}\next@
 \edef\tmp@{\expandafter\removePT@\the\dimen@}%
 \ifdim\B@<\A@ \dimen@=\tmp@\B@ \else \dimen@=\tmp@\A@ \fi
 \dimen@=-\dimen@
 \advance\X@c\cosDirection\dimen@
 \advance\Y@c\sinDirection\dimen@
 \edef\next@{\egroup \X@c=\the\X@c \Y@c=\the\Y@c}\next@ }%
\xydef@\ellipseOuter@{\Edge@c={\rectangleEdge}}
\xydef@\ellipseInner@{%
 \bgroup \X@p=\X@c \Y@p=\Y@c \ellipseEdge@
 \d@X=\X@c\advance\d@X-\X@p 
 \ifdim\d@X>\z@ \R@p=\d@X \L@p=\R@p 
 \ifdim\L@c=\R@c\else\advance\L@p\L@c \advance\L@p-\R@c \fi
 \else \L@p=-\d@X \R@p=\L@p 
 \ifdim\L@c=\R@c \else\advance\R@p\R@c \advance\R@p-\L@c \fi
 \fi
 \d@Y=\Y@c\advance\d@Y-\Y@p 
 \ifdim\d@Y>\z@ \U@p=\d@Y \D@p=\U@p 
 \ifdim\D@c=\U@c\else\advance\D@p\D@c \advance\D@p-\U@c \fi
 \else \D@p=-\d@X \R@p=\D@p 
 \ifdim\D@c=\U@c\else\advance\U@p\U@c \advance\U@p-\D@c \fi
 \fi 
 \edef\next@{\egroup
 \L@c=\the\L@p \D@c=\the\D@p \R@c=\the\R@p \U@c=\the\U@p}%
 \next@ \Edge@c={\rectangleEdge}}
\xydef@\rectangleEdge#1{\ifcase#1\expandafter\rectangleEdge@
 \or \expandafter\rectangleUnder@ \or \expandafter\rectangleDist@
 \or \expandafter\rectangleProp@
 \else \relax \fi}
\xydef@\rectangleEdge@{%
 \ifdim\d@Y<-.05\p@ \rectangleEdge@i
 \else\ifdim\d@Y<.05\p@ \rectangleEdge@ii
 \else \rectangleEdge@iii\fi\fi
 \resetupDirection@}
\xydef@\rectangleEdge@i{%
 \ifdim\d@X<-.05\p@ \settomin@\X@c+\R@c\U@c\d@X\d@Y \settomin@\Y@c+\U@c\R@c\d@Y\d@X
 \else\ifdim\d@X<.05\p@ \advance\Y@c\U@c
 \else \settomin@\X@c-\L@c\U@c\d@X\d@Y \settomin@\Y@c+\U@c\L@c\d@Y\d@X
 \fi\fi}
\xydef@\rectangleEdge@ii{%
 \ifdim\d@X<-.05\p@ \advance\X@c\R@c
 \else\ifdim\d@X<.05\p@
 \else \advance\X@c-\L@c
 \fi\fi}
\xydef@\rectangleEdge@iii{%
 \ifdim\d@X<-.05\p@ \settomin@\X@c+\R@c\D@c\d@X\d@Y \settomin@\Y@c-\D@c\R@c\d@Y\d@X
 \else\ifdim\d@X<.05\p@ \advance\Y@c-\D@c
 \else \settomin@\X@c-\L@c\D@c\d@X\d@Y \settomin@\Y@c-\D@c\L@c\d@Y\d@X
 \fi\fi}
\xydef@\settomin@#1#2#3#4#5#6{%
 \edef\nextii@{\A@=\the\A@ \B@=\the\B@}\quotient@\next@{#5}{#6}\nextii@
 \dimen@=\sd@X\sd@Y\next@#4\relax
 \ifdim#3<\dimen@ \dimen@=#3\fi \advance#1#2\dimen@}
\xydef@\rectangleUnder@{\Inside@false
 \ifdim\X@p=\X@c \ifdim\Y@p=\Y@c \Inside@true \fi\fi 
 \ifInside@ \else
 \dimen@=\X@p \advance\dimen@-\X@c 
 \ifdim \dimen@>-\L@c \relax \ifdim\dimen@<\R@c 
 \dimen@=\Y@p \advance\dimen@-\Y@c 
 \ifdim \dimen@>-\D@c \relax \ifdim\dimen@<\U@c 
 \Inside@true 
 \fi\fi\fi\fi\fi }
\xydef@\rectangleDist@{\let\next@=\rectangleDist@i
 \ifdim\d@X<-.05\p@ \dimen@=\R@c
 \else\ifdim\d@X<.05\p@ \dimen@=\z@ \DN@{\dimen@=\dimen@ii}%
 \else \dimen@=\L@c \fi\fi
 \ifdim\d@Y<-.05\p@ \dimen@ii=\U@c
 \else\ifdim\d@Y<.05\p@ \DN@{}%
 \else \dimen@ii=\D@c \fi\fi
 \next@}
\xydef@\rectangleDist@i{%
 \begingroup \quotient@\next\p@{\sd@X\cosDirection\p@}%
 \edef\next{\endgroup \dimen@=\next\dimen@}\next
 \begingroup \quotient@\next\p@{\sd@Y\sinDirection\p@}%
 \edef\next{\endgroup \dimen@ii=\the\dimen@ii}\next
 \ifdim\dimen@ii<\dimen@ \dimen@=\dimen@ii \fi}
\xydef@\rectangleProp@{%
 \enter@{\A@=\the\A@ \B@=\the\B@ \DirectionfromtheDirection@}%
 \reverseDirection@
 \dimen@=1\Direction \count@=\K@ \multiply\count@\tw@
 \ifnum \Direction>\count@
 \DN@{0}%
 \advance\dimen@-2\K@ \quotient@\nextii@{\dimen@}{2\K@}%
 \else\ifnum \Direction>\z@
 \dimen@=-\dimen@ \advance\dimen@2\K@ \quotient@\next@{\dimen@}{2\K@}%
 \DNii@{0}%
 \else\ifnum \Direction>-\count@
 \DN@{1}%
 \quotient@\nextii@{-\dimen@}{2\K@}%
 \else
 \advance\dimen@4\K@ \quotient@\next@{\dimen@}{2\K@}%
 \DNii@{1}%
 \fi\fi\fi
 \advance\X@c-\L@c \dimen@=\L@c \advance\dimen@\R@c 
 \ifdim\dimen@=\z@ \advance\X@c 2\L@c \else \advance\X@c\next@\dimen@ \fi
 \advance\Y@c+\U@c \dimen@=\D@c \advance\dimen@\U@c 
 \ifdim\dimen@=\z@ \advance\Y@c-2\U@c \advance\Y@c\Upness@\U@c
 \else \advance\Y@c-\nextii@\dimen@ \fi
 \leave@}
\message{connections;}
\xydef@\Creset@@{}
\xydef@\Cshavep@@{\noCshavep@@}
\xydef@\Cshavec@@{\noCshavec@@}
\xydef@\Cslidep@@{\noCslidep@@}
\xydef@\Cslidec@@{\noCslidec@@}
\xydef@\Calong@@{\noCalong@@}
\xydef@\noCshavep@@{\setupDirection@
 \enter@{\cfromthec@ \DirectionfromtheDirection@}%
 \reverseDirection@ \cfromp@ \the\Edge@c\z@
 \pfromc@ \leave@ \resetDirection@}
\xydef@\noCshavec@@{\setupDirection@ \the\Edge@c\z@ \resetDirection@}
\xydef@\noCslidep@@#1{\dimen@=#1\relax
 \advance\X@p\cosDirection\dimen@ \advance\Y@p\sinDirection\dimen@
 \resetDirection@}
\xydef@\noCslidec@@#1{\dimen@=#1\relax
 \advance\X@c\cosDirection\dimen@ \advance\Y@c\sinDirection\dimen@
 \resetDirection@}
\xydef@\noCalong@@#1{%
 \d@X=#1\d@X \d@Y=#1\d@Y \X@c=\X@p \Y@c=\Y@p \advance\X@c\d@X \advance\Y@c\d@Y
 \resetupDirection@}
\xydef@\Cintercept@@{\noCintercept@@}
\xydef@\noCintercept@@{\enter@{\pfromthep@}%
 \begingroup\Creset@@ \edef\tmp@{\endgroup
 \X@origin=\the\X@p \Y@origin=\the\Y@p \R@c=\the\d@X \U@c=\the\d@Y}\tmp@
 \loop@\dimen@=\ifdim\R@c<\z@-\fi\R@c \advance\dimen@\ifdim\U@c<\z@-\fi\U@c
 \ifdim\dimen@>10\p@ \advance\R@c \ifdim\R@c<\z@-\fi 16sp \divide\R@c\KK@ 
 \advance\U@c \ifdim\U@c<\z@-\fi 16sp \divide\U@c\KK@ \repeat@
 \intersect@ \leave@}
\xydef@\no@@{\setupDirection@
 \edef\Creset@@{\cfromthec@ \pfromthep@ \noexpand\setupDirection@}%
 \def\Cshavep@@{\noCshavep@@}\def\Cshavec@@{\noCshavec@@}%
 \def\Cslidep@@{\noCslidep@@}\def\Cslidec@@{\noCslidec@@}%
 \def\Calong@@{\noCalong@@}\def\Cintercept@@{\noCintercept@@}%
 \ifHidden@\else
 \ifdim\Y@c>\Y@max \Y@max=\Y@c \fi \ifdim\Y@p>\Y@max \Y@max=\Y@p \fi
 \ifdim\Y@c<\Y@min \Y@min=\Y@c \fi \ifdim\Y@p<\Y@min \Y@min=\Y@p \fi
 \ifdim\X@c>\X@max \X@max=\X@c \fi \ifdim\X@p>\X@max \X@max=\X@p \fi
 \ifdim\X@c<\X@min \X@min=\X@c \fi \ifdim\X@p<\X@min \X@min=\X@p \fi \fi}
\xydef@\Spread@@{}
\xydef@\checkoverlap@@{}
\xydef@\straight@#1{\setupDirection@ \def\Spread@@{#1}%
 \edef\Creset@@{\cfromthec@ \pfromthep@ \DirectionfromtheDirection@}%
 \DN@##1##2{\def\checkoverlap@@{%
 \ifdim##1\X@p>##1\X@c \let\next@=\relax \fi
 \ifdim##2\Y@p>##2\Y@c \let\next@=\relax \fi}}%
 \edef\nextii@{{\sd@X}{\sd@Y}}\expandafter\next@\nextii@
 \noCshavep@@\edef\Cshavep@@{\pfromthep@ \noexpand\resetDirection@}%
 \noCshavec@@\edef\Cshavec@@{\cfromthec@ \noexpand\resetDirection@}%
 \ifHidden@\else
 \ifdim\Y@c>\Y@max \Y@max=\Y@c \fi \ifdim\Y@p>\Y@max \Y@max=\Y@p \fi
 \ifdim\Y@c<\Y@min \Y@min=\Y@c \fi \ifdim\Y@p<\Y@min \Y@min=\Y@p \fi
 \ifdim\X@c>\X@max \X@max=\X@c \fi \ifdim\X@p>\X@max \X@max=\X@p \fi
 \ifdim\X@c<\X@min \X@min=\X@c \fi \ifdim\X@p<\X@min \X@min=\X@p \fi \fi
 \ifInvisible@\let\next@=\relax
 \else\ifdim 1\Direction<-2\K@ \let\next@=\straightv@
 \else\ifdim 1\Direction<\z@ \let\next@=\straighth@
 \else\ifdim 1\Direction<2\K@ \let\next@=\straightv@
 \else \let\next@=\straighth@ \fi\fi\fi\fi
 \checkoverlap@@ \next@
 \def\Cslidep@@{\noCslidep@@}\def\Cslidec@@{\noCslidec@@}%
 \def\Calong@@{\noCalong@@}\Creset@@}
\xydef@\straighth@{\setbox\z@=\hbox{%
 \A@=\wd\lastobjectbox@
 \B@=\dp\lastobjectbox@ \advance\B@\ht\lastobjectbox@
 \ifdim \A@=\z@ \count@@=\m@ne
 \else \dimen@=\sd@X\d@X \divide\dimen@\A@ \count@@=\dimen@ \fi
 \Spread@@
 \ifdim\d@X>\z@ \advance\X@c-\wd\lastobjectbox@ \fi
 \dimen@=-\sd@X\wd\lastobjectbox@
 \multiply\dimen@\K@dYdX \divide\dimen@\K@
 \ifdim\d@X>\z@ \advance\Y@c\dimen@ \advance\Y@c-\Leftness@\dimen@
 \else \advance\Y@c\Leftness@\dimen@ \fi
 \dimen@=\wd\lastobjectbox@ \A@=\sd@X\d@X \advance\A@-\dimen@
 \B@=\sd@X\dimen@ \multiply\B@\K@dYdX \divide\B@\K@
 \advance\B@-\d@Y \B@=\sd@Y\B@
 \count@=\count@@ \advance\count@\m@ne
 \ifnum\z@<\count@ \divide\A@\count@ \divide\B@\count@ \fi
 \A@=-\sd@X\A@ \B@=\sd@Y\B@ \wd\lastobjectbox@=\A@
 \kern\X@c \count@=\z@
 \loop@\ifnum\count@<\count@@ \advance\count@\@ne
 \raise\Y@c\copy\lastobjectbox@ \advance\Y@c\B@ \repeat@}%
 \ht\z@=\z@ \wd\z@=\z@ \dp\z@=\z@ {\Drop@@}}
\xydef@\straightv@{\setbox\z@=\vtop{%
 \A@=\wd\lastobjectbox@
 \B@=\dp\lastobjectbox@ \advance\B@\ht\lastobjectbox@
 \ifdim \B@=\z@ \count@@=\m@ne
 \else \dimen@=\sd@Y\d@Y \divide\dimen@\B@ \count@@=\dimen@ \fi
 \Spread@@
 \dimen@=\dp\lastobjectbox@ \advance\dimen@\ht\lastobjectbox@
 \B@=\sd@Y\d@Y \advance\B@-\dimen@
 \A@=\sd@Y\dimen@ \multiply\A@\K@dXdY \divide\A@\K@ \advance\A@-\d@X
 \A@=\sd@X\A@ \count@=\count@@ \advance\count@\m@ne
 \ifnum\z@<\count@ \divide\B@\count@ \divide\A@\count@ \fi
 \B@=\sd@Y\B@ \A@=\sd@X\A@ \ht\lastobjectbox@=\B@ \dp\lastobjectbox@=\z@
 \ifdim\d@Y<\z@ 
 \advance\Y@c\dimen@ \advance\Y@c\Upness@\B@
 \else
 \advance\dimen@\Upness@\B@ \advance\Y@c-\dimen@ \advance\Y@c\B@
 \fi
 \advance\Y@c\B@ 
 \ifdim\d@X<\z@ \else \advance\X@c-\wd\lastobjectbox@ \fi
 \null \kern-\Y@c \count@=\z@
 \loop@\ifnum\count@<\count@@ \advance\count@\@ne
 \nointerlineskip \moveright\X@c\copy\lastobjectbox@ \advance\X@c\A@
 \repeat@}%
 \ht\z@=\z@ \wd\z@=\z@ \dp\z@=\z@ {\Drop@@}}
\message{ Xy-pic loaded}\xyuncatcodes  

\xyoption{all}

\newcommand{\raw}{\rightarrow}
\newcommand{\Raw}{\Rightarrow}
\newcommand{\LRaw}{\Leftrightarrow}
\newcommand{\law}{\leftarrow}

\newcommand{\lraw}{\leftrightarrow}

\newcommand{\lto}{\mapsto}

\newcommand{\nex}{\raisebox{0.4mm}{\scriptsize $\bigcirc$}}

\newcommand{\orbar}{\ \big|\ }
\newcommand{\init}{\mbox{ \footnotesize \sc init }}
\newcommand{\stable}{\mbox{ \footnotesize \sc stable }}
\newcommand{\unless}{\mbox{ \footnotesize \sc unless }}

\newcommand{\becausec}{\mbox{ \footnotesize \sc because\_c }}
\newcommand{\because}{\mbox{ \footnotesize \sc because }}

\newcommand{\causes}{\mbox{ \footnotesize \sc leads\_to }}
\newcommand{\causesc}{\mbox{ \footnotesize \sc leads\_to\_c }}

\newcommand{\gop}{\mbox{ \footnotesize \sc op }}

\newcommand{\tcauses}{\mbox{\footnotesize \sc leads\_to}}
\newcommand{\tcausesc}{\mbox{\footnotesize \sc leads\_to\_c}}
\newcommand{\tinit}{\mbox{\footnotesize \sc init}}
\newcommand{\tstable}{\mbox{\footnotesize \sc stable}}
\newcommand{\tunless}{\mbox{\footnotesize \sc unless}}
\newcommand{\tbecausec}{\mbox{\footnotesize \sc because\_c}}
\newcommand{\tbecause}{\mbox{\footnotesize \sc because}}

\newcommand{\oikosadtl}{Oikos--{\em adtl\/} }

\newcommand{\mii}{\mbox{{\bf \sf m}$_{\mbox{\bf \sf i}}$} }
\newcommand{\mij}{\mbox{{\bf \sf m}$_{\mbox{\bf \sf j}}$} }

\newcommand{\mI}{\mbox{{\bf \sf m}$_{\mbox{\bf \sf 1}}$} }
\newcommand{\mII}{\mbox{{\bf \sf m}$_{\mbox{\bf \sf 2}}$} }

\newcommand{\mm}{\mbox{{\bf \sf m}} }
\newcommand{\tmm}{\mbox{{\bf \sf m}}}
\newcommand{\tmn}{\mbox{{\bf \sf n}}}
\newcommand{\mn}{\mbox{{\bf \sf n}} }
\newcommand{\moo}{\mbox{{\bf \sf o}} }

\newcommand{\mbb}{\mbox{{\bf \sf b}} }
\newcommand{\mtt}{\mbox{{\bf \sf t}} }
\newcommand{\muu}{\mbox{{\bf \sf u}} }

\newcommand{\tmii}{\mbox{{\bf \sf m}$_{\mbox{\bf \sf i}}$}}

\newcommand{\bari}{\bar{\mbox{{\bf \sf m}}}_{\mbox{\bf \sf i}}}

\newcommand{\bank}{\mbox{{\bf \sf bank}} }
\newcommand{\user}{\mbox{{\bf \sf user}} }

\begin{document}

\title{Distributed States Temporal Logic}

\author{Carlo Montangero \quad Laura Semini}

\institute{Dipartimento di Informatica, Universit\`a di Pisa.
\\
\{monta, semini\}@di.unipi.it}

\maketitle

\begin{abstract}
 We introduce a temporal logic to reason on global applications in an
asynchronous setting.  First, we define the Distributed States Logic
(DSL), a modal logic for localities that embeds the local theories of
each component into a theory of the distributed states of the
system. We provide the logic with a sound and complete
axiomatization. The contribution is that it is possible to reason
about properties that involve several components, even in the absence
of a global clock.  Then, we define the Distributed States Temporal
Logic (DSTL) by introducing temporal operators \`a la Unity.  We
support our proposal by working out a pair of examples: a simple
secure communication system, and an algorithm for distributed leader
election.

The motivation for this work is that the existing logics for
distributed systems do not have the right expressive power to reason
on the systems behaviour, when the communication is based on
asynchronous message passing. On the other side, asynchronous
communication is the most used abstraction when modelling global
applications.
\end{abstract}

\section{Introduction}

The current trend towards global computing needs software that works
in an open, concurrent, distributed, high--latency, security--sensitive
environment. Besides, this software must be reliable, scalable, and
``shipped today''.
Several trends are emerging in response to the challenges involved in
the development of software with so demanding requirements.

On one side, there is an increasing interest in the seamless
integration of asynchronous communication in programming,
coordination, and specification languages, since message--passing,
event--based programming, call--backs, continuations, dataflow models,
workflow models etc. are ubiquitous in global computing. Notable
examples in this direction can be found in the context of the
Microsoft .NET initiative, like the introduction of support for the
delegate--based asynchronous calling model in the libraries of the
Common Language Runtime~\cite{CLR2}, and the proposal of $chords$ in
Polyphonic C\# to accommodate asynchronous methods in
C\#~\cite{CSHARP}.  We provide an example of
coverage of asynchronous communication in coordination and
specification languages in~\cite{sgarrascico}.

Another significant trend is represented by Component--Oriented
Programming, that aims at producing software components for a software
market and for late composition. Composers are third parties, possibly
the end user, who are not able nor willing to modify components. This
trend emphasizes the need for high quality specifications that put the
composer into the position to decide what can be composed under which
conditions. In a previous work with
\oikosadtl\cite{coord99c,fmics01}, a specification language for
distributed systems based on asynchronous communications, we showed
how to accommodate asynchronous communication in the composition of
distributed systems specifications.

A notable example of component programming in the context of global
computing is offered by the Web Services~\cite{Kreger}, which leverage
the standard representation of data provided by XML to foster the
construction of new components (services) by the coordination of other
services. Since the cooperation is based on asynchronous protocols,
this is also an example of the convergence of asynchronous
communications and component programming.

Formal methods can play a major role in global computing. Precisely
because the actors are programmatically independent, they need to have
reliable ways to share precise knowledge of the artifacts they use or
produce, independently of the particular technology (programming
languages, middleware, \dots) they rely on. Formal methods offer
exactly this kind of independence and precision, since they provide
abstract models to share when operating or developing with
components.  They can provide ways to make precise the specifications
of the components and of their contextual dependencies, and to prove
in advance global properties, i.e.  that a composition will meet the
specifications it addresses.

\

\noindent 
In this paper we define DSTL (Distributed States Temporal Logic), an
extension of temporal logic to deal with distributed systems.
In~\cite{time02} we defined new modalities to name system
components. Here, we introduce the operators to causally relate
properties which might hold in distinguished components, in an
asynchronous setting.  A typical DSTL formula is:
\begin{eqnarray}
&&\mm  \,p \; \causes \; \mn  \,q \wedge \moo \, r \label{ex:intro}
\end{eqnarray}


\noindent 
where the operator $\tcauses$ is similar to Unity's $\lto$ (leads
to)~\cite{chmi88}, and \tmm, \tmn, and \moo express locality.
Formula~(\ref{ex:intro}) says that a property $p$ holding 
in component $m$, causes properties $q$ and $r$ to hold in future
states of components $n$ and $o$, respectively. An example is the
computation below. Horizontal arrows denote the sequence of states of
a component, oblique arrows denote the communications.

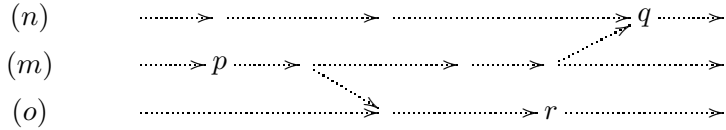
\vspace{-0.2cm}

\hspace{0.9cm}
 \xymatrix@R-=1pt{
 &&&&&&&&\\
  {(n)}& \ar@{.>}[r] &   \ar@{.>}[rr]  && 
                 \ar@{.>}[rrr] 
                 && &  {q} \ar@{.>}[r] & \\ 
  {(m)}&  \ar@{.>}[r] &  {p} \ar@{.>}[r] &  \ar@{.>}[rd]  
                  \ar@{.>}[rr]&& \ar@{.>}[r]  &
                                 \ar@{.>}[rr]\ar@{.>}[ru] & & &\\  
  {(o)}&  \ar@{.>}[rrr] &  &&  \ar@{.>}[rr] &&
         {r}\ar@{.>}[rr] &&& 
 }

\vspace{0.2cm}

\noindent
At this point a short philosophical note is needed. We tend to think
that our operators express {\em causality}, even though, strictly
speaking, they only define temporal relations, i.e. that their
consequences (right hand side operands) hold {\em after} (or before,
with past operators) their premises (left hand side operands).  In
fact, in our models, a state in a component is after one in another
component only if there has been a communication between the two.
Philosophically, this may not entail a causal relation, but our goal
is to specify systems: it is natural to think that the communication
carries the information needed to cause the intended effect.  It is in
this sense that we use the term causality.

A similar argument applies locally: the implementation will take care
that a state satisfying the consequences is reached, after one
satisfying the premises.

\vspace{0.5cm}

\noindent
From a technical point of view, the usual choices to build a Kripke
model for formulae like~(\ref{ex:intro}) are to consider the set of
worlds $W$ to be one of the following:
\begin{enumerate}
\item the set of the  states of a computation, i.e. the union of
all the states of the system components, like the circles in the
following figure.

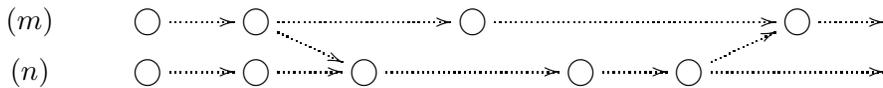
\vspace{-0.2cm}

\hspace{0.9cm}
 \xymatrix@R-=2pt{
 &&&&&&&&\\
 {(m)}&{\bigcirc} \ar@{.>}[r] &  {\bigcirc} \ar@{.>}[rd]   
	\ar@{.>}[rr]  && {\bigcirc}
                 \ar@{.>}[rrr] 
                 && &  {\bigcirc}  \ar@{.>}[r] & \\ 
 {(n})&{\bigcirc}  \ar@{.>}[r] & {\bigcirc} \ar@{.>}[r] & {\bigcirc}
                  \ar@{.>}[rr]&&{\bigcirc} \ar@{.>}[r]  &{\bigcirc}
                                 \ar@{.>}[rr]\ar@{.>}[ru] & & &\\  
 }

\vspace{0.3cm}

\noindent
This choice was adopted in \oikosadtl and has shown some
problems.  For instance, consequence weakening, or, more in general,
the possibility of reasoning on logical relations between formulae
like the premises or the consequences of~(\ref{ex:intro}), is not part
of the logic.  In particular, a formula like

\vspace{-0.8cm}

\begin{eqnarray}
&&(\mn \,q \wedge \mm \,r) \raw \mn \,q \label{ex:intro2}
\end{eqnarray}

\vspace{0.3cm}

\noindent 
which would permit to weaken the consequences of~(\ref{ex:intro})
would not be a legal formula, since no world can satisfy the conjunction
$\mn \,q \wedge \mm \,r$.

\item the set of global states, or snapshots, of the system, where
each world is a tuple of states, one for each component. These tuples must
satisfy some constraints to be coherent with the communications
between the subsystems. In the figure below, examples of worlds are
$\langle s_m^i ,\,s_n^j\rangle^{i=0,1}_{0\leq j \leq 2}$, while $\langle s_m^2
,\,s_n^1\rangle$ would not be a legal world.

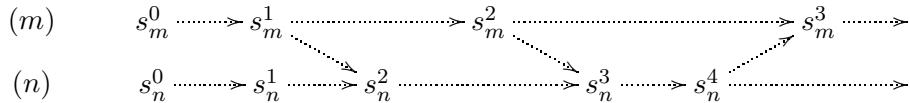
\vspace{-0.2cm}

\hspace{0.9cm}
 \xymatrix@R-=6pt{
 &&&&&&&&\\
 {(m)}&{s_m^0}   \ar@{.>}[r] &  {s_m^1} \ar@{.>}[dr]
\ar@{.>}[rr]  && \ar@{.>}[dr]{s_m^2} 
                 \ar@{.>}[rrr] 
                 && &  {s_m^3}  \ar@{.>}[r] & \\ 
 {(n)}&{ s_n^0}  \ar@{.>}[r] & { s_n^1}  
\ar@{.>}[r] &  { s_n^2}   
                  \ar@{.>}[rr]&&  { s_n^3} \ar@{.>}[r]  &{ s_n^4}
                                 \ar@{.>}[rr]\ar@{.>}[ru] & & &\\  
 }

\vspace{0.3cm}

\noindent 
This choice, adopted in many logics for distributed systems (see
Section~\ref{discussion}) is not applicable in the case of
asynchronous communication. Think of the case of property $p$ holding
{\em only} in state $s_m^1$ and $q$ holding {\em only} in states
$s_n^j$, for $0\leq j \leq 4$. The formula 
\vspace{-0.8cm}

\begin{eqnarray}
&&\mm \,p \raw   \mn \,q \label{ex:intro3}
\end{eqnarray}

\vspace{-0.2cm}

\noindent
would be valid in the model, inferring a remote instantaneous
knowledge which is meaningless in an asynchronous setting. Moreover,
it would be natural to say that world $\{s_m^2,\, s_n^3\}$ follows
$\{s_m^1,\,s_n^2\}$. In this case, one could assert that $\mn \,p
\causes \mm \,q$ holds, if $p$ and $q$ hold in $s_n^2$ and $s_m^2$,
respectively, even though not even a temporal relationship exists
between these two states.

\item a third possibility would be to consider all the $k$--tuples of
states (where $k$ is the number of the system components) as
worlds. But then, formula~(\ref{ex:intro3}) would be valid in the
model above if $q$ holds in all the states of component $n$. Even if
this is philosophically more acceptable, we claim that a better
solution can be found. What is more, this choice is not
adequate since if we let $p$ and $q$ hold in $s_m^1$ and $s_n^2$,
respectively, we would like the computation above to be a model for
$\mm \,p \causes \mn \,q$. On the contrary, world $\{s_m^1,\,s_n^3\}$
satisfies the premise but is not followed by any state satisfying the
consequence. 

\end{enumerate}

\noindent
The first contribution of our work is to introduce the distributed
state logic DSL, that carries over all meaningful propositional rules,
like {\em and} simplification, so that they can be exploited
orthogonally to any temporal operator.  A major consequence of the
introduction of DSL is that the exploitation of the local theories in
the proofs of the distributed properties becomes smooth and
robust. 

The second part of the paper defines DSTL: we add the temporal
operators, and the corresponding derivation rules.  The semantic
domain of DSL, the power--set of the set of all system states, even if
chosen for technical reasons, makes the full logic DSTL a very
expressive language, that meets the pragmatic expectations of a
designer fully (see Section~\ref{discussion} for a discussion).  The
achievement is that it is possible to reason about properties that
involve several components, even in the absence of a global clock, the
typical assumption in an asynchronous setting.

Section \ref{DSL} introduces the modal logic DSL, and its sound and
complete axiomatization.  Section \ref{logics} defines DSTL as an
extension of DSL with the temporal operators.  Sections \ref{example}
and \ref{example2} work out a pair of examples: a simple secure
communication system, and an algorithm for the leader election
problem. The last sections cover a discussion of the main design
issues, related work and future perspectives.

\section{DSL}
\label{DSL}

We assume a countable set of propositional letters $P$, with $p, q,
 \ldots$ ranging over $P$.  The DSL well--formed formulae over a
 finite set of components $\Sigma = \{m_1, m_2, \ldots, m_k\}$ are
 defined by:
\begin{eqnarray*}
F & \ :: = \ & p \orbar \bot \orbar \sim F \orbar F\wedge F^\prime
\orbar \mii F
\end{eqnarray*}

\noindent
where $\bot$ is the propositional constant {\em false}, and \mii for
$i=1\ldots k$ are  unary location operators.   With $\bari$ we
denote the dual of \tmii, i.e., $\bari F \equiv \,\sim \mii   \sim
F$. With $\top$ we denote {\em true}, i.e. $\top \equiv \:\sim \bot$.

\subsection{Semantics} 
A model ${\cal M}$ for DSL formulae is a tuple $(W, R_{1}, 
\ldots,R_{k}, V)$. Let $u,\, v,\, w$ range over $W$, the
reachability relations $R_i$ satisfy the following conditions:
\begin{eqnarray}
(u,\, v) \in R_i   & \mbox{ }\quad \raw \quad \mbox{ } 
& (v,\, v) \in R_i  \label{rc1}\\
(u,\, v) \in R_i   & \raw &  (v,\, w) \in R_i \raw v=w 
\quad \quad\quad \quad \mbox{ }\label{rc2}\\
(u,\, v) \in R_i   & \raw &  \not\exists w.\: (v,\, w) \in R_j \mbox{ for } j\neq i \label{rc3}
\end{eqnarray}

\noindent
To help the intuition, $W$ can be thought as having $k$ disjoint
subsets of worlds: we call these worlds {\em leaves}. Whenever $(u,\, v) \in
R_i$, then $v$ is a leaf for relation $R_i$, namely an {\em i--leaf}.
Condition~(\ref{rc1}) says that $R_i$ is reflexive on {\em i--leaves},
conditions~(\ref{rc2}) and~(\ref{rc3}) say that {\em i--leaves} are
actually leaves: no other world can be reached.  An example model is
in Section~\ref{anexamplemodel}, where the {\em i--leaves} are
singleton sets, having as unique element a state of component $m_i$.

\

\noindent
The semantics of the DSL formulae is given by:
\begin{eqnarray*}
&&({\cal M}, \, u) \models \top  \\
&&({\cal M}, \, u) \models p   \ \mbox{ iff } \ p\in V(u)\\
&&({\cal M}, \, u) \models \, \sim F   \ \mbox{ iff } \mbox{ not }  ({\cal M}, \, u) \models F\\
&&({\cal M}, \, u) \models F  \wedge F'  \ \mbox{ iff } \ ({\cal M}, \, u) \models F  \mbox{ and }   ({\cal M}, \, u) \models F' \\
&&({\cal M}, \, u) \models \mii F   \ \mbox{ iff } \ \exists  v.\; (u, v)\in R_{i} 
\; \mbox{and}\;  ({\cal M}, \, v) \models F
\end{eqnarray*}

\subsection{Axiom system} \label{dslaxsys}
  We propose the following axiomatization for
DSL. For the sake of readability, we use $\mm$ and  $\mn$, with $\mm\neq
\mn$, instead of  $\mii$ and  $\mij$. 

\

$\begin{array}{ll}
{\bf PC}  &  \mbox{axioms of the propositional calculus}\\
{\bf K}  & \bar{\mm} (F\raw F') \ \raw \ (\bar{\mm}F \raw \bar{\mm} F')\\
{\bf DSL1}   & \bar{\mm} (\bar{\mm}F \lraw F) \\
{\bf DSL2}  \ \  &  \bar{\mm}  \bar{\mn} \bot \\
{\bf MP} & \ {\small \prooftree F \quad F\raw F' \justifies F' \endprooftree}
\quad \quad {\bf Nec } \ \  {\small \prooftree F \justifies \bar{\mm}F \endprooftree}
 \end{array}
$

\vspace{0.2cm}

\begin{theorem} The DSL axiom system is sound and complete.

\vspace{0.2cm}

\noindent
{\bf Proof.} The soundness of the axioms is easy to see. We prove
completeness.

 Let $(W^{DSL}, R_{1}^{DSL}, \ldots,R_{k}^{DSL},
V^{DSL})$ be the canonical model for DSL: worlds in
$W^{DSL}$ are maximal consistent sets of DSL formulae ({\sc dsl--mcs}
in the following), and  $(u,v)\in R_{i}^{DSL}$ if and only if
$\bar{\mii} F \in u \; \raw \; F \in v$.  We need to show
that, for all $i$, $R_{i}^{DSL}$ satisfies
conditions~(\ref{rc1})--(\ref{rc3}).

\vspace{-0.3cm}

\begin{description} 
\item[{\rm \em Cond. (\ref{rc1}):}]  we prove that    $(u,v)\in
R_{i}^{DSL} \; \raw \; (v,v)\in 
R_{i}^{DSL}$
\\
Suppose $\bar{\mii} F \in v$. $u$ is a {\sc dsl--mcs} and hence (see
DSL1) $ \bar{\mii} (\bar{\mii}F \raw F)  \in u$.  But $(u,v)\in
R_{i}^{DSL}$, hence  $ \bar{\mii}F \raw F \in v $. Thus, by modus
ponens, $F\in v$.

\item[{\rm \em Cond.  (\ref{rc2}):}] we prove that $(u,v)\in
R_{i}^{DSL}$ and $(v,w)\in 
R_{i}^{DSL}$ imply $v=w$ 
\\ 
It is sufficient to prove  that $v\subseteq w$. In fact, $v$ and $w$ are  
{\sc dsl--mcs} and it is not the case that $v
\subset w$, thus $v=w$. 
Let $F\in v$. $u$ is a {\sc dsl--mcs} and hence (see
DSL1) it includes $ \bar{\mii} (F \raw \bar{\mii}F)$. But $(u,v)\in
R_{i}^{DSL}$, hence  $ F \raw \bar{\mii} F \in v $. 
Thus, by modus
ponens, $\bar{\mii} F\in v$. As $(v,w)\in R_{i}^{DSL}$, we conclude
that $F\in w$. 

\item[{\rm \em Cond.  (\ref{rc3}):}] we prove that $(u,v)\in
R_{i}^{DSL}$ implies 
$\not\exists w. $ $(v,w)\in R_{j}^{DSL}$, for $j\neq i$ 
\\ 
Assume $(v,w)\in R_{j}^{DSL}$.  As $u$ is a {\sc dsl--mcs}, it
includes $\bar{\mii}\bar{\mij} \bot$ (DSL2). As $(u,v)\in R_{i}^{DSL}$, then
$\bar{\mij} \bot \in v$.  As $(v,w)\in R_{j}^{DSL}$, then $\bot \in w$,
which is an absurd. \mbox{ }\hfill $\Box$

\end{description}

\end{theorem}

\begin{example} \label{toprove}
The following formulae can be derived. Formulae are followed by the
list of axioms or rules used in their proof. The proofs are in the
appendix.

\

$\begin{array}{lll}
{\bf axiom \; 4} \ & \bar{\mm} F \ \raw \ \bar{\mm}\bar{\mm}F & [DSL1,K]\\
 {\bf D1} & \mm\mm F \lraw \mm F \quad\quad \quad\quad\quad\quad\quad\quad& [DSL1,K,PC]
\\
 {\bf D2} \quad\quad & \mm (F\wedge F') \raw(\mm F\wedge \mm F') & [PC,Nec,K]\\
 {\bf D3} & \bar{\mm}(F\raw F') \ \raw \ (\mm F \raw \mm F^\prime)
& [Nec,K,MP,PC] \\ 
{\bf D4} & \bar{\mm} F \raw (\mm \top \raw \mm F ) & [D3] \\
 {\bf D5}  & \bar{\mm} (\mm F \lraw F)  & [DSL1,PC] \\
 {\bf D6} &  ( \bar{\mm} (F\raw F') \wedge  \bar{\mm}
(F'\raw F'') )\raw  \bar{\mm} (F\raw F'') \quad\quad\quad & [Nec,K]\\
 {\bf D7}  & \mm (F\vee F' )\lraw (\mm F \vee \mm F') & [D3,Nec,K,PC]\\
 {\bf D8}  & \bar{\mm} ((\mm F\wedge \mm F')\raw \mm (F\wedge F')) &
[D5,D6,D7,Nec,K]\\

 \end{array}$

\end{example}

\subsection{A frame of distributed states} 
\label{anexamplemodel} 

\noindent
Let $S_{i}$ be the set of states of component $m_i$, with $S_{i}\cap
S_{j}= \emptyset$ for $i\neq j$, $S = \bigcup^{i=1,}_{k} S_{i}$, $DS =
2^{S}$, and $ds,\, ds'\in DS$.  Let $(ds, ds')\in R_{i}$ if and only
if $ds'$ is a singleton set $\{s\}$, with $s\in S_{i} \cap ds$. The
frame $(DS, R_{1}, \ldots,R_{k})$, satisfies
conditions~(\ref{rc1})--(\ref{rc3}) above. We call these frames
{\em frames on} $DS$, and call $DS$ the set of {\em distributed states},
from which the name of the logic DSL. The frames on $DS$ play a
central role in the paper, since they are used to build the models for
DSTL formulae.  

Some examples follow.

\begin{example}\label{Frame}
Let the set $DS$ be built on $S_1 = \{s, s^{\prime}\}$ and $S_2 =
\{s^{\prime\prime}\}$, then the frame on $DS$ can be
represented as:

\vspace{-0.7cm}

{\small
\mbox{ } \quad \quad\quad \quad \quad\quad \quad \quad\quad \mbox{ }  
\xymatrix@R-=10pt@C-=6pt{
&&&&&&\\
         && & \{s,s^{\prime\prime}\}\ar[dl]^{R_1}\ar[dr]_{R_2} 
        &&&   \\
& & \{s\}\ar@(ur,ul)[]_{R_1} &&  \{s^{\prime\prime}\}\ar@(ur,ul)[]_{R_2} &&\\
\{s,s^{\prime}\}\ar[urr]_{R_1}\ar[ddrrr]_{R_1}
&&& 
 \{s,s^{\prime},s^{\prime\prime}\}\ar[dd]_{R_1}\ar[ul]_{R_1} \ar[ur]^{R_2}&&&
\{s^{\prime},s^{\prime\prime}\}\ar[ull]^{R_2}\ar[ddlll]^{R_1}\\
&&&&&&\\
&&&  \{s^{\prime}\}\ar@(dl,dr)[]_{R_1} &&&\\
&&&&&&\\
}
}

\end{example}

\begin{note}
For the sake of readability, when we use $\mm$ and $\mn$, we also use
$S_m$, and $S_n$.
\end{note}

\begin{example}
If we take $s\in S_m$, $s^\prime \in S_n$

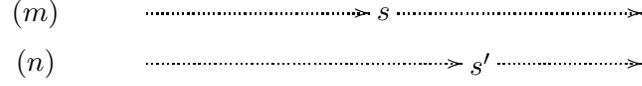
\vspace{-0.2cm}

\hspace{2.2cm}
 \xymatrix@R=2pt{
 &&&&&&&\\
 {(m)}& \ar@{.>}[rrr] &&& {s} \ar@{.>}[rrr] &&& \\
 {(n)}&  \ar@{.>}[rrrr] &&&& {s^\prime}  \ar@{.>}[rr] && \\
 }

\vspace{0.2cm}

\noindent
with $V(\{s\})=\{p\}$, $V(\{s^\prime\})=\{q\}$, then the distributed state
$\{s,s^\prime\}$ satisfies $\mm p \wedge \mn q$.

\end{example}

\begin{example}
The implication $\mm (F \wedge F') \raw \mm F \wedge \mm F'$ holds,
while the converse does not. Indeed, for $ds = \{s, s'\} \subseteq
S_m$

\vspace{-0.2cm}

\hspace{2.2cm}
 \xymatrix@R=2pt{
 &&&&&&&\\
 {(m)}& \ar@{.>}[rr] && {s}\ar@{.>}[rr] && {s^\prime} \ar@{.>}[rr] && \\
 }

\vspace{0.2cm}

\noindent
and $V(\{s\}) = \{p\}$, $V(\{s'\}) = \{q\}$, we have $ds\models
\mm p \wedge \mm q$, but not $ds\models \mm \, (p \wedge q)$.  With an
eye to the full logic DSTL, this non--equivalence is useful to specify
that an event can have different future effects in a component,
without constraining them to occur in the same state (see
Section~\ref{discussion} for further discussion).

\end{example}

\begin{example}
The formula $\mm \mn F$ is false. In fact, $ds \models \mm \mn F$ if
and only if there exists an $s\in S_{n} \cap S_{m} \cap ds$ such that
$\{s\} \models F$, but no such $s$ can exist since $S_{m}$ and $S_{n}$
are disjoint. Conversely, $\mm \mm F$ is satisfiable, and it is
equivalent to $\mm F$.
\end{example}

\begin{example}
The formula  $\mm \top$ is satisfied by all the distributed states
$ds$ such that  $ds \cap S_{m} \neq \emptyset$. 
\end{example}

\section{DSTL}
\label{logics}

DSTL extends DSL adding temporal operators.  Formulae are built as
follows:

\vspace{-0.5cm}

\begin{eqnarray*}
\phi & \ :: = \ & F \orbar F\: \causes \: F^\prime  \orbar F\:
\because \: F^\prime  \orbar F\: \causesc \: F^\prime  \orbar F\:
\becausec \: F^\prime  \orbar  \\
&&F\: \unless \: F^\prime \orbar \init F
\end{eqnarray*}

\vspace{-0.1cm}

\noindent
where $F,\, F'\in DSL$.  Operator $\tcauses$ expresses a liveness
condition, and is similar to Unity's $\lto$ (leads to): $F$ is surely
followed by $F'$.  Operator $\tbecause$ expresses a safety condition, and
says that $F$ must be preceded by $F^\prime$.

Suffix {\footnotesize \sc c} stands for {\em closely}, $\tcausesc$
requires formula $F'$ to hold in the distributed states in which $F$
holds, or in the next ones. Dually, $\tbecausec$ says that $F'$ must hold
in the states immediately preceding those satisfying $F$, or in the
same ones.

Operator $\tunless$ extends Unity's $\tunless$, and $\tinit$ permits to
describe the initial state. A special case of $\tunless$ is stability:

$$ \stable F \stackrel{def}{=} \, F \unless \bot$$

\subsection{Semantics} 
\label{logics.semantics}

The models for DSTL formulae are built on structures like the one in
the following figure, which describes the computation of a system with
three components ($m,\, n,\, o$).  We call $s_m^{i}$ the $i^{th}$
state of component $m$. We call $ds^0$ the set of the
initial states $\{s^0_m, s^0_n, s^0_o\}$.

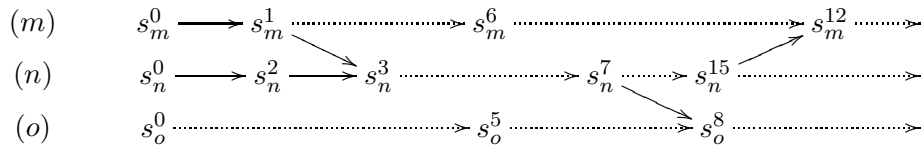
\vspace{-0.2cm}

\hspace{0.5cm}
 \xymatrix@R-=2pt{
 &&&&&&&&\\
 {(m)}& {s_m^0} \ar[r] & {s_m^1} \ar[rd] \ar@{.>}[rr]  &&  {s_m^6}
                 \ar@{.>}[rrr] 
                 && &  {s_m^{12}} \ar@{.>}[r] & \\ 
 {(n)}& {s_n^{0}}  \ar[r] &  {s_n^{2}}\ar[r] &  {s_n^{3}}
                  \ar@{.>}[rr]&&{s_n^{7}}\ar@{.>}[r] \ar[rd]  &{s_n^{15}}
                                 \ar@{.>}[rr]\ar[ru] & & &\\  
 {(o)}& {s_o^{0}}  \ar@{.>}[rrr] &  && {s_o^{5}} \ar@{.>}[rr] &&
         {s_o^{8}}\ar@{.>}[rr] &&& 
 }

\

\noindent
In the figure, plain arrows denote atomic state
transitions and communications, dotted arrows denote sequences of
them.

\begin{definition} {($R, \; R^=, \; R^*$)}

\noindent
State transitions and communications define a next state relation $R$,
where $(s,s')\in R$ if and only if $s$ and $s'$ are states of the same
component, with $s'$ immediately following $s$, or if there is a
communication from $s$ to $s'$. For example, in the computation above,
$(s_m^0,s_m^1),\, (s_m^1,s_n^{3}),\, (s_n^{2},\, s_n^{3})\in R$. The
plain arrows between pairs of states denote relation $R$.

We call $R^=$ the reflexive closure of $R$, and $R^*$ its reflexive
and transitive closure.  For example, in the computation above,
$(s_m^0,s_m^1),\, (s_m^0,s_n^{3}),\, (s_m^1,s_n^{3}),\, (s_n^{2},\,
s_m^{12})\in R^*$.  We say that $s'$ causally depends on $s$ when
$(s,s')\in R^*$. Causal dependency has to be read as the partial order
relationship between states of a distributed computation, defined by
state transitions and communications~\cite{lamport78}. If
neither $(s,s')\in R^*$ nor $(s',s)\in R^*$, states $s$
and $s'$ are concurrent.

\end{definition}

\begin{definition} {(Models, $\leq,\; \leq_c$)}

 \noindent
A model ${\cal M}$  is a tuple $(DS, R_{1},
\ldots,R_{k}, \leq, \leq_c, V)$, where:
\begin{eqnarray*}
 ds \leq ds^{\prime} & \  {\rm iff} \ &   \forall s\in ds, \: \exists
s^{\prime}\in ds^{\prime}.\; (s, s^{\prime})\in R^* 
 \  {\rm and} \ 
\forall s^{\prime}\in ds^{\prime}, \: \exists s\in ds.\; (s, s^{\prime})\in R^*
\\
 ds \leq_c ds^{\prime} & \  {\rm iff} \ &   \forall s\in ds, \: \exists
s^{\prime}\in ds^{\prime}.\; (s, s^{\prime})\in R^=
 \  {\rm and} \ 
\forall s^{\prime}\in ds^{\prime}, \: \exists s\in ds.\; (s, s^{\prime})\in R^=
\end{eqnarray*}

\noindent
We also  ask that the valuation function
$V: DS\raw 2^{P}$ satisfies $V(ds)= \bigcap_{s\in ds} V(\{s\})$.

\end{definition}

\begin{definition} \label{DSTLsemantics} {(Semantics)}

\noindent
Let ${\cal M}$ be a model, and $ds^0$ the set
of its initial states. We define:

\

$\begin{array}{lcl}
{\cal M} \models_T F 
      &  \quad {\rm iff}  \quad  &
	\forall  ds.\: ds\models F \\
{\cal M} \models_T F \causes  F^\prime  &   {\rm iff} &
         \forall  ds.\: ds \models F
        \mbox { implies } 
         \exists \: ds^{\prime} \geq ds.\;
          ds^{\prime} \models F^\prime\\
{\cal M} \models_T F \because  F^\prime  &   {\rm iff} &
         \forall  ds.\: ds \models F
        \mbox { implies } 
         \exists \: ds^{\prime} \leq ds.\;
          ds^{\prime} \models F^\prime\\
{\cal M} \models_T F \causesc  F^\prime  &   {\rm iff} &
         \forall  ds.\: ds \models F
        \mbox { implies } 
         \exists \: ds^{\prime} \geq_c ds.\;
          ds^{\prime} \models F^\prime\\
{\cal M} \models_T F \becausec  F^\prime  &   {\rm iff} &
         \forall  ds.\: ds \models F
        \mbox { implies } 
         \exists \: ds^{\prime} \leq_c ds.\;
          ds^{\prime} \models F^\prime\\
{\cal M} \models_T F \unless  F^\prime  &   {\rm iff} &
         \forall  ds.\: ds \models F
        \mbox { implies } \exists  ds^{\prime} \geq_c ds. \\
	&& \quad\quad ( ds^{\prime} \not\supseteq ds \: 
	\wedge\:    ds^{\prime} \models F) \ \vee \ ds^{\prime} \models
	F^\prime\\
{\cal M} \models_T  \init  F  &   {\rm iff} &
          ds^0 \models F
 \end{array}
$

\

\noindent
where $\models$ is the DSL satisfiability relation. 

The next section discusses this definition using some examples. In
particular, the side condition $ ds^{\prime}\not\supseteq ds $ for
$\tunless$ is illustrated in Example~\ref{ex:subset}.

\end{definition}

\subsection{Examples}

To exemplify the definition of the DSTL semantics, we choose some
formulae and discuss whether they are satisfied or not by a model
${\cal M}$ (a computation of a system made up of two components, $m$
and $n$). In the examples we can only present the initial fragments,
the discussion on satisfiability is done with respect to the given
fragment. From now on, we label the states with the predicates holding
in them instead of a name. 

We recall that, according to the definition in Section~\ref{DSL}, a
distributed state is any set of states. This means that when we have
to check a condition like $\forall ds \ldots \exists ds' \ldots$, we
need to consider all possible sets of states as $ds$. This may lead to
counter-intuitive choices, like picture (c) of
Table~\ref{tabex} to reason on the first formula of
Example~\ref{ex:temp}. We consider these choices in the examples to
clarify the semantic details. However, the specifier can be safely
guided by the natural interpretation of the operators. Anyhow, our
definition of distributed state is exactly what was needed to overcome
the problems with the existing models, as discussed in the
introduction.

\begin{table} \label{tabex}

\mbox{}\hrule\mbox{}

\

\

 \xymatrix@R-=1pt@C-=16pt{ 
{(m)}& {p} \ar[r] & {q} \ar[rd]
 \ar@{.>}[rr] && {r} \ar@{.>}[rrr] && & *+[F]{u,z} \ar[r] & {z}
 \ar@{.>}[r] & {z} \ar@{.>}[r] & && {\;\:(a)}\\ 
{(n)}& {p,t} \ar[r] & {u}\ar[r] &
 {v} \ar@{.>}[rr]&& {p,u} \ar@{.>}[r] & *++[o][F]{w,t}
 \ar@{.>}[rrr]\ar[ru] & & & {t}\ar@{.>}[r] &&& } 

\

\


 \xymatrix@R-=1pt@C-=16pt{
 {(m)}& {p} \ar[r] & {q} \ar[rd] 
	\ar@{.>}[rr]  &&  
	*++[o][F-]{r}      \ar@{.>}[rrr] 
                 && &  
	*+[F]{u,z} \ar[r] & {z} \ar@{.>}[r] & {z} \ar@{.>}[r] & && (b)\\ 
 {(n)}& {p,t}  \ar[r] &  
	{u}\ar[r] &  
	{v}  \ar@{.>}[rr]&&
	*++[o][F-]{p,u} \ar@{.>}[r] &
	{w,t}     \ar@{.>}[rrr]\ar[ru] & & & {t}\ar@{.>}[r] & &&
}

\

\


 \xymatrix@R-=1pt@C-=16pt{
 {(m)}& {p} \ar[r] & {q} \ar[rd] 
	\ar@{.>}[rr]  &&  
	{r}      \ar@{.>}[rrr] 
                 && &  
	*+[F]{u,z} \ar[r] & 
	*++[o][F-]{z} \ar@{.>}[r] & *+[F]{z} \ar@{.>}[r] & && (c)\\ 
 {(n)}& {p,t}  \ar[r] &  
	{u}\ar[r] &  
	{v}  \ar@{.>}[rr]&&
	*++[o][F-]{p,u} \ar@{.>}[r] &
	{w,t}     \ar@{.>}[rrr]\ar[ru] & & & {t}\ar@{.>}[r] & &&
} 

\

\


 \xymatrix@R-=1pt@C-=16pt{ 
 {(m)}& *++[o][F-]{p} \ar[r] & {q} \ar[rd] 
	\ar@{.>}[rr]  &&  
	{r}      \ar@{.>}[rrr] 
                 && &  
	*+[F]{u,z} \ar[r] & {z} \ar@{.>}[r] & {z} \ar@{.>}[r] &&& {\;\:(d)}\\ 
 {(n)}& {p,t}  \ar[r] &  
	{u}\ar[r] &  
	*++[o][F-]{v}  \ar@{.>}[rr]&&
	{p,u} \ar@{.>}[r] &
	  *+[F]{w,t}     \ar@{.>}[rrr]\ar[ru] & & & {t}\ar@{.>}[r] & &&
} 

\

\

 \xymatrix@R-=1pt@C-=16pt{ 
 {(m)}& {p} \ar[r] & *++[o][F-]{q} \ar[rd] 
	\ar@{.>}[rr]  &&  
	{r}      \ar@{.>}[rrr] 
                 && &  
	{u,z} \ar[r] & {z} \ar@{.>}[r] &  *+[F]{\bigcirc\!\!\!\!\!z\;}
 \ar@{.>}[r] & && {\!\!\!(e)} \\  
 {(n)}& {p,t}  \ar[r] &  
	{u}\ar[r] &  
	*+[F]{v}  \ar@{.>}[rr]&&
	{p,u} \ar@{.>}[r] &
	 *++[o][F]{w,t}     \ar@{.>}[rrr]\ar[ru] & & & {t}\ar@{.>}[r] & &&
}

\

\

 \xymatrix@R-=1pt@C-=16pt{
 {(m)}& *++[o][F-]{p} \ar[r] & *+[F]{q} \ar[rd] 
	\ar@{.>}[rr]  &&  
	{r}      \ar@{.>}[rrr] 
                 && &  
	{u,z} \ar[r] & {z} \ar@{.>}[r] & {z} \ar@{.>}[r] &&& {\;\;\:(f)}\\ 
 {(n)}& {p,t}  \ar[r] &  
	{u}\ar[r] &  
	*+[F]{\bigcirc\!\!\!\!\!v\;}  \ar@{.>}[rr]&&
	{p,u} \ar@{.>}[r] &
	  {w,t}     \ar@{.>}[rrr]\ar[ru] & & & {t}\ar@{.>}[r] & &&
}

\

\

 \xymatrix@R-=1pt@C-=16pt{ 
 {(m)}& {p} \ar[r] & {q} \ar[rd] 
	\ar@{.>}[rr]  &&  
	{r}      \ar@{.>}[rrr] 
                 && &  
	{u,z} \ar[r] & {z} \ar@{.>}[r] &  {z}
 \ar@{.>}[r] &&& {\;\:(g)} \\  
 {(n)}& {p,t}  \ar[r] &  
	{u}\ar[r] &  
	{v}  \ar@{.>}[rr]&&
	*+[F]{p,u} \ar@{.>}[r] &
	 *++[o][F]{w,t}     \ar@{.>}[rrr]\ar[ru] & & & {t}\ar@{.>}[r] & &&
}

\

\

 \xymatrix@R-=1pt@C-=16pt{ 
 {(m)}& {p} \ar[r] & {q} \ar[rd] 
	\ar@{.>}[rr]  &&  
	{r}      \ar@{.>}[rrr] 
                 && &  
	{u,z} \ar[r] & {z} \ar@{.>}[r] &  {z}
 \ar@{.>}[r] &&& {\;\;(h)} \\  
 {(n)}& *+[F]{p,t}  \ar[r] &  
	*+[F]{u}\ar[r] &  
	{v}  \ar@{.>}[rr]&&
	{p,u} \ar@{.>}[r] &
	 *++[o][F]{w,t}     \ar@{.>}[rrr]\ar[ru] & & & {t}\ar@{.>}[r] & &&
}

\

\mbox{}\hrule\mbox{}

\

\caption{We provide various representations of a computation, to
outline the distributed states of interest for the examples.}
\end{table}

\

\begin{example} {\em (Invariants.)}  
We consider, as model ${\cal M}$, the computation in
Table~\ref{tabex}. We refer to picture (a), and call $s$ and
$s^{\prime}$ the states outlined by the circle and the rectangle,
respectively. We show that $w\raw t$, $\bar{\mn} (w\raw t)$, and $\mn
\top \raw \mn (w\raw t)$ are invariants of the computation,
while $\mn (w\raw t)$ is not invariant. 

\vspace{-0.4cm}

\begin{description}
\item[${\cal M} \models_T w\raw t$.]  This formula reads: in any
distributed state of the computation, $w\raw t$ holds. 

State $s$ is the only one satisfying $w$. Take $ds = \{s\}$, then $ds
\models w \wedge t$, and thus $ds \models w\raw t$. For any
distributed state $ds' \neq ds$ we have that $ ds'\not\models w$ (even
though $s\in ds'$), and thus $ds' \models w\raw t$.

\item[${\cal M} \models_T \bar{\mn} (w\raw t)$.]  This formula reads:
in any distributed state $w\raw t$ holds in all the states of $n$, or,
in short, $w\raw t$ holds in any state of $n$.

We have to show that for all $ds$, $ds \models \bar{\mn} (w\raw t)$,
that is for all $s_n \in ds\cap S_n$, $\{s_n\} \models w\raw t$.
This, in turn, holds since $\{s\} \models w\wedge t$, and for all $s_n
\neq s$, $\{s_n\} \not\models w$. By the way, this result follows by
{\small \bf Nec} from the previous one.

\item[${\cal M} \models_T \mn \top \; \raw \; \mn (w\raw t)$.]  This
formula reads: in any distributed state of the computation that
contains at least one state of $n$, there is a state of $n$ where
$w\raw t$ holds.
 
Any distributed state $ds$ satisfying the premise $\mn \top$
includes a state in $S_n$, and all states in  $S_n$ satisfy $w\raw
t$. So $ds\models \mn (w\raw t)$.  

\item[${\cal M} \not \models_T \mn (w\raw t)$.]  The formula reads: in
any distributed state of the computation, there is a state of $n$
where $w\raw t$ holds, and it is false in ${\cal M}$.

For ${\cal M} \models_T \mn (w\raw t)$ to be true, we would need that
for all $ds$, $ds \models \mn (w\raw t)$, which is true only if a
state $s_n \in ds\cap S_n$ exists, and satisfies $(w\raw t)$. However,
there are distributed states not including any state $s_n \in S_n$,
e.g. $\{s^{\prime}\}$. In the practice, formulae like $\mm F$ are used
only as subformulae of larger formulae, e.g. as premises and
conclusions of an implication.

\end{description}

\end{example}

\begin{example} \label{ex:temp}
{\em (Temporal operators.)}  
In the example, we refer to pictures (b)--(h) in
Table~\ref{tabex}. The distributed state $ds$ satisfying the premise
is the set of states outlined with a circle, and the distributed state
$ds'$ satisfying the consequence is the set of states outlined with a
rectangle.

\vspace{-0.4cm}

\begin{description}

\item[${\cal M} \models_T \mn u\causes \mm u$.] It is enough to
consider those distributed states that contain the last state of $n$
where $u$ holds. Pictures (b) and (c) show two relevant cases: in the
second case we need to consider a larger distributed state to evaluate
the consequence, just to satisfy the ``follows'' relation.

Picture (c) also shows that DSTL overcomes the problems discussed
at point 3 in the introduction: a distributed state satisfying the
consequence and following $ds$ exists.

\item[${\cal M}\models \mm p \wedge \mn v \causes \mm z \wedge \mn
t$.]  See picture (d): the distributed state satisfying $ \mm p \wedge
\mn v$ is followed by a distributed state satisfying $\mm z \wedge \mn t$.

\item[${\cal M}\models \mm q\causes \mn v$.]  See picture (e): the
distributed state satisfying the premise includes states which are
irrelevant with respect to property $ \mm q$, for them we only need to
check that the ``follows'' relation is satisfied.  The state
satisfying $z$ belongs both to $ds$ and $ds^{\prime}$.

\item[${\cal M}\models \mm p \wedge \mn v \causesc \mm q$.]  See
picture (f): the state where $q$ holds immediately follows the one
satisfying $p$. Then any state equal or immediately following the one
satisfying $v$ is fine to build the distributed state satisfying the
consequence, and the ``closely'' relation.

\item[${\cal M}\models \mn w \because \mn p \wedge \mn u $.]  Here it
is enough to consider those distributed states that contain the first
state of $n$ where $w$ holds. Then, in the example model, we show two
distributed states that satisfy the consequence: see pictures (g) and
(h).

\item[${\cal M}\models \mn w \because \mn (p \wedge u)$.] See picture
(g).  Note that we need a singleton state satisfying both $p$ and
$q$. Hence, in this case, the distributed state $ds^{\prime}$ in
picture (h) does not satisfy the consequence. 
\end{description}

\end{example}

\begin{table}[t] \label{tabunl}

\mbox{}\hrule\mbox{}

\


 \xymatrix@R-=1pt@C-=16pt{ 
 {(m)}& {p} \ar[r] & {q,z} \ar[rd] 
	\ar@{.>}[rr]  &&  
	{r,z}      \ar@{.>}[rrr] 
                 && &  
	{u,z} \ar[r] & {z} \ar@{.>}[r] &  {z}
 \ar@{.>}[r] & &&& (i)\\  
 {(n)}& *++[o][F]{p,t}  \ar[r] &  
	*++[o][F]{u,p}\ar[r] &  
	*++[o][F]{v,p}  \ar@{.>}[rr]&&
	*++[o][F]{p,u} \ar@{.>}[r] &
	*+[F]{w,t}     \ar@{.>}[rrr]\ar[ru] & & & {t}\ar@{.>}[r] &  &&&
}

\

\


 \xymatrix@R-=20pt@C-=16pt{ 
 {(m)}& *++[o][F]{p} \ar[r] \ar@{.}[d] \ar@{.}[rd] \ar@{.}[rrd]
 \ar@{.}[rrrrd] & *+[F]{q,z}\ar@{.}[rrrrd] \ar[rd] 
	\ar@{.>}[rr]  &&  
	{r,z}      \ar@{.>}[rrr] 
                 && &  
	{u,z} \ar[r] & {z} \ar@{.>}[r] &  {z}
 \ar@{.>}[r] & &&& (l) \\  
 {(n)}& *++[o][F]{p,t}  \ar[r] &  
	*++[o][F]{u,p}\ar[r] &  
	*++[o][F]{v,p}  \ar@{.>}[rr]&&
	*++[o][F]{p,u} \ar@{.>}[r] &
	*+[F]{w,t}     \ar@{.>}[rrr]\ar[ru] & & & {t}\ar@{.>}[r] &  &&&
}

\

\mbox{}\hrule\mbox{}

\

\caption{We provide a pair of representations of a computation, to
outline the distributed states of interest for example~\ref{ex:unl}.}
\end{table}

\begin{example} \label{ex:unl} ($\tunless$ formulae.)   
We consider, as a model, the computation in
Table~\ref{tabunl}. 

\begin{description}

\item[${\cal M}\models \mn p \unless \mn \, t$.]  See picture (i): we
take singleton sets for $ds$ and $ds'$, and outline with a sequence of
circles the sequence of distributed states satisfying the formula
premise, and use a rectangle to outline the distributed state
satisfying the formula consequence.

\item[${\cal M}\models p \unless q \vee t$.] The sequence of
distributed states in picture (i) provides a first demonstration. We
also consider, in picture (l), the distributed states in the sequence
to be pairs of states: each distributed state is made of the two
states related by a dotted line, circles outline the states satisfying
the formula premise, rectangles the  states satisfying the
formula consequence.  For instance, the initial state is the first
distributed state we consider, followed by the set $\{$first state of
$m$, second state of $n\}$, and so on. 

\end{description}

\end{example}

\vspace{0.2cm}

\begin{example} ($ ds^{\prime}\not\supseteq ds $ in the definition of
the semantics of $\tunless$.)
\label{ex:subset}
Assume we did not require condition $ ds^{\prime}\not\supseteq ds $ in
the definition of the semantics of $\tunless$, then the following
computation would have been a model for $\mn p \unless \mn q$, in
discrepancy with the intended meaning for \tunless. We consider the
sequence $ds,\, ds',\, ds'',\, ds''',\ldots$ of distributed states,
where $ds$ contains the first state of component $n$, $ds'$ contains
the first two states of component $n$, $ds''$ contains the first
three, and so on: all these distributed states satisfy $\mn p$.

\vspace{0.2cm}

 \xymatrix@R-=2pt@C-=16pt{ 
 {(m)}& {p} \ar[r]  & {q,z} \ar[rd] 
	\ar@{.>}[rr]  &&  
	{r,z}      \ar@{.>}[rrr] 
                 && &  
	{u,z} \ar[r] & {z} \ar@{.>}[r] &  {z}
 \ar@{.>}[r] & \\  
 {(n)}& {p,t}  \ar[r] 
	\ar@{.}@(dl,dr)[]|{ds}
	\ar@{.}@/_3pc/[r]|{ds'}
	\ar@{.}@/_2pc/[rr]|{ds''} 
	\ar@{.}@/_1pc/[rrrr]|{ds'''}
	&  
	{u}\ar[r] 
	&  
	{v}  \ar@{.>}[rr]
	&&
	{r} \ar@{.>}[r] &
	{t}     \ar@{.>}[rrr]\ar[ru] & & & {t}\ar@{.>}[r] & 
}

\end{example}

\begin{example} ($\tstable$.) The following computation is a model
 for$\stable p$.

 \xymatrix@R-=1pt@C-=16pt{ {(m)}& {p} \ar[r] & {p} \ar[rd]
 \ar@{.>}[rr] && {r,z} \ar@{.>}[rrr] && & {u,z} \ar[r] & {z}
 \ar@{.>}[r] & {z} \ar@{.>}[r] & \\ {(n)}& {p,t} \ar[r] & {u,p}\ar[r]
 & {v,p} \ar@{.>}[rr]&& {p,u} \ar@{.>}[r] & {p,w,t}
 \ar@{.>}[rrr]\ar[ru] & & & {p,t}\ar@{.>}[r] & }

\vspace{0.2cm}

\noindent
Notice that, unlike in Unity, $p$ is not an invariant of the
computation, even though $\init p$ and $\stable p$ hold. In the next
section, we provide the correct derivation rule (SE) that can be used
in DSTL.
\end{example}

\subsection{Axioms and Rules}
\label{DSTLax}

We present the most useful axioms and rules of the logic.  Among
them, temporal operators introduction, strengthening of premises and
weakening of consequences, transitivity.

\subsubsection{Necessitation.}
First, we observe that the definition for ${\cal M} \models_T F$
entails that a necessitation rule holds (we use $\vdash_{T}$ for the
sake of comprehension).

\begin{center}
{\small
\prooftree
\vdash \ F
\justifies
\vdash_{T} \ F
\using{\bf Nec}
\endprooftree 
}
\end{center}

\vspace{0.2cm}

\subsubsection{Operators introduction and elimination.}

Rules and axioms {\small \bf LcI}, {\small \bf BcI}, {\small \bf LI},
{\small \bf BI}, {\small \bf UI}, {\small \bf InI}, {\small \bf SI}
introduce $\tcausesc$, $\tbecausec$, $\tcauses$, $\tbecause$, $\tunless$, 
$\tinit$, $\tstable$ respectively. Rule {\small \bf SE} eliminates
$\tstable$.

\begin{center}
{\small
{\bf LcI:} \  $F\causesc F$ \hfill
{\bf BcI:} \  $F\becausec F$ \hfill
\prooftree
F\causesc G
\justifies
F\causes G
\using{\bf LI}
\endprooftree  
}
 \hfill{\small
\prooftree
F\becausec G
\justifies
F\because G
\using{\bf BI}
\endprooftree
}
\end{center}

\

\

\begin{center}
{\small
{\bf UI:} \  $F\unless F$
 \hfill
\prooftree
F 
\justifies
\init F 
\using{\bf InI}
\endprooftree
 \hfill 
\prooftree
F 
\justifies
\stable F 
\using{\bf SI}
\endprooftree
 \hfill 
\prooftree
\init \mm F \quad \stable  \mm F 
\justifies
\bar{\mm}F 
\using{\bf SE}
\endprooftree
}
\end{center}

\subsubsection{Transitivity.}

{\small \bf LTR} and {\small \bf BTR} are the rules for $\tcauses$ and
$\tbecause$ transitivity.

\begin{center}
{\small
\prooftree
F \causes   F^{\prime}  \quad
 F^{\prime}     \!\causes G
\justifies
F \; \causes G
\using{\bf LTR}
\endprooftree
} \quad \quad{\small
\prooftree
F \because   F^{\prime}  \quad
 F^{\prime}     \!\because G
\justifies
F \; \because G
\using{\bf BTR}
\endprooftree
}
\end{center}

\

\noindent
No transitivity rule holds for $\tcausesc$ and $\tbecausec$.  In the
case of $\tunless$, there is a weaker result (a weak form of the rule
called {\em cancellation} in~\cite{chmi88}):

\begin{center}
{\small
\noindent
\prooftree
\mm F \; \unless \; \mm F^{\prime}  \quad
\mm F^{\prime} \; \unless \; \mm G  \quad
\justifies
\mm F \vee \mm F^{\prime} \;  \unless  \;  \mm G
\using{\bf UC}
\endprooftree 
}
\end{center}

\subsubsection{Premises and consequences strengthening and weakening.}

{\small \bf *SW} permits the strengthening of the premise, and the
weakening of the consequences, and {\small \bf *PD} and {\small \bf
*CC} stay for premise disjunction and consequence conjunction,
respectively. Actual rules {\small \bf LSW}, {\small \bf LPD} and
{\small \bf LCC} are obtained by substituting \gop with $\tcauses$.
Similarly, {\small \bf BSW}, {\small \bf BPD}, and {\small \bf BCC}
are obtained by substituting \gop with $\tbecause$; {\small \bf LcSW},
{\small \bf LcPD}, and {\small \bf LcCC} are obtained by substituting
\gop with $\tcausesc$; {\small \bf BcSW}, {\small \bf BcPD}, and {\small
\bf BcCC} are obtained by substituting \gop with $\tbecausec$.

\begin{center}

{\small
\prooftree
G \raw F \quad
F \; \gop \; F^{\prime}  \quad
F^{\prime}  \raw G^{\prime}
\justifies
G \; \gop G^{\prime}
\using{\bf *SW}
\endprooftree 
} \quad \quad 
{\small
\prooftree
F \; \gop \; G  \quad
F^{\prime} \; \gop \; G  
\justifies
F\vee F^{\prime} \; \gop G
\using{\bf *PD}
\endprooftree 
 \quad \quad
\prooftree
G \; \gop \; F  \quad
G \; \gop \; F^{\prime}  
\justifies
G \; \gop F \wedge F^{\prime}
\using{\bf *CC}
\endprooftree 
}
\end{center}

\noindent
In the case of $\tunless$ and $\tinit$:

\begin{center}
{\small
\noindent
\prooftree
F \; \unless \; F^{\prime}  \quad
F^{\prime}  \raw G
\justifies
F \; \unless \; G
\using{\bf UCW}
\endprooftree 
 \quad \quad
\prooftree
F \; \unless \; F^{\prime}  \quad
 G \; \unless \; G^{\prime}  \quad
\justifies
F \vee G \unless F^{\prime} \vee
G^{\prime}
\using{\bf UD}
\endprooftree 
 \quad \quad
\prooftree
\init F  \quad F \raw G
\justifies
\init G 
\using{\bf IW}
\endprooftree 
}
\end{center}

\subsubsection{Notification.}  
Some future remote assertions can be made on the bases of a message
received.

\begin{center}
{\small
\noindent
\prooftree
F \because G \quad G \causes \mm G^{\prime} \quad \stable \mm G^{\prime}
\justifies
F \wedge \mm \top \causes \mm G^{\prime}
\using{\bf Notif}
\endprooftree 
}
\end{center}

\

\noindent
Explicit reference to the name $\mm$ of the component where the remote
effect $G^{\prime}$ takes place, and the extra premise $\mm \top$ are
needed to guarantee that the state satisfying the consequence follows
the state satisfying the premise, even in the absence of a
communication towards $\mm$.

\

\noindent
To help the intuition, we consider an instance of the rule: 

\begin{center}
{\small
\noindent
\prooftree
\mn p \because \mm q \quad \mm q \causes \mm r \quad \stable \mm r
\justifies
\mn p \wedge \mm \top \causes \mm r
\endprooftree 
}
\end{center}

\

\noindent
condition $p$ can be established in $\mn$ only if previously $q$ has
held in $\mm$. The second and the third premises guarantee that if $q$
holds somewhere in $\mm$, then eventually $r$ will hold, and it will
continue holding forever. Thus, for any $ds$ satisfying $\mn p \wedge
\mm \top$ we can find a state $s_m$ of $S_{m}$, such that $\{s_m\}\geq
ds$ and $\{s_m\}\models \mm r$. Conversely, in the absence of
communications from $\mn$ to $\mm$, if we take a state $s_n$ of
$S_{n}$ such that $\{s_n\}\models \mn p$, we cannot find any
distributed state following $\{s_n\}$ and including a state of $
S_{m}$, as needed to satisfy $\mm r$.

\subsubsection{Confluence.} 
The converse of DSL axiom {\small \bf D2} holds, under appropriate
stability conditions:

\begin{center}
{\small
\noindent
\prooftree
\stable  \mm F \quad \stable  \mm F^{\prime} 
\justifies
\mm F\wedge \mm F^{\prime}  \ \raw \ \mm (F\wedge F^{\prime})
\using{\bf Conf}
\endprooftree 
}
\end{center}

\subsubsection{Properties of the initial  state.} The following rules are a
consequence of the fact that the initial distributed state $ds^0$
contains exactly one state for each component.

\

{\small
{\bf I1:} \  $\init \mm \top$ \quad \quad \quad  \quad \quad  
\prooftree
\init \mm F
\justifies
\init \bar{\mm} F 
\using{\bf I2}
\endprooftree \quad \quad \quad  \quad \quad  
\prooftree
\init \bar{\mm} F 
\justifies
\init \mm F
\using{\bf I3}
\endprooftree 
}

\vspace{0.3cm}

\begin{example} (SE)
The following computation satisfies $\init \mm p$  and $\stable \mm p$.

\vspace{0.2cm}

\xymatrix@R-=1pt@C-=16pt{ 
 {(m)}& {p} \ar[r] & {p} \ar[rd] 
	\ar@{.>}[rr]  &&  
	{p,r,z}      \ar@{.>}[rrr] 
                 && &  
	{p,u,z} \ar[r] & {p,z} \ar@{.>}[r] &  {p,z}
 \ar@{.>}[r] & \\  
 {(n)}& {p,t}  \ar[r] &  
	{u}\ar[r] &  
	{v,p}  \ar@{.>}[rr]&&
	{u} \ar@{.>}[r] &
	{w,t}     \ar@{.>}[rrr]\ar[ru] & & & {p,t}\ar@{.>}[r] & 
}

\vspace{0.3cm}

\noindent 
 Hence, applying rule SE, we obtain that the
computation satisfies $\bar{\mm}p$, i.e. that $p$ is invariantly true
in component $\mm$.

\end{example}

\noindent
It is also interesting to discuss why the cancellation rule 

\begin{center}
{\small
\noindent
\prooftree
F \; \unless \; F^{\prime}  \quad
F^{\prime} \; \unless \; G  \quad
\justifies
F \vee F^{\prime} \;  \unless  \;  G
\endprooftree 
}
\end{center}

\noindent
does not hold in general. We consider, as rule premises, $\mm
p\unless \mm p \wedge \mn q$ and $\mm p \wedge \mn q \unless \mm r
\wedge \mn s$. The following computation is a model of the premises,
but not of the consequence $\mm p \vee (\mm p \wedge \mn q) \unless
\mm r \wedge \mn s$.

\xymatrix@R-=1pt@C-=16pt{ 
 {(m)}& {p} \ar@{.>}[rr] && {p} \ar[rd] 
	\ar[r]  &
	{r}      \ar@{.>}[rrr] 
                 && &  
	{u,z} \ar[r] & {z} \ar@{.>}[r] &  {z}
 \ar@{.>}[r] & \\  
 {(n)}& {p,t}  \ar[r] &  
	{u}\ar@{.>}[rr] &  &
	{q}  \ar[r]&
	{s} \ar@{.>}[r] &
	{w,t}     \ar@{.>}[rrr]\ar[ru] & & & {p,t}\ar@{.>}[r] & 
}

\

\

\noindent
{\bf Theorems.}  We introduce two rules we need in
the case study of Section~\ref{example2}. They are derived by the
rules above, as shown in the appendix.

\begin{center}
{\small
\noindent
\prooftree
F  \causes  G\vee G^{\prime}  \quad
G \causes F^{\prime} 
\justifies
F \causes F^{\prime} \vee G^{\prime} 
\using{\mbox{\small Cor1}}
\endprooftree 
 \quad \quad \quad
\prooftree
F  \causes  G\wedge G^{\prime}  \quad
G \causes F^{\prime} 
\justifies
F \causes F^{\prime} \wedge G^{\prime} 
\using{\mbox{\small Cor2}}
\endprooftree 
}
\end{center}

\

\

\noindent
{\bf Correctness and completeness.}  The soundness of the DSTL proof
system can be immediately proved applying Def.~\ref{DSTLsemantics}. In
the appendix we provide the proof of the most complex rules, namely
{\small \bf Notif} and {\small \bf Conf}.

Unfortunately, the proof system is not complete. Let us consider a
system satisfying $\bar{\mm}_i p$, for all $i$. The system also
satisfies $p$, as a consequence of the property $V(ds)=
\bigcap_{s\in ds} V(\{s\})$, but we cannot find a general rule to
derive it. Indeed, the rule

\vspace{0.2cm}

\begin{center}
{\small
\prooftree
\bar{\mm}_1 F \quad \bar{\mm}_2 F  \quad \ldots  \quad \bar{\mm}_k F 
\justifies
F
\endprooftree 
}
\end{center}

\vspace{0.2cm}

\noindent 
is not correct. It holds for $F=p$, or $F=p\wedge q$, but not, for
instance, for $F=p\vee q$. In fact, consider a very simple system
composed of a unique component $m$, with states $s_0,s_1, s_2\ldots$,
and $p\in V(s_0)$, $q\in V(s_1)$, $q\in V(s_2)$, $\ldots$. All
distributed states satisfy $\bar{\mm} \, (p\vee q)$, while the
distributed states including $s_0$ do not satisfy $p\vee q$. Take
$ds=\{s_0,s_1\}$, we have that $ds \models p\vee q$ iff $ds \models p$
or $ds \models q$, iff $p\in V(ds)$ or $q\in V(ds)$, iff $p\in
V(s_0)\cap V(s_1)$ or $q\in V(s_0)\cap V(s_1)$. Hence, since
$V(s_0)\cap V(s_1)=\emptyset$, we have that $ds \not \models p\vee q$.

Thus, a complete proof system, if any, would likely be unmanageable,
and we do not pursue the issue further. On the other side, the
consequence of relaxing the constraint on the valuation function,
would be as unpractical as explicitly specifying the truth value of
all predicated on all distributed states.


\section{An  Example: Private Keys}
\label{example}

Consider the system $\{b,t,u\}$, where $b$ is a component that
broadcasts the encrypted version of a message to all the other
components in the system, i.e. $t$ (trusted) and $u$ (untrusted). We
assume that these components try to decrypt the message. We represent
with predicate $p$ the fact that the message is readable, and with
predicate $dep$ the fact that a decryption has been
attempted. However, the decryption yields $p$ if and only if the key
is held. Predicate $key$ represents the property of holding the key.

\subsection{Reasoning on distributed states: DSL} 
\label{ex1}

The properties of the distributed states of the system are described
by the following DSL formulae:

\vspace{-0.6cm}

\begin{eqnarray}
&&(\sim \mbb \top) \raw ((key \wedge dep) \lraw p) \label{k1}\\
&& \bar{\mtt} \, key\label{k2}\\
&& \bar{\muu} \sim key\label{k3}
\end{eqnarray}

\noindent
Formula~(\ref{k1}) tells that in all components, with the exception of
$b$, $p$ and $dep$ are equivalent only if the key is held. Indeed, if
(\ref{k1}) holds, as required, in all $ds\in DS$, it holds in particular
in all $ds$ which are singleton sets. So, it holds for all $\{s_t\}$
and $\{s_u\}$. Since in these states the premise of (\ref{k1}) is
satisfied, so it is the conclusion, i.e. in all states of $t$ and $u$:
$(key \wedge dep) \lraw p$. We derive the property for $t$:

\

\begin{center}

{\small
\prooftree
\prooftree
(\sim \mbb \top) \raw ((key \wedge dep) \lraw p)
\justifies
\bar{\mtt} (\sim \mbb \top) \raw \bar{\mtt} ((key \wedge dep) \lraw p)
\using{\mbox{\small Nec, K}}
\endprooftree
\quad
\prooftree
\bar{\mtt} \, \bar{\mbb} \bot
\justifies
\bar{\mtt} (\sim \mbb \top)
\endprooftree
\justifies
\bar{\mtt} ((key \wedge dep) \lraw p)
\using{\mbox{\small MP}}
\endprooftree
}

\end{center}

\

\noindent
Component $t$ holds the key (\ref{k2}), while component $u$ does not
(\ref{k3}).  We derive that $t$ is able to correctly decrypt the
message.  We pick one of the implications, i.e. $(key \wedge dep) \raw
p$ and prove that $\bar{\mtt} \, (dep \raw p)$ (the top leftmost
formula is a tautology of the propositional calculus):

\

\begin{center}

{\small
\prooftree
\prooftree
\prooftree
((key \wedge dep) \raw p ) \raw (key\raw(dep\raw p))
\quad \quad \bar{\mtt} ((key \wedge dep) \lraw p)
\justifies
 \bar{\mtt} \, (key\raw(dep\raw p))
\using{\mbox{\small Nec, K, MP}}
\endprooftree
\justifies 
\bar{\mtt} \, key\raw \bar{\mtt} \,(dep\raw p)
\using{\mbox{\small K}}
\endprooftree
\quad
\bar{\mtt} \, key
\justifies 
 \bar{\mtt}\, (dep\raw p)
\using{\mbox{\small MP}}
\endprooftree
}

\end{center}

\

\noindent
We now consider component $u$ and prove that $\bar{\muu} \sim p$
holds, i.e. that the untrusted component is not able to read the
message. We consider the implication $p\raw (key \wedge dep)$ (the top
leftmost formula is a tautology of the propositional calculus):

\

\begin{center}

{\small
\prooftree
\prooftree
(p\raw (key \wedge dep) ) \raw (\sim key \raw \sim p)
\quad\quad \bar{\muu}\, ((key \wedge dep) \lraw p)
\justifies
 \bar{\muu}\, (\sim key \raw \sim p)
\using{\mbox{\small Nec, K, MP}}
\endprooftree
\quad \bar{\muu}\, \sim key
\justifies
 \bar{\muu}\,  \sim p
\using{\mbox{\small K,MP}}
\endprooftree
}

\end{center}

\subsection{Reasoning on distributed computations: DSTL}

 We now add some constraints on the temporal behaviour of the private
keys system: as soon as the message is readable in $b$, $b$ broadcasts
its encrypted version (\ref{L1}); $t$ and $u$ try to decrypt the
message (\ref{L2}, \ref{L3}). 
\begin{eqnarray}
&&\mbb \,p  \; \causes  \; \mtt \, ep \wedge \muu \,ep  \mbox{\hspace{6cm}}
		 \label{L1}\\
&&  \mtt\, ep  \; \causes  \; \mtt \,dep \label{L2}\\
&& \muu \, ep  \; \causes \; \muu \, dep \label{L3}
\end{eqnarray}

\noindent
To prove that $u$ will not correctly decrypt the message, we need to
prove that $ \bar{\muu} \sim p$. This is immediately obtained by
applying {\small \bf Nec} to the corresponding DSL formula derived in
Section~\ref{ex1}. We prove that $\mbb \, p \: \causes \: \mtt \,
p$. We exploit the conclusion of Section~\ref{ex1}: $\bar{\mtt} \,
(dep\raw p)$

\

{\small
\prooftree
\prooftree
\prooftree
\mbb \,p  \; \causes  \; \mtt \, ep \wedge \muu \,ep   \quad (\mtt \, ep \wedge \muu\,
ep) \raw  \mtt\,  ep
\justifies
\mbb\, p  \: \causes \: \mtt \, ep
\using{\mbox{\small LSW}}
\endprooftree 
  \mtt\, ep  \; \causes  \; \mtt \,dep
\justifies
\mbb  \, p  \: \causes \: \mtt  \,dep 
\using{\mbox{\small LTR}}
\endprooftree 
\quad\quad\quad \prooftree
\bar{\mtt} \,  (dep\raw p)
\justifies
\mtt  \,dep\raw \mtt \, p
\using{\mbox{\small D3}}
\endprooftree 
\justifies
\mbb  \,  p \: \causes \: \mtt  \, p
\using{\mbox{\small LSW}}
\endprooftree 
}

\


\section{An  Example: Leader Election}
\label{example2}

The leader election problem is a typical example of distributed
consensus. It is well known that in an asynchronous setting, no
algorithm can guarantee that a distributed consensus is reached (see,
for instance~\cite{CatSynch}). The solution we discuss here leads to
the election of a leader, or to the agreement that no leader has been
chosen, in this case a new election round can take place.

Initially all the $k$ participants are eligible. They toss a coin:
those who get head are no longer eligible and acknowledge the other
participants; those who get tail toss the coin again. The election
round ends when either only one participant is still eligible and
becomes the leader, or nobody is eligible.

\

\noindent
Predicate $e_i$ says that participant $i$ is still eligible: initially
all participants agree that they are all eligible; each participant
falsify his $e_i$ when acknowledged that participant $i$ got a head.

In the following we list the local properties satisfied by each
participant and derive the global property of the proposed solution:
eventually  all participants agree that either nobody is eligible,
i.e. $\sim e_i$ holds for all $i$ and for all participants, or only
one participant is still eligible, i.e. there exists a $j$ such that
for all participants $e_j$ holds while $e_k$ is false for all $k\neq
j$. Formally:

\vspace{-0.6cm}

\begin{eqnarray*}
\bigwedge_i \mii \top & \causes & \quad 
\bigwedge_i  \mii   \bigwedge_j \sim e_j 
\quad \vee \quad
\bigvee_j \bigwedge_i  \mii  ( e_j \wedge  \bigwedge_{k\neq j} \sim e_k)
\end{eqnarray*}

\noindent
In the case of two participants:

\vspace{-0.6cm}

\begin{eqnarray*}
\mI \top \wedge \mII \top & \causes & 
\mI (\sim e_1 \wedge \sim e_2)  \wedge
 \mII (\sim e_1 \wedge \sim e_2) \quad \quad\quad \mbox{(no leader elected)}\\
&&\vee \quad   \mI ( e_1 \wedge  \sim e_2)  \wedge
 \mII ( e_1 \wedge  \sim e_2) \quad \quad \quad \mbox{ ($e_1$ elected)}\\
&&\vee \quad   \mI (e_2 \wedge  \sim e_1 )  \wedge
 \mII (e_2 \wedge  \sim e_1 )\quad \quad \quad \mbox{ ($e_2$ elected)}
\end{eqnarray*}

\noindent
The local properties follow.

\begin{enumerate}

\item Fairness of the toss up: nobody can spin the coin infinite times
and nether get a head. So, either a participant eventually stops
spinning the coin or he gets a head.  For all $i$:

\vspace{-0.6cm}

\begin{eqnarray*}
&& \mii \top \causes \mii (stop \vee h)
\end{eqnarray*}

\item Participant $i$ stops if and only if the other participants
are no longer eligible:

\vspace{-0.6cm}  

\begin{eqnarray*}
&& \bar{\mm}_{\bf \sf i} (stop \lraw \bigwedge_{j\neq i} \sim e_j )
\end{eqnarray*}

\item When  participant $i$ gets a head, he sends an ack to all 
participants, who declare $i$ non eligible.

\vspace{-0.6cm}

\begin{eqnarray*}
&& \mii h \causes \bigwedge_j \mij \sim e_i
\end{eqnarray*}

\item A participant can be  declared non eligible only if he got a
head:

\vspace{-0.6cm}

\begin{eqnarray*}
&& \mii  \sim e_j \because \mij h
\end{eqnarray*}

\item Initially all participants are eligible. 

\vspace{-0.6cm}

\begin{eqnarray*}
&& \init \bigwedge_i \mii  (\bigwedge_j e_j \wedge \sim h)
\end{eqnarray*}

\item Non eligibility is stable:

\vspace{-0.6cm}

\begin{eqnarray*}
&& \stable \mii \sim e_j 
\end{eqnarray*}

\end{enumerate}

\noindent
We prove that the global property holds in the case of two
participants.  The proofs for the other cases are similar. 
In the first step of the proof, we exploit properties 1 and 2:

\

\begin{center}
{\small
\noindent
\prooftree
\prooftree
\mI \top \causes \mI (stop \vee h)
\justifies
\mI \top \causes \mI stop \vee \mI h
\using {\mbox{\small D7}}
\endprooftree 
\  \prooftree
\bar{\mm}_{\bf \sf 1}  (stop \lraw \sim e_2)      
\justifies
\mI stop \ \lraw \ \mI \sim e_2
\using {\mbox{\small D3}}
\endprooftree 
\justifies \mI \top \causes \mI h \vee \mI \sim e_2
\using {\mbox{\small LSW}}
\endprooftree  
}
\end{center}

\

\noindent
The same holds for $\mII$. We apply {\small \bf LSW} and {\small \bf
LCC} and obtain:

\vspace{-0.6cm}

\begin{eqnarray}
\mI \top \wedge \mII \top & \causes & \quad \mI h  \wedge \mII h \label{hh} \\
&&  \vee  \quad  \mI h  \wedge  \mII \sim e_1 \label{he} \\
&&  \vee  \quad  \mI \sim e_2 \wedge  \mII h \label{eh} \\
&&  \vee  \quad  \mI \sim e_2 \wedge \mII \sim e_1 \label{ee} 
\end{eqnarray}

\noindent
In the remaining part of the section we prove that: 

\vspace{-0.6cm}

\begin{eqnarray*}
(\ref{hh})  & \causes & 
	\mI (\sim e_1  \wedge \sim e_2 ) \wedge   
	\mII (\sim e_1 \wedge \sim e_2) 
	\quad \quad \quad \quad \mbox{(no leader elected)} \\
(\ref{he}) & \causes & 
	\mI (\sim e_1 \wedge e_2) \wedge \mII (\sim e_1 \wedge e_2) \\
	&&  \vee \ \mI (\sim e_1 \wedge \sim e_2) 
	\wedge \mII (\sim e_1 \wedge \sim e_2)
	\quad \quad \quad \mbox{($e_2$ elected or no leader)} \\
(\ref{eh}) & \causes &
	\mI (e_1 \wedge \sim e_2) \wedge \mII (e_1 \wedge \sim e_2) \\
	&&  \vee \ \mI (\sim e_1 \wedge \sim e_2) 
	\wedge \mII (\sim e_1 \wedge \sim e_2) 
	\quad \quad \quad \mbox{($e_1$ elected or no leader)} \\
(\ref{ee}) & \causes & 
	\mI (\sim e_1 \wedge \sim e_2)
 	\wedge \mII (\sim e_1\wedge \sim e_2)
	\quad \quad \quad \quad  \mbox{(no leader elected)} 
\end{eqnarray*}

\noindent
So, we can apply {\small \bf Cor1} and conclude.

\

\noindent
{\bf Proof of} $(\ref{hh}) \causes$ no leader elected 

\

\noindent 
We exploit hypothesis 3:  

\vspace{-0.6cm}

\begin{eqnarray*}
\mI h  \wedge \mII h & \causes & 
\mI \sim e_1  \wedge  \mII \sim e_1 \wedge 
\mI \sim e_2  \wedge  \mII \sim e_2 
\end{eqnarray*}

\noindent
We apply {\small \bf Conf}:

\begin{center}
{\small
\noindent
\prooftree
   \stable \mI \sim e_1     \quad     \stable \mI \sim e_2
   \justifies 
   \mI \sim e_1 \wedge \mI \sim e_2 \ \raw \  \mI (\sim e_1 \wedge \sim e_2) 
\endprooftree  
}
\end{center}

\noindent
A similar implication holds for $\mII$, hence:

\vspace{-0.6cm}

\begin{eqnarray*}
\mI h  \wedge \mII h & \causes & 
\mI (\sim e_1  \wedge \sim e_2)   \wedge  
\mII (\sim e_1 \wedge  \sim e_2)
\end{eqnarray*}

\

\noindent
{\bf Proof of} $(\ref{he}) \causes e_2$ elected or no leader elected
(the case for (\ref{eh}) is symmetric).

\

\noindent
We exploit again hypothesis 3 and obtain, using
{\small \bf Cor2}, that:

\vspace{-0.6cm}

\begin{eqnarray*}
\mI h  \wedge  \mII \sim e_1  & \causes & \mI \sim e_1  \wedge  \mII \sim e_1 
\end{eqnarray*}

\noindent
Now, since we don't know anything on the truth of $e_2$, we need to
consider all the possibilities:

\vspace{-0.6cm}

\begin{eqnarray}
\mI \sim e_1  \wedge  \mII \sim e_1 &\LRaw & 
	\quad \mI (\sim e_1 \wedge e_2) \wedge \mII (\sim e_1 \wedge e_2) 
		\label{cc1}\\
&& \vee  \mI (\sim e_1 \wedge \sim e_2) \wedge \mII (\sim e_1 \wedge e_2)
	\label{cc2}\\
&& \vee  \mI (\sim e_1 \wedge e_2) \wedge \mII (\sim e_1 \wedge \sim e_2)
	\label{cc3}\\
&&\vee  \mI (\sim e_1 \wedge \sim e_2) \wedge \mII (\sim e_1 \wedge \sim e_2)
	\label{cc4}
\end{eqnarray}

\noindent
In case~(\ref{cc1}) an agreement is reached that $e_2$ is the
leader. In case~(\ref{cc4}) the participants agree that no leader has
been elected. The other two cases are symmetric: 
we consider case~(\ref{cc2}) and show that it leads to a state where
no leader has been elected. We first show that a state
is reached where participant $2$ agrees that he cannot be the leader:

\begin{center}
{\small
\noindent
	\prooftree
		\prooftree
		\mI \sim e_2 \because \mII h
		\justifies 
		\mI (\sim e_1 \wedge \sim e_2) \because \mII h
		\using{\mbox{\small BSW}}
		\endprooftree
		  \mII h \causes  \mII \sim e_2 
		\quad \quad \stable  \mII \sim e_2 \quad \quad \mbox{ }
	\justifies 
\prooftree
	\mI (\sim e_1 \wedge \sim e_2) \wedge \mII  \top \causes \mII \sim e_2 
\justifies 
 \mI (\sim e_1 \wedge \sim e_2) \wedge \mII (\sim e_1 \wedge e_2) 
\causes \mII \sim e_2 
	\using{\mbox{\small LSW}}
\endprooftree  
	\using{\mbox{\small Notif}}
	\endprooftree
}
\end{center}

\

\noindent
where the last step ({\small LSW}) exploits the following implication:

\begin{center}
{\small
\noindent
	\prooftree
	\sim e_1 \wedge e_2 \raw \top
	\justifies 
	 \mII (\sim e_1 \wedge e_2) 
	\raw 
	 \mII  \top
	\using{\mbox{\small Nec,D3}}
	\endprooftree
}
\end{center}

\noindent
We carry on some calculation:

\begin{center}
{\small
\noindent
\prooftree
  \prooftree
	\mI (\sim e_1 \wedge \sim e_2) \wedge \mII (\sim e_1 \wedge e_2) 
   \causes \mII \sim e_2  
   \justifies 
   \mI (\sim e_1 \wedge \sim e_2) \wedge \mII (\sim e_1 \wedge e_2) 
   \causes \mI (\sim e_1 \wedge \sim e_2) \wedge \mII (\sim e_1 \wedge e_2) 
              \wedge \mII \sim e_2
\using{\small LCC ({\tiny F\causes F})}
\endprooftree  
\justifies 
   \mI (\sim e_1 \wedge \sim e_2) \wedge \mII (\sim e_1 \wedge e_2) 
   \causes \mI (\sim e_1 \wedge \sim e_2) \wedge \mII \sim e_1 
  \wedge \mII \sim e_2
\using{\small D2, LSW (twice)}
\endprooftree  
}
\end{center}

\

\noindent
We now apply {\small \bf Conf} and conclude:

\

$\mI (\sim e_1 \wedge \sim e_2) \wedge \mII (\sim e_1 \wedge e_2) 
\causes \mI (\sim e_1 \wedge \sim e_2) \wedge \mII (\sim e_1 \wedge \sim e_2) 
$

\

\noindent
{\bf Proof of} $(\ref{ee}) \causes$ no leader elected

\

\noindent
We apply the proof schema above ({\small \bf
Notif} and then {\small \bf Conf}) twice and conclude.


\section{Discussion and Related Work}
\label{discussion}

{\bf The semantic domain of DSL.} The choice of $2^S$ as a semantic
domain of the distributed state logic formulae, and the
non--equivalence between $\mm\, (F \wedge F)$ and $ \mm \, F \wedge
\mm\, F'$ are useful to specify that a given condition can have
different future effects, without constraining them to occur in the
same state. Similarly, we can express complex preconditions in a
temporal formula. For instance, assume we want to specify and reason
on the delivery of credit cards to customers. The bank, for security
reasons, sends the card and the code separately.  Once the customer
has got both of them, he is allowed to withdraw money from an ATM
machine:
\begin{eqnarray}
&& \bank\, new\_card \; \causes \; \user\, receive\_card \wedge 
	\user \,receive\_code 
	\label{bank1}\\
 &&  \user\,
	can\_withdraw \; \because   \; \user \, receive\_card \wedge 
	\user\, receive\_code
\label{bank2}
\end{eqnarray}

\noindent
The equivalence between $\mm\, (F \wedge F)$ and $ \mm\, F \wedge \mm\, F'$
would have required the following specification, somewhat less intuitive: 
\begin{eqnarray}
&& \bank\, new\_card \;\causes \; \user \,receive\_card
	\label{banku1}\\
&& \bank\, new\_card \;\causes \; \user \,receive\_code 
	\label{banku2}\\
&& \user\, can\_withdraw  \; \because   \; \user \, receive\_card 
	\label{banku3}\\
&& \user\, can\_withdraw  \; \because   \; \user\, receive\_code
	\quad	\quad \mbox{ }\quad\quad \mbox{ }\quad\quad\quad \mbox{ }
	\label{banku4}
\end{eqnarray}

\noindent
since~(\ref{bank1}),~(\ref{bank2}) would be too restrictive, asking
for card and code to be received at the same time.

Last, but not least, with an eye to a $1^{st}$ order extension, a
formula like~(\ref{bank1}) makes it easier to bind variables in $card$
and $code$ than with the unrelated
formulae~(\ref{banku1}),~(\ref{banku2}).

\

\noindent
{\bf Classical Logic.} Another point of discussion is why we
need a modality ($\mm$) rather than a distinguished propositional
symbol $here_m$, to replace systematically each sub--formula $\mm\, F$
with $here_m \wedge F$. One motivation is that we do not want the
equivalence between $\mm\, (F \wedge F'$) and $ \mm\, F \wedge \mm\,
F'$, as discussed previously. On the contrary, the two translations
$here_m \wedge F\wedge F'$ and $here_m \wedge F\wedge here_m \wedge
F'$ would be equivalent. 

More importantly, $(\mm F \wedge \mn F') \causes \moo G$ would be
translated in a formula with a false premise, namely $(here_m \wedge F
\wedge here_n \wedge F') \causes (here_o \wedge G)$. 

\

\noindent
{\bf Hybrid Logic.}  Hybrid logic allows the specifier to directly
refer to specific points (states) in the model, through the use of
{\em nominals}~\cite{tesiAreces}.  A nominal $i$ is an atom which is
true at exactly one point in any model.  The operator $@_i$ permits to
jump to the point named by nominal $i$.  We might consider defining an
hybrid signature including distinguished sets of state variables, one
for each component, and translate $\mm\, F$ in $\exists x.\; @_{x}F$,
where $x$ is a state variable in the appropriate set. Likely, the
resulting setting would be more complex than that offered by DSTL.

\

\noindent
{\bf Metric and Layered Temporal Logic.}  Some similarities can be
found between our location operator and the $MLTL$ operators defined
in~\cite{tesiAmontanari}, that make it possible to compose formulae
associated with different time granularities and to switch from one
granularity to another.  Time instants are organized in temporal
domains, and the set of temporal domains is totally ordered with
respect to the coarseness of the domain elements. To look for an
embedding of DSTL, we can consider three domains: {\em system}, with a
unique element; {\em components}, whose elements are the components
$m_1, \ldots, m_k$; {\em states}, the domain of the states. Then the
formulae are translated using an appropriate combination of $MLTL$
operators. For instance, the translation of $\mm\, F$ should be
$\diamond \Delta_m^{components} \diamond \exists \alpha
\Delta_\alpha^{states} F$. Since the full expressive power of $MLTL$
is likely not needed, the simpler framework of DSTL is of pragmatical
interest.

\

\noindent
{\bf  Other logics for distributed systems.}  Various extensions
of temporal logic have been defined in the literature to deal with
distributed
systems. 

 In TTL~\cite{masini92}, for each local state of the system, a {\em
 visibility} function specifies which remote information is
 accessible. The visibility function is defined on the basis of a
 relation among states which is {\em symmetric} in the case of states
 belonging to distinguished components.

 A trace based extension of linear time temporal logic, called {\em
 TrPTL}, has been defined in ~\cite{Thiaga94} (see also
 ~\cite{Thiaga98DVL}). The logic has been designed to be interpreted
 over infinite traces, i.e., labelled partial orders of actions, which
 respect some dependence relations associated to the alphabet of
 actions.  

 In~\cite{lprm95}, a temporal logic, StepTL, is defined and interpreted
 over multistep transition systems. These are a well known extension of
 transition systems, permitting to describe as concurrent the steps of
 computation that can actually be executed in parallel.  A multistep
 transition system thus contains transitions of the form $s\,A\,s'$,
 where $A$ is a set of actions, instead of a single one.

 Three distinguished logics are presented in~\cite{Rama96} to describe
 systems composed of sets of communicating {\em agents}. The logics
 differentiate on the amount of information each agent can have on the
 other agents running on the system, but share a common setting: agents
 communicate via common actions.  The models for these logics are runs
 of networks of synchronizing automata.  The logics {\bf D}$_0$ and
 {\bf D}$_1$ presented in~\cite{EhrichDalLibro} are based on a similar
 approach. 

In all these proposals, components communicate via some form of
synchronization, and logic formulae are interpreted on models shaping:

\vspace{0.2cm}

\hspace{1cm}\xymatrix@R-=9pt{
{(m)}&\ar@{.>}[rr]&& {a} \ar@{.>}[r] & {b} \ar@{.>}[rr] && {c} \ar@{.>}[r] &\\
{(n)}&\ar@{.>}[r]& {d} \ar@{.>}[rr] && {e} \ar@{<->}[u] \ar@{.>}[r] &
{f} \ar@{.>}[rr] &&
}

\

\noindent
Therefore, in any logic defined over these models, it is not possible
to express the asymmetric nature of causality we are interested in when
modelling the behaviour of agents communicating asynchronously by
message passing. Indeed, in the previous model we can both assert that
$a \causes f$ and that $d \causes c$.

A logic closer to DSTL is proposed in~\cite{LodRamThi92Corto}, where a
        branching time temporal logic for asynchronously communicating
        sequential agents (ACSAs) is defined. ACSAs communicate
        asynchronously via message passing. The logic contains
        temporal modalities indexed with a local point of view of one
        agent and allows an agent ``i'' to refer to local properties
        of another agent ``j'' according to the latest message
        received: an agent can gain information about another agent by
        receiving messages but not by sending them.  We allow agents
        to make remote future assertions: therefore it is easier to
        express global liveness properties.

\

\noindent
{\bf  Knowledge Logic.} 
A logic to reason on asynchronous message passing systems is proposed
in~\cite{fhmv95}. The language used, ${\cal L}^U _n$, is obtained by
extending their language of knowledge with the modal operators $U$ and
$\nex$. Formulae in ${\cal L}^U _n$ permit to express how the $n$ {\em
agents} in a system gain knowledge over time.
A set of characteristic formulae valid in the logic are presented, but
a sound and complete axiom system is not defined.
The authors focus their attention on systems based un-reliable
communications, while only state that properties of reliable
communications can be expressed. 
A major difference with our work relies on the models used to
interpret ${\cal L}^U _n$ formulae. Even if the knowledge of the
agents is limited to their current local histories, i.e. sequences of
messages sent or received and of internal actions, interpretation
structures are based on global time and state.

\

\noindent
{\bf Partial Order Temporal Logics.}
Partial Order Temporal Logics (POTL)~\cite{PiWo} permits to deal with
the causal relationships between the events of a set of processes
executing concurrently. The Interleaving Set Temporal Logic
(ISTL)~\cite{LogKatz} extends POTL with features form linear temporal
logic and branching temporal logic. The Kripke structures for both
logics are very different from the one defined in this paper.

We are addressing a specific class of systems that we consider very
relevant nowadays, that is distributed systems with asynchronous
message passing. These systems have a few notable characteristics:
there is no global state, and interactions among components occur only
via messages. As a consequence, a specification is essentially devoted
to describing the causal relationships among the components. We think
that these characteristics are so important that the designer
working on a specification will greatly benefit if they are naturally
embedded in the basic model he is using. Hence, the
investigation in Kripke's structures presented in this paper.

\

\noindent
{\bf Logics for Mobile Systems.} 
 Often mobile systems are specified using a process calculus with
 primitives for mobility, and some logics have been defined, tailored
 for these calculi. This is the case, for instance, of the Ambient
 Logic~\cite{CG00:logic-for-MA}, studied for the Ambient
 Calculus~\cite{CG00:mobile-ambients}, the logic for Klaim~\cite{loretiTOCL},
 and the Spatial Logic for Concurrency~\cite{CardCair1e2}, whose
 underlying computational model is the asynchronous
 $\pi$--calculus. These logics include modalities for describing the
 evolution over time and the location of the system processes. They
 are inspired by the Hennessy--Milner logic: they are conceived
 for model checking rather than for specifying and reasoning on the system
 properties.

In particular, the Spatial Logic for Concurrency can express
properties of freshness, secrecy, structure, and behavior of
concurrent systems.  Spatial operations correspond to composition,
local name restriction, and a primitive fresh name quantifier.  The
logical treatment of the notion of freshness can prove useful in
extending DSTL to reason on the dynamic creation of components.

A linear--time logic for specifying mobile systems is
MTLA~\cite{merz03}, which extends Lamport's Temporal Logic of Actions
with spatial modalities to deal with mobile systems. The main
difference with DSTL is that a synchronous computational model is
assumed.

\

\noindent
{\bf Oikos--{\em adtl}.}  The work reported here stems from our
experience with Oikos--{\em adtl}, a specification language for
distributed systems based on asynchronous communications, designed to
support the composition of specifications~\cite{coord99c}.  \oikosadtl
is intended to give designers a language to express the properties of
interest in a natural way, and it is associated with a refinement
method which supports the gradual introduction of details, as design
proceeds.  It has been used to specify software architectures and
patterns~\cite{isawCorto} and to analyse security issues in
mobile systems~\cite{wits00Corto,adhoc,denver}. It is supported by a
proof assistant, Mark~\cite{mark}, that deploys a number of proof
strategies that partially automate property verification.

Coming back to our motivating example in the introduction, in
Oikos--{\em adtl} it is possible to weaken the consequences of a
formula like~(\ref{ex:intro}) including operator $\tcauses$, but the
rule shapes

\

\prooftree
\mm  \,p \; \causes \; \mn  \,q \wedge \moo \, r 
\justifies
\mm  \,p \; \causes \; \mn  \,q  
\endprooftree

\

\noindent
since a formula like~(\ref{ex:intro2}) is not part of the
 logic. So, the price is writing one rule for each possible weakening
 relation.


\section{Conclusions}


To reason on global applications, we have introduced the temporal
logic DSTL. Models for DSTL are space--time diagrams describing the
behaviour of a set of components communicating asynchronously. The
logic has been introduced in two steps.  First, we have defined DSL, a
modal logic for localities that embeds the theories describing the
local states of each component into a theory of the distributed states
of the system. No notion of time or state transition is present at
this stage. To support reasoning in the logic, we have presented a
sound and complete axiom system. Then, we have added the temporal
operators, and the corresponding derivation rules. The contribution is
that it is possible to reason about properties that involve several
components, even in the absence of a global clock, which is a
meaningless notion in an asynchronous setting. The logic has been used
to reason on the properties of a simple secure communication system
and on an algorithm for the leader election.

Future work includes the extension of DSTL to predicate logic, the
introduction of an {\em event} operator, the study of compositionality
results, and a revision of the theorem prover Mark.  We foresee that
formulae in the 1$^{st}$ order extension will shape $\mm\,p(x)\causes
\mn\, q(x,y)$, and be interpreted as $\;\forall x. [\mm\,p(x)\causes
\exists y. \,\mn \, q(x,y)]$. This way, the semantics should
smoothly extend that of DSTL.  Compositionality results will permit to
derive the properties satisfied by a system from the properties
satisfied by its components when executed in isolation.  This requires
reasoning on the possible interferences due to communications from the
added components.


\newpage 

\appendix

\noindent
{\Large \bf Appendix}

\

\noindent
{\large \bf Proofs from Section~\ref{dslaxsys}}

\

\noindent
{\bf Axiom 4}

\

{\small
\prooftree
\bar{\mm} (\bar{\mm}  F \law   F)
\justifies
\bar{\mm}\bar{\mm}   F \law \bar{\mm}  F
\using{K}
\endprooftree
}

\

\

\noindent
{\bf D1}

\

{\small
\prooftree
\prooftree
\prooftree
\prooftree
\bar{\mm} (\bar{\mm} \sim F \lraw  \sim F)
\justifies
\bar{\mm}\bar{\mm}  \sim F \lraw \bar{\mm}  \sim F
\using{K}
\endprooftree
\justifies
\sim \mm  \sim \sim \mm  \sim  \sim F \lraw \sim \mm  \sim  \sim F
\endprooftree
\justifies
\sim \mm   \mm   F \lraw \sim \mm   F
\endprooftree
\justifies
\mm F \ \lraw \ \mm\mm F
\endprooftree
}

\

\

\noindent
{\bf D2} 
We show that  $\mm  (F\wedge F') \: \raw \:  \mm  F$,  $\mm  (F\wedge F')
\: \raw \:  \mm  F'$ is proved analogously. 

\

{\small
\prooftree
\prooftree
\prooftree
\prooftree
\prooftree
F\wedge F' \: \raw \: F
\justifies
\sim (F\wedge F') \: \law \: \sim  F
\endprooftree
\justifies
\bar{\mm} (\sim (F\wedge F') \: \law \: \sim  F)
\using{Nec}
\endprooftree
\justifies
\bar{\mm} \sim (F\wedge F') \: \law \: \bar{\mm}  \sim  F
\using{K}
\endprooftree
\justifies
\sim \mm \sim \sim (F\wedge F') \: \law \: \sim \mm  \sim \sim  F
\endprooftree
\justifies
\mm  (F\wedge F') \: \raw \:  \mm  F
\endprooftree
}

\

\

\noindent
{\bf D3}

\

{\small
\prooftree 
\prooftree 
\prooftree 
\prooftree 
\bar{\mm} (F\raw F') \quad
\prooftree 
(F\raw F') \raw (\sim F' \raw \sim F)
\justifies 
\bar{\mm} (F\raw F') \raw \bar{\mm}(\sim F' \raw \sim F)
\using{\mbox{\small Nec,  K}} 
\endprooftree 
\justifies 
 \bar{\mm}(\sim F' \raw \sim F)
\using{\mbox{\small MP}} 
\endprooftree 
\justifies 
 \bar{\mm}\sim F' \raw \bar{\mm}\sim F
\using{\mbox{\small  K}} 
\endprooftree 
\justifies 
 \sim \mm F' \raw \sim \mm F
\using{\mbox{\small  Def}} 
\endprooftree 
\justifies 
  \mm F \raw  \mm F'
\using{\mbox{\small  PC}} 
\endprooftree 
}

\

\

\noindent
{\bf D5}

\

{\small
\prooftree
\prooftree
\bar{\mm} (\bar{\mm} \sim F \lraw \sim F)
\justifies
\bar{\mm} (\sim {\mm} \sim \sim F \lraw \sim F)
\endprooftree
\justifies 
\bar{\mm} ( {\mm}  F \lraw  F)
\endprooftree
}

\newpage

\noindent
{\bf D6} For $A=F\raw F'$ , $B=F'\raw F''$, and $C=F\raw F''$

\

{\small
\prooftree
\prooftree
\prooftree
\prooftree
\prooftree
\prooftree
\prooftree
A\wedge B \raw C
\justifies
\sim A \vee \sim B \vee C
\endprooftree
\justifies
 A \raw (B \raw C)
\endprooftree
\justifies
\bar{\mm} A \raw \bar{\mm} (B \raw C)
\using{Nec,K}
\endprooftree
\justifies
\bar{\mm} A \raw  (\bar{\mm}B \raw \bar{\mm}C)
\using{K}
\endprooftree
\justifies
\sim \bar{\mm} A \vee \sim \bar{\mm}B \vee  \bar{\mm}C
\endprooftree
\justifies
\sim (\bar{\mm} A \wedge \bar{\mm}B) \vee  \bar{\mm}C
\endprooftree
\justifies
(\bar{\mm} A \wedge \bar{\mm}B) \raw  \bar{\mm}C
\endprooftree
}

\

\

\noindent
{\bf D7}  We prove  $\mm(F\vee F' )\raw (\mm F \vee \mm F')$ on the left, 
and $\mm(F\vee F' )\law (\mm F \vee \mm F')$ on the right.

\

\

\noindent
{\small
\prooftree
\prooftree
\prooftree
\prooftree
\prooftree
\prooftree
\prooftree
\prooftree
\prooftree
\prooftree
\prooftree
(A\wedge B) \raw (A\wedge B)
\justifies
\sim A \vee \sim B \vee (A\wedge B)
\endprooftree
\justifies
 A \raw (B \raw (A\wedge B))
\endprooftree
\justifies
\bar{\mm} A \raw \bar{\mm} (B \raw (A\wedge B))
\using{Nec,K}
\endprooftree
\justifies
\bar{\mm} A \raw  (\bar{\mm}B \raw \bar{\mm}(A\wedge B))
\using{K}
\endprooftree
\justifies
\sim \bar{\mm} A \vee \sim \bar{\mm}B \vee  \bar{\mm}(A\wedge B)
\endprooftree
\justifies
\sim (\bar{\mm} A \wedge \bar{\mm}B) \vee  \bar{\mm}(A\wedge B)
\endprooftree
\justifies
(\bar{\mm} A \wedge \bar{\mm}B) \raw  \bar{\mm}(A\wedge B)
\endprooftree
\justifies
(\sim \mm \sim A \wedge \sim \mm \sim B) \raw  \sim \mm \sim (A\wedge B)
\endprooftree
\justifies
\sim (\mm \sim A \vee \mm \sim B) \raw  \sim \mm (\sim A \vee \sim B)
\endprooftree
\justifies
(\mm \sim A \vee \mm \sim B) \law   \mm (\sim A \vee \sim B)
\endprooftree
\justifies
\mm(F\vee F' )\raw (\mm F \vee \mm F')
\using{\begin{array}{l}F\equiv \sim A\\ F'\equiv \sim B\end{array}}
\endprooftree
}
\hspace{-1cm}
{\small
\prooftree
\prooftree
\prooftree
F \raw (F\vee F')
\justifies
\bar{\mm} (F \raw (F\vee F'))
\using{Nec}
\endprooftree
D3
\justifies
\mm F \raw \mm(F\vee F')
\using{MP}
\endprooftree
\prooftree
\prooftree
F' \raw (F\vee F')
\justifies
\bar{\mm} (F' \raw (F\vee F'))
\using{Nec}
\endprooftree
 D3
\justifies
\mm F' \raw \mm(F\vee F')
\using{MP}
\endprooftree
\justifies
(\mm F \vee \mm F') \raw \mm(F\vee F')
\endprooftree
}

\

\

\noindent
{\bf  D8}
If we prove  $\bar{\mm} ((\mm F\wedge \mm F')\raw (F\wedge F'))$ 
and 
$\bar{\mm} ((F\wedge F')\raw \mm (F\wedge F'))$ 
then we can apply D6 and conclude.
The second formula  is an instance of D5, we prove the first one:

\

{\small
\prooftree
(\mm F\raw F \; \wedge \; \mm F'\raw F') \raw (\mm F\wedge \mm F'  \; \raw \; F\wedge F')
\justifies
\bar{\mm} (\mm F\raw F \; \wedge \; \mm F'\raw F') \raw  \bar{\mm}  (\mm F\wedge \mm F'  \; \raw \; F\wedge F')
\using{Nec,K}
\endprooftree
}

\

\

{\small
\prooftree
\prooftree
\prooftree
D5
\justifies
\bar{\mm} (\mm F\raw F) \; \wedge \; \bar{\mm}(\mm F'\raw F')
\endprooftree
\quad \prooftree 
	D7
	\justifies
	(\bar{\mm}F \wedge \bar{\mm}F') \raw \bar{\mm}(F\wedge F')
	\endprooftree
\justifies
\bar{\mm} (\mm F\raw F \; \wedge \; \mm F'\raw F')
\using{MP}
\endprooftree
\justifies
\bar{\mm} (\mm F\wedge \mm F' \; \raw F \wedge F')
\using{MP, \mbox{with the conclusions of the previous derivation} }
\endprooftree
}

\newpage

\noindent
{\large \bf Proof of the Notification Rule}

\

\begin{center}
{\small
\noindent
\prooftree
F \because G \quad G \causes \mm G^{\prime} \quad \stable \mm G^{\prime}
\justifies
F \wedge \mm \top \causes \mm G^{\prime}
\using{\bf Notif}
\endprooftree 
}
\end{center}

\

\noindent
Let $ds$ be a distributed state satisfying $F\wedge \mm\top$, we have
that:

\vspace{-0.6cm}

\begin{eqnarray*}
ds \models F\wedge \mm\top & \Raw & ds \models F\\
 & \Raw & \exists ds^{\prime} \leq  ds. \  ds^{\prime}\models G
				\quad \quad\quad \quad (since F \because G)\\
 & \Raw & \exists ds^{\prime\prime} \geq ds^{\prime}. \
	ds^{\prime\prime}\models \mm G^{\prime} \;\quad \quad (since G
	\causes \mm G^{\prime})
\end{eqnarray*}

\noindent
Summing up, $\forall ds\models F \ \exists ds^{\prime\prime}\models
\mm G^{\prime}$. 

Now, $ds^{\prime\prime}\models \mm G^{\prime}$ implies that $ \exists
s \in ds^{\prime\prime} \cap S_m$ with $\{s\} \models \mm G^{\prime}$.
Stability of $\mm G^{\prime}$ guarantees that for any state
$s^{\prime}\in S_m$ that follows $s$, $\{s^{\prime}\} \models \mm
G^{\prime}$. So, we can build a distributed state which follows any
$ds$ satisfying $F\wedge \mm\top$ and satisfies $\mm G^{\prime}$.

\

\

\noindent
{\large \bf Proof of the Confluence Rule}

\begin{center}
{\small
\noindent
\prooftree
\stable  \mm F \quad \stable  \mm F^{\prime} 
\justifies
\mm F\wedge \mm F^{\prime}  \ \raw \ \mm (F\wedge F^{\prime})
\using{\bf Conf}
\endprooftree 
}
\end{center}

\

\noindent
Let $ds$ be a distributed state satisfying $\mm F\wedge \mm F^{\prime}$:

\vspace{-0.6cm}

\begin{eqnarray*}
ds \models \mm F\wedge \mm F^{\prime} & \LRaw & 
	ds \models \mm F \mbox{ and } ds \models \mm F^{\prime}\\
 & \LRaw & \exists s \in ds \cap S_m. \ \{s\} \models  F   \mbox{ and }
   \exists   s^{\prime}\in ds \cap S_m. \ \{s^{\prime}\} \models  F^{\prime}
\end{eqnarray*}

\noindent
Let $\{s\} \geq \{s^{\prime}\}$ (the case $\{s\} \leq \{s^{\prime}\}$ is
symmetric), for the stability of $F^{\prime}$ we have that also
$\{s\}$ satisfies $F^{\prime}$:

\vspace{-0.6cm}

\begin{eqnarray*}
\{s\} \models F   \mbox{ and } \{s\} \models F^{\prime} & \LRaw & 
	\{s\} \models F \wedge  F^{\prime} \\
 & \LRaw & ds \models \mm (F\wedge F^{\prime})
\end{eqnarray*}

\

\noindent
{\large \bf Proof of Cor1 and Cor2}

\

\begin{center}
{\small
\noindent
\prooftree
F\causes G\vee  G^{\prime} \quad 
  \prooftree
	\prooftree
	G\causes F^{\prime} \quad F^{\prime} \raw F^{\prime} \vee G^{\prime}
	\justifies
	G\causes F^{\prime} \vee G^{\prime}
	\using{\bf LSW}
	\endprooftree
	\quad 
	\prooftree
	G^{\prime} \causes G^{\prime} \quad 
		G^{\prime} \raw F^{\prime} \vee G^{\prime}
	\justifies
	G^{\prime} \causes F^{\prime} \vee G^{\prime}
	\using{\bf LSW}
	\endprooftree
  \justifies
  G \vee G^{\prime} \causes F^{\prime} \vee G^{\prime}
   \using{\bf LPD}
  \endprooftree
\justifies
F\causes F^{\prime} \vee G^{\prime}
\using{\bf LTR}
\endprooftree
}
\end{center}

\

\begin{center}
{\small
\noindent
\prooftree
  \prooftree
	\prooftree
	F \causes G \wedge G^{\prime} \quad  G \wedge G^{\prime}\raw G
	\justifies
	F \causes G
	\using{\bf LSW}
	\endprooftree
	\quad 
	G \causes F^{\prime} 
  \justifies
  F \causes  F^{\prime} 
   \using{\bf LTR}
  \endprooftree
\quad 
  \prooftree
  F \causes G \wedge G^{\prime} \quad  G \wedge G^{\prime}\raw G^{\prime}
   \justifies
  F \causes  G^{\prime}
  \using{\bf LSW}
   \endprooftree
\justifies
F\causes F^{\prime} \wedge G^{\prime}
\using{\bf LCC}
\endprooftree
}
\end{center}

\end{document}